\documentclass[11pt]{article}

\usepackage[final]{acl}
\usepackage{float}

\usepackage{times}
\usepackage{latexsym}
\usepackage{makecell}
\usepackage{tablefootnote}

\usepackage[table]{xcolor}
\usepackage{booktabs}
\usepackage{multirow}
\usepackage{graphicx}
\usepackage[T1]{fontenc}
\usepackage[utf8]{inputenc}

\usepackage{microtype}

\usepackage{inconsolata}

\usepackage{graphicx}

\usepackage{amsmath,amsfonts,bm}

\def\eqref#1{equation~\ref{#1}}
\def\1{\bm{1}}

\DeclareMathAlphabet{\mathsfit}{\encodingdefault}{\sfdefault}{m}{sl}
\SetMathAlphabet{\mathsfit}{bold}{\encodingdefault}{\sfdefault}{bx}{n}

\usepackage{hyperref}
\usepackage{url}
\usepackage[most]{tcolorbox}
\usepackage[x11names]{xcolor}
\usepackage{mdframed}
\usepackage{graphicx}
\usepackage{enumitem}
\usepackage[normalem]{ulem}
\usepackage{tabularx} % 导言区
\usepackage{booktabs} % 更好看的表格线

\usepackage{easyReview}
\usepackage{siunitx} 
\usepackage{multirow}
\usepackage{amsmath}
\usepackage{pifont}
\usepackage[table]{xcolor}
\usepackage{subcaption}
\usepackage{caption}  % 提供了 \captionof 命令
\usepackage{wrapfig}
\usepackage{hyperref}
\usepackage{longtable}
\usepackage{listings}
\usepackage{adjustbox}

\usepackage{tabularx} 
\usepackage{booktabs}
\usepackage{arydshln}
\usepackage{colortbl}

\usepackage{listings}
\usepackage{xcolor}
\usepackage[normalem]{ulem}
\usepackage{algpseudocode}
\usepackage{needspace}

\definecolor{darkblue}{rgb}{0, 0, 0.5}
\definecolor{deepred}{rgb}{1, 0.5, 0.5}
\definecolor{medred}{rgb}{1, 0.7, 0.7}
\definecolor{lightred}{rgb}{1, 0.9, 0.9}
\definecolor{deepgreen}{rgb}{0.5, 0.9, 0.5}
\definecolor{medgreen}{rgb}{0.7, 0.9, 0.7}
\definecolor{lightgreen}{rgb}{0.9, 1, 0.9}
\definecolor{lightgrey}{rgb}{0.937, 0.937, 0.937}

\usepackage[linesnumbered,ruled,vlined]{algorithm2e}
\newcommand{\cmark}{\color{green}\ding{51}}%
\newcommand{\xmark}{\color{red}\ding{55}}%

\title{E2EDev: Benchmarking Large Language Models in End-to-End Software Development Task}
\author{
Jingyao Liu$^{1,4}$\quad  Chen Huang$^{2}$\thanks{Corresponding author.} \quad Zhizhao Guan$^{1,4}$\quad 
\textbf{Wenqiang Lei}$^{1,4}$\quad \textbf{Yang Deng}$^{3}$ \\
$^{1}$ College of Computer Science, Sichuan University  \\
$^{2}$ Institute of Data Science, National University of Singapore  \\
$^{3}$ CHAT NLP Group, Singapore Management University  \\
$^{4}$ Engineering Research Center of Machine Learning and Industry Intelligence,\\
Ministry of Education, China  \\
\texttt{liujingyao1@stu.scu.edu.cn, huang\_chen@nus.edu.sg}
}

\begin{document}
\maketitle
\begin{abstract}
The rapid advancement in large language models (LLMs) has demonstrated significant potential in End-to-End Software Development (E2ESD). However, existing E2ESD benchmarks are limited by coarse-grained requirement specifications and unreliable evaluation protocols, hindering a true understanding of current framework capabilities. To address these limitations, we present E2EDev, a novel benchmark grounded in the principles of Behavior-Driven Development (BDD) to assess whether the generated software meets user needs through mimicking real user interactions. 
E2EDev comprises (i) {a fine-grained set of user requirements for each target software project} (ii) multiple BDD test scenarios with corresponding Python step implementations for each requirement, and (iii) a fully automated testing pipeline built on the Behave framework.
By evaluating various E2ESD frameworks and LLM backbones with E2EDev, our analysis reveals a persistent struggle to effectively solve these tasks, underscoring the critical need for more effective and cost-efficient E2ESD solutions. Our codebase and benchmark are available at \url{https://github.com/SCUNLP/E2EDev}.

\end{abstract}

\section{Introduction}

The effectiveness of large language models (LLMs) in understanding user needs and demonstrating logical reasoning has been validated in code generation \citep{jimenez2024swe,gao2024current, zhang2025pretraincode}.
LLMs can generate code snippets based on provided code context and user query \citep{yu2024codereval,chen2021evaluating,hui2024qwen2}, and debug code through feedback from either human or system \citep{zhang2025eyemulator,zhong2024debug,yang-etal-2025-elaboration,li2022automating,yang2024swe}. The success in generating isolated functions is fueling further research into advanced software development automation. This has led to a shift from function-level code synthesis to \textbf{End-to-End Software Development (E2ESD)}, where complete software is automatically generated from user requirements (e.g., \textit{`generate an HTML game of Angry Birds'}). By this means, E2ESD could enhance both the code effectiveness and efficiency by removing the manual assembly bottleneck. Current LLM-based E2ESD frameworks can be broadly categorized into multi-agent approaches, inspired by traditional software engineering principles ~\citep{dong2024self,hong2024metagpt,du2024multi,royce1987managing,qian2024chatdev}, and single-agent approaches, where one LLM manages the complete development pipeline~\citep{gptengineer,cognition2024devin}.

\begin{figure*}[t]
  \centering
  \includegraphics[width=1.0\textwidth]{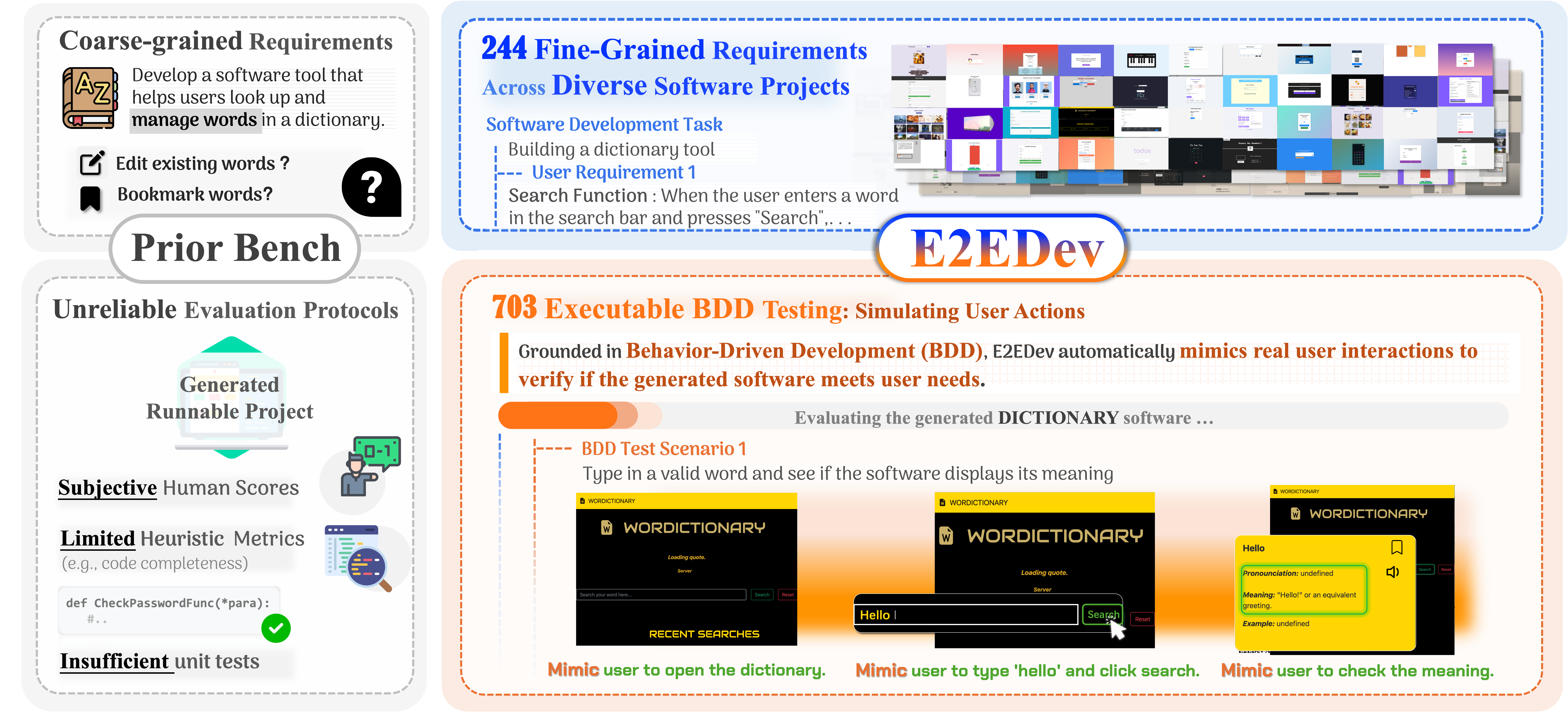}
  \caption{Comparing existing work with ours, existing benchmarks use coarse-grained requirements with unreliable evaluations whereas E2EDev provides fine-grained requirements with executable BDD tests.}
  \label{fig:intro}
  % \vspace{-3mm}
\end{figure*}

As LLM-driven approaches advance, there is growing demand for robust benchmarks to automate the evaluation of E2ESD frameworks \citep{hong2024metagpt,qian2024chatdev,he2024llm,huself}. However, existing E2ESD benchmarks, such as SoftwareDev \citep{hong2024metagpt} and SRDD \citep{qian2024chatdev}, suffer from two critical limitations. (1) \textbf{Coarse-grained Requirement Specifications}: Current benchmarks substitute user requirements with ambiguous software descriptions as input, making it difficult to verify whether the generated software aligns with actual user needs. As shown in Figure~\ref{fig:intro} (left), the vague requirement ``manage words'' for a dictionary tool can refer to various actions, such as editing existing entries or bookmarking words. Without precise operational specifications of such requirements, systematic testing becomes impractical. (2) \textbf{Unreliable Evaluation Protocols}: Evaluations in these benchmarks rely heavily on labor-intensive human assessments \citep{hong2024metagpt} and lack standardized evaluation methodologies grounded in established software engineering principles \citep{qian2024chatdev}. This results in inconsistent and unreliable performance comparisons across frameworks. 

To this end, we introduce the \textbf{E2EDev} benchmark for evaluating the performance of LLM-based frameworks on E2ESD tasks, along with a human-in-the-loop annotation framework designed to ease the annotation burden. Specifically, E2EDev is derived from real-world open-source web application projects. {Following the principles of Behavior-Driven Development (BDD), which specifies and validates software behavior from a user perspective, E2EDev evaluates whether the generated software meets user requirements by mimicking real user interactions, as illustrated in Figure~\ref{fig:intro} (right). E2EDev consists of: (1) a fine-grained list of user requirements for each software, {(2) for each user requirement, we provide multiple BDD test scenarios to verify its correctness, each corresponding to an executable Python code implementation, and (3) an automated testing pipeline built on Behave}\footnote{Behave is a Python framework for behavior-driven development using natural language tests (Gherkin). See \url{https://github.com/behave/behave} for details.}.
 To alleviate the annotation burden while ensuring data quality, we propose the Human-in-the-Loop Multi-Agent Annotation Framework (HITL-MAA). In this framework, specialized agents analyze the project source code to generate candidate requirements and {executable tests}, with human supervisors involved at key points to avoid errors. 

% {\yao{To this end, we introduce \textbf{E2EDev}\add{[Maybe change this name, too]}, a benchmark for evaluating LLM-based frameworks on R2E-REP tasks. E2EDev is derived from real-world open-source web application projects, serving as a practical starting point for R2E-REP evaluation.} \yao{And following Behavior-Driven Development (BDD), \textbf{E2EDev automatically evaluates generated software by executing predefined tests that mimic real user interactions {for requirement verification}}}, as illustrated in Figure~\ref{fig:intro} (right). E2EDev consists of: (1) a fine-grained list of user requirements for each software, {(2) for each user requirement, we provide multiple BDD test scenarios to verify its correctness, each corresponding \yao{with} an executable Python code implementation, and (3) an automated testing pipeline built on Behave}\footnote{Behave is a Python framework for behavior-driven development using natural language tests (Gherkin). See \url{https://github.com/behave/behave} for details.}. To alleviate the annotation burden while ensuring data quality, we propose the Human-in-the-Loop Multi-Agent Annotation Framework (HITL-MAA). In this framework, specialized agents analyze the project source code to generate candidate requirements and {executable tests}, with human supervisors involved at key points to avoid errors. 

{Our evaluation of various E2ESD frameworks and LLM backbones reveals a significant struggle to effectively solve E2ESD tasks. Even the Claude series, among the strongest coding-oriented LLMs, fails to exceed 60\% performance. These failures stem from the difficulty of simultaneously performing high-level planning (i.e., deciding what to build) and faithfully following fine-grained functional details.
In multi-agent staged pipelines, this challenge is often addressed by explicitly separating
planning from detailed implementation across different phases. However, the effectiveness of such designs critically depends on how intermediate information is communicated. Excessive context sharing tends to dilute key information,
whereas overly restricted communication weakens the influence of earlier design choices
on later implementation, resulting in high interaction costs with limited performance gains.
We believe this benchmark underscores the need for more effective and cost-efficient E2ESD framework design.}

\section{Related Work}

% \textbf{Benchmarks for Software Development}. To evaluate the capabilities of LLM-based methods in software development, a range of benchmarks have been introduced. Most existing benchmarks focus on function-level tasks \citep{chen2021evaluating,austin2021program,hendrycks2021measuring,zhuo2024bigcodebench}, as summarized in the upper part of Table~\ref{tab:benchmark-comparison}. In contrast, only a limited number of benchmarks address project-level R2E-REP, shown in the lower part of the table. Specifically, SoftwareDev \citep{hong2024metagpt} relies on human ratings to assess requirement satisfaction across 70 samples, while SRDD \citep{qian2024chatdev} employs subjective LLM-based metrics such as Completeness (absence of unfinished or placeholder code), Executability (compilation and runtime success), and Consistency (semantic similarity between requirements and code). However, these project-level benchmarks face two main limitations. First, the input specifications are often generated by LLMs or presented as high-level textual descriptions \citep{qian2024chatdev}, which are typically vague or underspecified. This makes it difficult to verify whether the generated software truly satisfies the intended requirements. Second, the absence of standardized evaluation methods hinders objective and reliable performance comparison across different frameworks.
\begin{table*}[t]
\centering
\small
\setlength{\tabcolsep}{4.5pt} % 内容精简了，间距可以稍微宽一点点，更舒服
\renewcommand{\arraystretch}{1.25}

\begin{tabularx}{\linewidth}{l c c l l X}
\toprule
\textbf{{Benchmark}} & \textbf{\# Tasks} & \textbf{\# Reqs} & \textbf{Test Type / Scale} & \textbf{Source} & \textbf{Key Features} \\
\midrule

% Group 1: Generic
SoftwareDev~\citep{hong2024metagpt}
& 70
& Amb.
& {\xmark} Human Eval.
& Real
& Private benchmark; human ratings only \\

SRDD~\citep{qian2024chatdev}
& 1,200
& Amb.
& {\xmark} Heuristics
& Gen.
& Heuristic indicators; no execution \\

% 虚线
\cdashline{1-6} \noalign{\vskip 3pt}

% Group 2: Task-Level
GitTaskBench~\citep{ni2025gittaskbench}
& 18
& 54
& {\cmark} Repo Ops.
& Real
& Repo operation execution \\

Mle-Bench~\citep{chan2024mle}
& 72
& 72
& {\cmark} Pipeline Exec.
& Real
& Machine Learning pipeline execution\\

rSDE-Bench~\citep{huself}
& 53
& Amb.
& {\cmark} 616 Unit Tests
& Gen.
& Function-level output checks \\

% 虚线
\cdashline{1-6} \noalign{\vskip 3pt}

% Group 3: Ours
\rowcolor{gray!10}
\textbf{E2EDev}
& \textbf{46}
& \textbf{244}
& \textbf{{\cmark} 703 BDD Tests}
& \textbf{Real}
& \textbf{{Requirement Verification via BDD}} \\
% Requirement-level BDD tests
\bottomrule
\end{tabularx}
% \vspace{-3mm}
\caption{Comparison of Repo-Level Code Generation Benchmarks.
\textbf{Gen.}: LLM-Generated.
\textbf{Real}: Real-World Repositories.
\textbf{Amb.}: Ambiguous or unspecified number of requirements.
}
\label{tab:benchmark-comparison}
% \vspace{-4mm}
\end{table*}

{
\textbf{Benchmarks for Software Development}. Table~\ref{tab:benchmark-comparison} summarizes representative benchmarks for E2ESD. These benchmarks, which involve project-level code generation, are scarce due to the inherent complexity of their construction, often resulting in limited dataset sizes~\citep{hong2024metagpt, huself}. They typically face two primary limitations. First, vague or underspecified textual specifications~\citep{qian2024chatdev} make it difficult to verify whether the generated software aligns with user requirements. Second, a reliance on subjective human evaluation or heuristic metrics lacks grounding in established software development principles~\citep{hong2024metagpt, qian2024chatdev}, thereby hindering reliable performance assessment. Although some benchmarks incorporate executable evaluation, they are typically confined to specific commands~\citep{ni2025gittaskbench}, workflows~\citep{chan2024mle}, or individual functions~\citep{huself}, failing to verify whether the overall system behavior aligns with the user's specific intent.
}

% \textbf{Benchmarks for Repo level code generation}. 
% All Repo level code genreation bench suffer from 数据少，due to it high construction成本 
% This scarcity stems from the inherent complexity of their construction, often resulting in limited dataset sizes  \citep{hong2024metagpt, huself}. 

% These benchmarks face two primary limitations. First, the often vague or underspecified textual input specifications  \citep{qian2024chatdev} make it difficult to objectively verify if the generated software truly meets the user requirements. Second, they rely on subjective human evaluation or heuristic metrics that are not grounded in established software development principles \cite{hong2024metagpt, qian2024chatdev}, which hinders reliable performance evaluation. Although rSDE-Bench \cite{huself} offers 616 test cases, its reliance on Unit Tests\footnote{Appendix \ref{whatisbdd} distinguishes the BDD Test from Unit Tests.} restricts assessment to the function level, thereby overlooking the overall behavior of LLM-generated projects and failing to validate actual user needs.

\textbf{LLM-based Frameworks for E2ESD}. 
% Recent advances primarily fall into two categories: multi-agent and single-agent frameworks. 
Most frameworks employ a multi-agent paradigm, decomposing the development process into {explicit} subtasks following established software engineering workflows~\citep{sommerville2011software, hong2024metagpt,du2024multi,rasheed2024codepori,sami2024experimenting}. These workflows can be either predefined~\citep{dong2024self, islam2024mapcoder} or {adaptively constructed during interaction}~\citep{qian2024chatdev, lin2024think}. In contrast, {some} work employs single-agent frameworks~\citep{gptengineer}, where a single LLM manages the entire development process without explicit task decomposition.
\section{E2EDev Benchmark Construction}

\begin{table}[t]
\centering
\setlength{\abovecaptionskip}{5pt}
\setlength{\belowcaptionskip}{0pt}

% 1. 设置字号为 \small (或者 \footnotesize 如果还觉得大)
\small 
% 2. 稍微增加一点行高，防止字太密
\renewcommand{\arraystretch}{1.1} 

% 3. 使用 tabular* 并设置宽度为 \linewidth
% @{\extracolsep{\fill}} 会自动撑开列间距
\begin{tabular*}{\linewidth}{@{\extracolsep{\fill}} ll rrrr }
\toprule
& & \textbf{Max} & \textbf{Min} & \textbf{Mean} & \textbf{Total}\\
\midrule

% Repo
\textbf{Repo.} & Prompt Len. & 1907 & 206 & 712.7 & \textbf{46} \\
\midrule

% Req
\multirow{2}{*}{\textbf{Req.}} 
& Reqs / Repo & 11 & 2 & 5.3 & \multirow{2}{*}{\textbf{244}} \\
& \color{black!80}\textit{Words / Req.} & \color{black!80}{113} & \color{black!80}{20} & \color{black!80}{64.1} & \\
\midrule

% Tests
\multirow{8}{*}{\textbf{Tests}} 
& Tests / Req. & 7 & 1 & 2.9 & \multirow{8}{*}{\textbf{703}} \\
& Tests / Repo. & 34 & 2 & 15.3 & \\

% Group 1
& \multicolumn{4}{c}{\cellcolor{gray!15}\textbf{\color{black!80}{Gherkin Scenarios}}} & \\ 

& \color{black!80}\textit{Lines / Scen.} & \color{black!80}{30} & \color{black!80}{7} & \color{black!80}{11.2} & \\
& \color{black!80}\textit{Words / Scen.} & \color{black!80}{360} & \color{black!80}{53} & \color{black!80}{109.0} & \\

% Group 2
& \multicolumn{4}{c}{\cellcolor{gray!15}\textbf{\color{black!80}{Step Definitions}}} & \\

& \color{black!80}\textit{Lines / Def.} & \color{black!80}{157} & \color{black!80}{13} & \color{black!80}{61.4} & \\
& \color{black!80}\textit{Words / Def.} & \color{black!80}{700} & \color{black!80}{74} & \color{black!80}{217.4} & \\
\bottomrule
\end{tabular*}
\caption{Benchmark Statistics}
\label{tab:bench_statistic}
% \vspace{-15pt}
\end{table}

\begin{figure*}[t]
  \centering
  \includegraphics[width=0.97\textwidth]{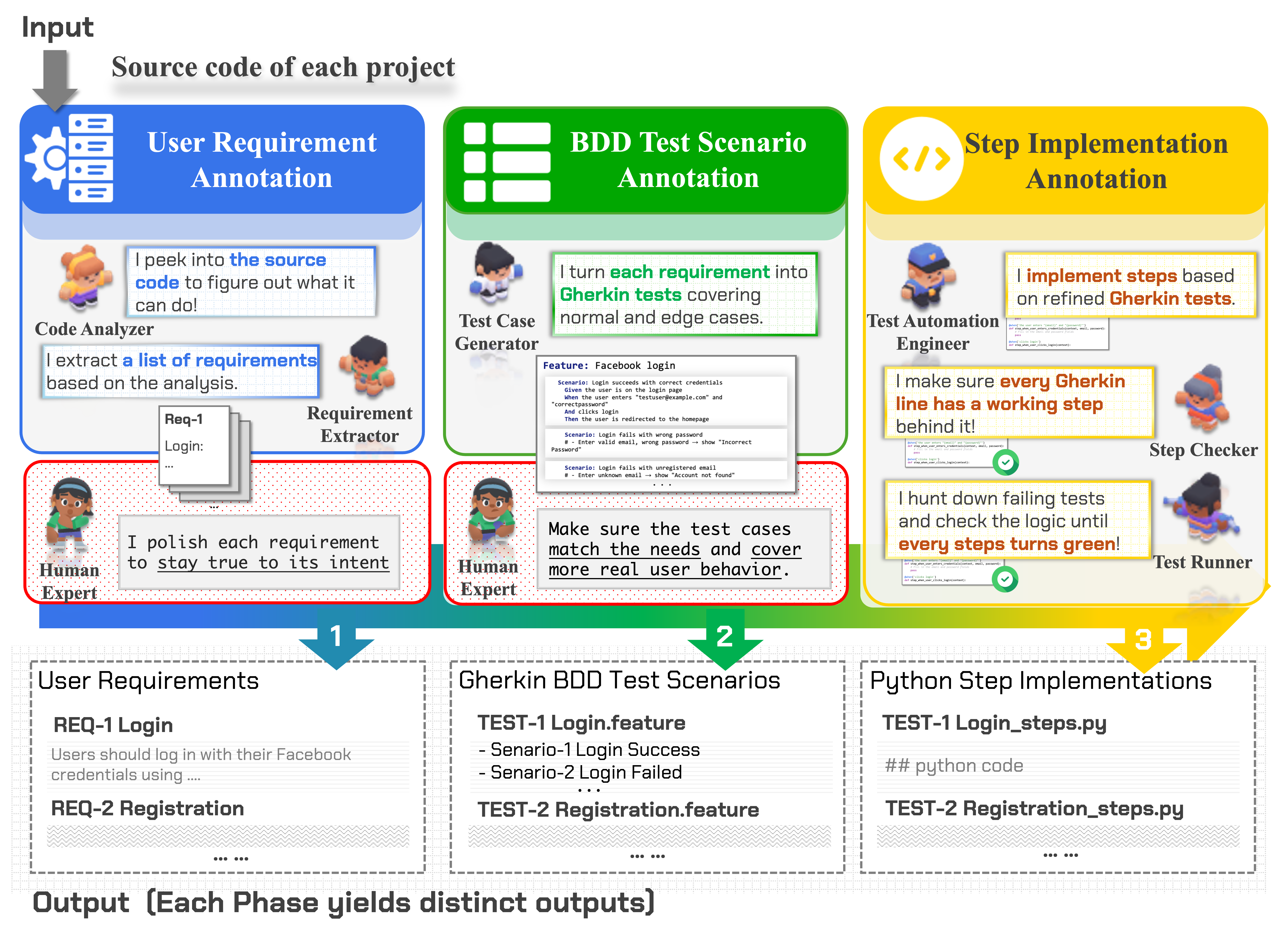}
  \caption{HITL-MAA framework for semi-automated dataset construction. Given source code, the framework extracts user requirements and generates corresponding BDD test scenarios (Gherkin format) along with Python step implementations, with human verification or agent-based refinement at each stage to ensure dataset quality.}
  \label{fig:framework}
  % \vspace{-3mm}
\end{figure*}

\textbf{Overview}. E2EDev is constructed by transforming real-world software projects into fine-grained user requirements and corresponding executable tests, using our HITL-MAA framework for efficient annotation and validation. Each requirement is associated with a set of \textbf{BDD test scenarios} written in Gherkin\footnote{Gherkin is a structured language using Given-When-Then statements to describe software behavior.}. Each scenario is accompanied by a corresponding \textbf{step implementation} in Python, which specifies how to execute each step for automated evaluation using the Behave framework. Table~\ref{tab:bench_statistic} summarizes dataset statistics, and example entries are provided in Appendix~\ref{ap:real_case_06}.

\subsection{Project Construction}  

% {
% \noindent\textbf{Project Selection}. we focus on Web applications. We select highly starred GitHub projects that are executable, functionally complete, and suitable for rapid evaluation. This process yields 46 high-quality projects, with detailed selection procedures provided in Appendix~\ref{app:selection}.
% }
\noindent\textbf{Project Selection}. 
Focusing on Web applications, we selected 46 highly starred GitHub projects that are executable, functionally complete, and suitable for rapid evaluation (details in Appendix~\ref{app:selection}).

\noindent\textbf{Features}. These projects cover a broad range of Web application types, such as calculators and mini games, and support diverse user interactions, such as typing, clicking, and dragging (cf. Figure~\ref{fig:statistic}, Appendix~\ref{ap:e2edev_feature}). They also include technically complex features, ranging from simple DOM manipulation to more advanced tasks such as local storage usage and external API integration. Overall, the diversity, technical depth, and realism of these projects make them a qualified and representative ground for evaluating E2E-REP methods.

\subsection{HITL-MAA Annotation Framework}

{Grounded in BDD-based software engineering practices, HITL-MAA is employed for each collected project to annotate fine-grained user requirements, along with corresponding BDD test scenarios and Python step implementations, to ensure reliable evaluation.}
As illustrated in Figure \ref{fig:framework}, HITL-MAA involves the following three steps, with human supervisors involved at key points in each step to avoid errors. 
Before generating requirements and test cases, we use GPT-4o to assign unique test IDs to key UI components. 
{These test IDs serve as stable, structure-invariant DOM anchors, enabling consistent component references across requirements, BDD tests, and different projects generated from the same requirement.}
All implementation details are provided in \textbf{Appendix \ref{app:maa}}.

\noindent\textbf{User Requirement Annotation}.
This process transforms a source project into a list of fine-grained user requirements through the collaborative effort of two LLM agents under human supervision. (1) The Code Analyzer Agent analyzes the project’s core functionalities and their interactions with UI elements by reading the source code. (2) The Requirement Extractor Agent then automatically generates candidate user-facing requirements based on this analysis. Human supervisors review the generated requirements to ensure accuracy, clarity, and functional consistency by verifying them against the actual system behavior.

\noindent\textbf{{BDD Test Scenario} Annotation}. 
For each human-validated requirement, HITL-MAA generates a set of Gherkin-style BDD test scenarios. Each scenario describes a specific condition or user interaction to be tested and follows the Given–When–Then format, clearly specifying the preconditions (\textit{Given}), the user action (\textit{When}), and the expected outcome (\textit{Then}) in natural language. To support this process, HITL-MAA employs a Test Case Generation Agent that analyzes both validated requirements and source code to produce relevant scenarios. The generation process is guided to cover a wide range of test cases, including both typical behaviors and unexpected or edge cases. Since the agent relies solely on source code—which does not explicitly capture all user interaction patterns—some unconventional scenarios may be overlooked and require supplementation by human experts. To ensure comprehensive coverage and high quality, five software testing experts collaboratively review and refine the generated scenarios. Meanwhile, HITL-MAA incorporates a final expert review step to resolve any remaining inconsistencies between requirements and test cases, guaranteeing reliable and high-quality annotations.

\noindent\textbf{Step Implementation Annotation}.
This step focuses on converting each {BDD test scenario} into an executable Python script. To achieve this, HITL-MAA utilizes a Test Automation Engineer Agent to generate initial Python implementations corresponding to the Gherkin-style test steps. 
{The generated scripts then undergo an iterative verification process to ensure both executability and correctness. Two agents provide feedback to support self-correction: the \textit{Dry Run Verifier} checks for missing or syntactically invalid step definitions, while the \textit{Test Runner} executes the script to detect logical inconsistencies or failed assertions. Scripts unresolved after maximum self-correction attempts are flagged for human refinement. Empirically, this self-correction mechanism resolves nearly all dry-run errors autonomously and corrects logical issues in over 80\% of cases without human intervention. Validation on the generated dataset achieved a 100\% pass rate across all executable tests, demonstrating the high quality and reliability of E2EDev.}

\noindent\textbf{Automated Evaluation}.
E2EDev adopts the \texttt{behave} framework for fully automated evaluation. Gherkin scenarios are executed through their corresponding Python step implementations to test the generated project against defined requirements, enabling deterministic pass/fail verification of model-generated code. (Details in Appendix~\ref{app:how_to_test}.)

\section{Experiment}
\begin{table*}[t!]
  \centering
  \scriptsize 
  % [关键调整] 增加列间距 (4pt -> 8pt)，让表格横向更宽，视觉上更舒展
  \setlength{\tabcolsep}{8pt} 
  \renewcommand{\arraystretch}{1.0}

  % =======================================================
  % 颜色定义
  % =======================================================
  \definecolor{rowgray}{HTML}{F5F5F5}
  \definecolor{bgred}{HTML}{FFF0F0}
  \definecolor{bggreen}{HTML}{F0FFF0}
  \definecolor{sdtext}{HTML}{999999}

  % =======================================================
  % 宏定义 (修复双重符号问题)
  % =======================================================
  
  % 1. \sd: 核心样式 (灰色、小号、自带±)
  \newcommand{\sd}[1]{\textcolor{sdtext}{\tiny $\pm$#1}}

  % 2. 样式宏: 
  % [关键修复] 去掉了 #2 前面的 \sd，只负责背景色和加粗
  % 这样 \base{48.69}{\sd{1.30}} 就会输出 "48.69 ±1.30" 而不是 "48.69 ±±1.30"
  
  % Effectiveness - High (Red)
  \newcommand{\high}[2]{\cellcolor{bgred}#1 #2}
  \newcommand{\highb}[2]{\cellcolor{bgred}\textbf{#1} #2}
  
  % Effectiveness - Low (Green)
  \newcommand{\low}[2]{\cellcolor{bggreen}#1 #2}
  \newcommand{\lowb}[2]{\cellcolor{bggreen}\textbf{#1} #2}
  \newcommand{\lowno}[1]{\cellcolor{bggreen}#1}

  % Vanilla (Gray)
  \newcommand{\base}[2]{#1 #2}
  \newcommand{\baseul}[2]{\uline{#1} #2}
  
  % Efficiency (Plain)
  \newcommand{\eff}[2]{#1 #2} 

  % =======================================================

  \resizebox{\linewidth}{!}{%
    \begin{tabular}{ll ccc ccc} 
      \toprule
      \multirow{2}{*}{\textbf{Backbone}} & \multirow{2}{*}{\textbf{Method}} & \multicolumn{3}{c}{\textbf{Effectiveness ($\uparrow$, \%)}} & \multicolumn{3}{c}{\textbf{Efficiency ($\downarrow$)}} \\
      \cmidrule(lr){3-5} \cmidrule(lr){6-8}
      & & Req. Acc. & Test Acc & Bal. Score & Cost (\$) & CO$_{2}$ (g) & Time (s) \\
      \midrule

      % ================= Claude-Haiku 4.5 =================
      \multirow{6}{*}{\textbf{{Claude-Haiku 4.5}}} 
      & \cellcolor{rowgray}Vanilla LLM & \cellcolor{rowgray}\base{48.69}{\sd{1.30}} & \cellcolor{rowgray}\base{63.08}{\sd{0.50}} & \cellcolor{rowgray}\base{54.45}{\sd{0.98}} & \cellcolor{rowgray}0.015 & \cellcolor{rowgray}0.028 & \cellcolor{rowgray}53 \\
      & GPT-Engineer & \high{\textbf{53.75}}{\sd{0.43}} & \high{\textbf{69.41}}{\sd{1.07}} & \high{\textbf{60.01}}{\sd{0.69}} & 0.016 & 0.038 & 59 \\
      & Self-Collab. & \high{49.01}{\sd{1.74}} & \low{61.50}{\sd{1.26}}  & \low{54.01}{\sd{1.55}}  & 0.014 & 0.033 & 75 \\
      & MapCoder     & \high{\uline{49.61}}{\sd{0.50}} & \high{\uline{65.65}}{\sd{0.44}} & \high{\uline{56.03}}{\sd{0.48}} & 0.099 & 0.081 & 129 \\
      & ChatDev      & \low{44.73}{\sd{1.50}}  & \low{58.09}{\sd{1.03}}  & \low{50.07}{\sd{1.31}}  & 0.174 & 0.188 & 335 \\
      & MetaGPT      & \low{5.39}{\sd{0.76}}   & \low{10.66}{\sd{1.77}}  & \low{7.50}{\sd{1.16}}   & 0.090 & 0.091 & 366 \\
      \cmidrule(lr){1-8} 

      % ================= GPT-4o =================
      \multirow{6}{*}{\textbf{GPT-4o}} 
      & \cellcolor{rowgray}Vanilla LLM & \cellcolor{rowgray}\base{45.95}{\sd{1.25}} & \cellcolor{rowgray}\base{60.88}{\sd{1.97}} & \cellcolor{rowgray}\base{51.92}{\sd{1.54}} & \cellcolor{rowgray}0.0160 & \cellcolor{rowgray}0.083 & \cellcolor{rowgray}28 \\
      & GPT-Engineer & \high{\textbf{50.83}}{\sd{1.33}} & \high{\textbf{66.59}}{\sd{2.43}} & \high{\textbf{57.13}}{\sd{1.77}} & 0.0198 & 0.132 & 21 \\
      & Self-Collab. & \high{46.83}{\sd{1.12}} & \high{61.15}{\sd{1.47}} & \high{52.56}{\sd{1.26}} & 0.0155 & 0.109 & 37 \\
      & MapCoder     & \high{\uline{47.70}}{\sd{2.57}} & \high{\uline{63.97}}{\sd{2.25}} & \high{\uline{54.21}}{\sd{2.44}} & 0.1091 & 0.750 & 93 \\
      & ChatDev      & \low{42.71}{\sd{2.44}}  & \low{58.93}{\sd{2.66}}  & \low{49.20}{\sd{2.53}}  & 0.1947 & 1.910 & 114 \\
      & MetaGPT      & \low{0.00}{\sd{0.00}}   & \low{0.17}{\sd{0.05}}   & \low{0.07}{\sd{0.02}}   & 0.0951 & 0.794 & 66 \\
      \cmidrule(lr){1-8} 

      % ================= GPT-4o-mini =================
      \multirow{6}{*}{\textbf{GPT-4o-mini}} 
      & \cellcolor{rowgray}Vanilla LLM & \cellcolor{rowgray}\base{\textbf{44.82}}{\sd{2.01}} & \cellcolor{rowgray}\base{\textbf{60.65}}{\sd{2.49}} & \cellcolor{rowgray}\base{\textbf{51.15}}{\sd{2.20}} & \cellcolor{rowgray}0.0010 & \cellcolor{rowgray}0.003 & \cellcolor{rowgray}16 \\
      & GPT-Engineer & \low{\uline{42.13}}{\sd{1.81}}  & \low{57.51}{\sd{1.98}}  & \low{\uline{48.28}}{\sd{1.88}}  & 0.0012 & 0.005 & 18 \\
      & Self-Collab. & \low{37.90}{\sd{1.92}}  & \low{52.99}{\sd{2.51}}  & \low{43.94}{\sd{2.16}}  & 0.0009 & 0.004 & 25 \\
      & MapCoder     & \low{41.30}{\sd{1.49}}  & \low{\uline{57.87}}{\sd{0.85}}  & \low{47.93}{\sd{1.23}}  & 0.0072 & 0.033 & 88 \\
      & ChatDev      & \low{33.16}{\sd{2.71}}  & \low{48.40}{\sd{2.75}}  & \low{39.26}{\sd{2.73}}  & 0.0118 & 0.071 & 157 \\
      & MetaGPT      & \low{0.00}{\sd{0.00}}   & \low{0.23}{\sd{0.05}}   & \low{0.09}{\sd{0.02}}   & 0.0067 & 0.040 & 63 \\
      \cmidrule(lr){1-8} 

      % ================= Qwen-Max =================
      \multirow{6}{*}{\textbf{Qwen-Max}} 
      & \cellcolor{rowgray}Vanilla LLM & \cellcolor{rowgray}\base{43.33}{\sd{0.66}} & \cellcolor{rowgray}\base{58.57}{\sd{0.73}} & \cellcolor{rowgray}\base{49.43}{\sd{0.69}} & \cellcolor{rowgray}0.0025 & \cellcolor{rowgray}0.043 & \cellcolor{rowgray}53 \\
      & GPT-Engineer & \high{\textbf{49.61}}{\sd{0.60}} & \high{63.70}{\sd{1.41}} & \high{\textbf{55.25}}{\sd{0.92}} & 0.0030 & 0.067 & 66 \\
      & Self-Collab. & \low{42.30}{\sd{1.88}}  & \high{\uline{60.44}}{\sd{1.33}} & \high{49.56}{\sd{1.66}} & 0.0022 & 0.053 & 75 \\
      & MapCoder     & \high{\uline{48.83}}{\sd{1.68}} & \high{\textbf{64.00}}{\sd{1.82}} & \high{\uline{54.90}}{\sd{1.74}} & 0.0186 & 0.424 & 366 \\
      & ChatDev      & \high{43.93}{\sd{1.94}} & \high{59.44}{\sd{0.99}} & \high{50.13}{\sd{1.56}} & 0.0249 & 0.790 & 300 \\
      & MetaGPT      & \low{1.65}{\sd{0.25}}   & \low{2.80}{\sd{0.58}}   & \low{2.11}{\sd{0.38}}   & 0.0207 & 0.544 & 312 \\
      \cmidrule(lr){1-8} 

      % ================= Qwen-70B =================
      \multirow{6}{*}{\textbf{Qwen-70B}} 
      & \cellcolor{rowgray}Vanilla LLM & \cellcolor{rowgray}\base{35.75}{\sd{0.00}} & \cellcolor{rowgray}\base{53.14}{\sd{0.00}} & \cellcolor{rowgray}\base{42.71}{\sd{0.00}} & \cellcolor{rowgray}0.0029 & \cellcolor{rowgray}0.050 & \cellcolor{rowgray}39 \\
      & GPT-Engineer & \high{\uline{42.08}}{\sd{0.00}} & \highb{58.78}{\sd{0.00}}& \high{\uline{48.76}}{\sd{0.00}} & 0.0037 & 0.080 & 45 \\
      & Self-Collab. & \high{\textbf{42.61}}{\sd{0.06}} & \high{55.63}{\sd{0.10}} & \high{47.82}{\sd{0.08}} & 0.0030 & 0.066 & 69 \\
      & MapCoder     & \high{40.59}{\sd{0.26}} & \high{57.49}{\sd{0.07}} & \high{47.35}{\sd{0.18}} & 0.0293 & 0.624 & 294 \\
      & ChatDev      & \high{43.15}{\sd{0.51}} & \high{\uline{57.62}}{\sd{0.08}} & \high{\textbf{48.94}}{\sd{0.34}} & 0.0387 & 1.006 & 341 \\
      & MetaGPT      & \low{0.00}{\sd{0.00}}   & \low{1.73}{\sd{0.06}}   & \low{0.69}{\sd{0.02}}   & 0.0290 & 0.786 & 151 \\
      \cmidrule(lr){1-8} 

      % ================= Qwen-7B =================
      \multirow{6}{*}{\textbf{Qwen-7B}} 
      & \cellcolor{rowgray}Vanilla LLM & \cellcolor{rowgray}\baseul{22.37}{\sd{0.00}}& \cellcolor{rowgray}\baseul{34.75}{\sd{0.00}}& \cellcolor{rowgray}\baseul{27.32}{\sd{0.00}}& \cellcolor{rowgray}0.0003 & \cellcolor{rowgray}0.005 & \cellcolor{rowgray}32 \\
      & GPT-Engineer & \highb{24.03}{\sd{0.00}} & \highb{39.94}{\sd{0.00}}& \highb{30.39}{\sd{0.00}}& 0.0003 & 0.007 & 31 \\
      & Self-Collab. & \low{20.65}{\sd{1.00}}  & \low{33.37}{\sd{0.63}}  & \low{25.74}{\sd{0.85}}  & 0.0003 & 0.006 & 52 \\
      & MapCoder     & \low{11.90}{\sd{1.61}}  & \low{26.96}{\sd{1.12}}  & \low{17.92}{\sd{1.41}}  & 0.0024 & 0.049 & 236 \\
      & ChatDev      & \low{10.96}{\sd{1.73}}  & \low{21.35}{\sd{2.10}}  & \low{15.12}{\sd{1.88}}  & 0.0075 & 0.187 & 344 \\
      & MetaGPT      & \low{0.00}{\sd{0.00}}   & \low{1.98}{\sd{0.28}}   & \low{0.79}{\sd{0.11}}   & 0.0207 & 0.095 & 301 \\
      \bottomrule
    \end{tabular}%
  }
  
  \caption{Benchmark analysis of effectiveness and efficiency.
The \colorbox{rowgray}{grey row} denotes the Vanilla baseline.
\colorbox{bgred}{Red} and \colorbox{bggreen}{green} indicate performance better and worse than Vanilla, respectively.
Standard deviations are shown in small grey text.
For effectiveness, the best result is shown in \textbf{bold} and the second-best is \uline{underlined}.}
  % \vspace{-5mm}
  \label{tab:model_performance}
\end{table*}
This section enables automated benchmark analysis. E2EDev is an automated testing benchmark suite, accessible even to researchers less familiar with software testing procedures\footnote{Cf. Appendix~\ref{app:software} for background knowledge of project/software testing.}.

% # 实验设置 为什么选这几个framework 以及 backbone,以及metric
\subsection{Experiment Setup}
\label{exp_setting}

% \textbf{Baseline \& LLM Backbones}.
% \yao{We select a diverse set of representative \textbf{requirement-to-executable software development} frameworks across various LLM backbones.} These include: \uline{(1) Vanilla LLM} that generates the project via simply prompting LLM. \uline{(2) Single-Agent Framework}, \textit{GPT-Engineer}~\citep{gptengineer}, which builds the complete project in a single reasoning pass. \uline{(3) Multi-Agent Frameworks}, including \textit{Self-Collaboration~\citep{dong2024self}}, \textit{MetaGPT~\citep{hong2024metagpt}}, \textit{MapCoder~\citep{islam2024mapcoder}}, and \textit{ChatDev~\citep{qian2024chatdev}}. \yao{We additionally include two widely used coding agents, \textit{OpenHands}~\citep{wang2024openhands} and \textit{SWE-Agent}~\citep{yang2024swe}, as reference systems.
% These agents are designed for general-purpose software engineering scenarios with extensive tool and environment interaction, which differ from the requirement-to-executable software generation setting considered in E2EDev. Their results are reported in the appendix.}
% Finally, to enables controlled analysis of framework versus model backbone effects, we consider LLM backbones with various scales and architectures \citep{hurst2024gpt,hui2024qwen2}, including: \yao{Claude-Haiku-4.5}, GPT-4o, GPT-4o-mini, Qwen2.5-7B, Qwen2.5-70B, and the mixture-of-experts Qwen2.5-Max. Baseline introduction and implementation details are in \textbf{Appendix \ref{app:baselines}}.
{
\textbf{Baselines and LLM Backbones.}
We consider a representative set of E2ESD methods, covering three categories:
(1) \textit{Vanilla LLM}, which directly generates an executable project via prompting;
(2) \textit{Single-agent frameworks}, represented by \textit{GPT-Engineer}~\citep{gptengineer}, where a single LLM constructs the entire project in a single reasoning process;
and (3) \textit{Multi-agent frameworks}, including \textit{Self-Collaboration}~\citep{dong2024self}, \textit{MetaGPT}~\citep{hong2024metagpt}, \textit{MapCoder}~\citep{islam2024mapcoder}, and \textit{ChatDev}~\citep{qian2024chatdev}.
We additionally include two widely used coding agents, \textit{OpenHands}~\citep{wang2024openhands} and \textit{SWE-Agent}~\citep{yang2024swe}, as reference systems, with results reported in Appendix~\ref{app:openhands_swe} since they target a different task setting from E2EDev.
To disentangle the effects of framework design from model capacity, we run all baselines across diverse LLM backbones with varying scales and architectures~\citep{hurst2024gpt,hui2024qwen2}, including Claude-Haiku~4.5, GPT-4o, GPT-4o-mini, Qwen2.5-7B, Qwen2.5-70B, and the mixture-of-experts Qwen2.5-Max.
Implementation details are provided in Appendix~\ref{app:baselines}.
}

\noindent\textbf{Evaluation Metrics}.
For comprehensive evaluation, we assess \textbf{code effectiveness} (how well the generated code meets user requirements) and \textbf{generation efficiency} (the cost and time of code generation). {Effectiveness is averaged over three runs per task, while efficiency is measured from a single run due to negligible variance.}
% As for code effectiveness, the evaluation metrics include: (1) \uline{Req. Acc}, the proportion of requirements successfully validated by all their corresponding test cases across a project, which measures the requirement-level accuracy; (2) \uline{Test Acc}, the proportion of all passed test cases across a project, which measures the test-level accuracy; (3) \uline{Balanced Score}, a weighted combination of \textit{Req. Acc} and \textit{Test Acc} to mitigate bias from varying test case granularity per requirement.
% Additionally, generation efficiency is assessed using the following three metrics: (1) \uline{Cost (USD)}, the average API pricing for LLM token usage; (2) \uline{Carbon Footprint (CO\textsubscript{2})}, the average carbon footprint for generating one project; (3) \uline{Duration}, the average wall-clock time for project generation. Metric details are presented in \textbf{Appendix \ref{impl_metric}}.
For \textbf{code effectiveness}, we evaluate how well a generated project satisfies requirements at two levels.
{
(1) \uline{Req. Acc} measures the fraction of requirements that are fully satisfied across a project, i.e., all test cases associated with a requirement pass;
(2) \uline{Test Acc} measures the fraction of all test cases that pass across a project;
}
(3) \uline{Balanced Score}, a weighted combination of \textit{Req. Acc} and \textit{Test Acc} to mitigate bias from varying test case granularity per requirement.
Additionally, generation efficiency is assessed using the following three metrics:
(1) \uline{Cost (USD)}, the average API pricing for LLM token usage \textbf{per project};
(2) \uline{Carbon Footprint (CO\textsubscript{2})}, the average carbon footprint for generating one project;
(3) \uline{Duration}, the average wall-clock time for project generation.
Metric details are presented in Appendix~\ref{impl_metric}.

%\paragraph{Implementation Details}
%For Single-Agent and Multi-Agent frameworks, we implemented them using their official code available on Github. For fair comparison, we adapt Self-Collaboration and MAPCoder to better suit front-end web application generation (HTML, JS, CSS) by refining system prompts. Since these frameworks rely on test-case-based correctness verification, we limit their iterative debugging steps to a single pass for consistency. For Vanilla LLMs, we organize their generated files by code blocks and store them as a runnable project. Finally, all experiments are conducted with fixed temperature settings (\textit{temperature=0.2}), and each model is executed once to reflect realistic.

% Results for the primary benchmark are presented in Table~\ref{tab:model_performance}. To further validate the generality of our findings, we additionally evaluate on HumanEval and HumanEval-ET datasets using the same effectiveness and efficiency metrics. For effectiveness, we report \texttt{pass@1}; for efficiency, we reuse the aforementioned three indicators. Results are shown in Table~\ref{tab:model_performance_humaneval}.

\subsection{Main Results}
\label{exp:main}

Table~\ref{tab:model_performance} presents our benchmark analysis on off-the-shelf methods across various LLM backbones regarding
their effectiveness and efficiency. Our key observations are detailed below.

% \chen{You can refer to the following writing style to make your content more structured and highlighted: \textbf{\textit{question?} -- answer, which is your main observations/findings}. As shown in Table [What table or figure reader should focus on in order to comprehend your observations], [describe how you draw your observation], [describe that the above analysis draw a observation. Here, you must paraphrase the way you state your observation], [describe significance of your observations]} \chen{This structure provides a clear and concise narrative. Feel free to include concrete examples to aid understanding, address any outliers, or incorporate other interesting details, while maintaining consistency with the main observation. If you have multiple observations, present each in a separate paragraph. Also, It would be better if you present them in a progressive way. If you have a single observation, the "Question" component at the beginning can be omitted.}

% \noindent\textbf{\textit{%What is the effectiveness of
% How effective are existing methods on E2ESD?} -- Despite meeting broad project requirements, current performance remains significantly below acceptable standards due to a lack of proficiency in handling detailed specifics}. 
% \noindent \colorbox{gray!10}{\textit{\textbf{How effective are existing methods on E2ESD?}}}    \textbf{Despite meeting broad project requirements, performance remains below acceptable standards due to lack of proficiency in handling specifics.} 
\noindent\textbf{\underline{Limited effectiveness}: While existing methods meet broad project requirements, their performance remains below acceptable standards due to limited proficiency in handling specifics.
}
Off-the-shelf models excel at function-level tasks (e.g., HumanEval \citep{chen2021evaluating} in Appendix \ref{humaneval}) but underperform on E2ESD, achieving only 30\%–50\% Req. Acc. on the proposed E2EDev dataset (Table~\ref{tab:model_performance}), {with even advanced models like Claude-Haiku 4.5 and GPT-4o failing to exceed 60\%}.
Furthermore, {Soft Req.} Acc., which allow partial fulfillment, suggest failures stem from their lack of proficiency in handling detailed functional specifics during implementation. This phenomenon, visualized in Figure~\ref{fig:cb}, persists across different LLM backbones and E2ESD frameworks. The observed discrepancy of over 25\% in both metrics suggests that while models can implement the required functionality, they often fail to address complex edge cases. Also, this issue is reflected in $\sim$10\% performance gap observed on HumanEval and HumanEval-ET (cf. Appendix \ref{humaneval}).

\begin{figure*}[t]
  \centering
  \includegraphics[width=0.98\textwidth]{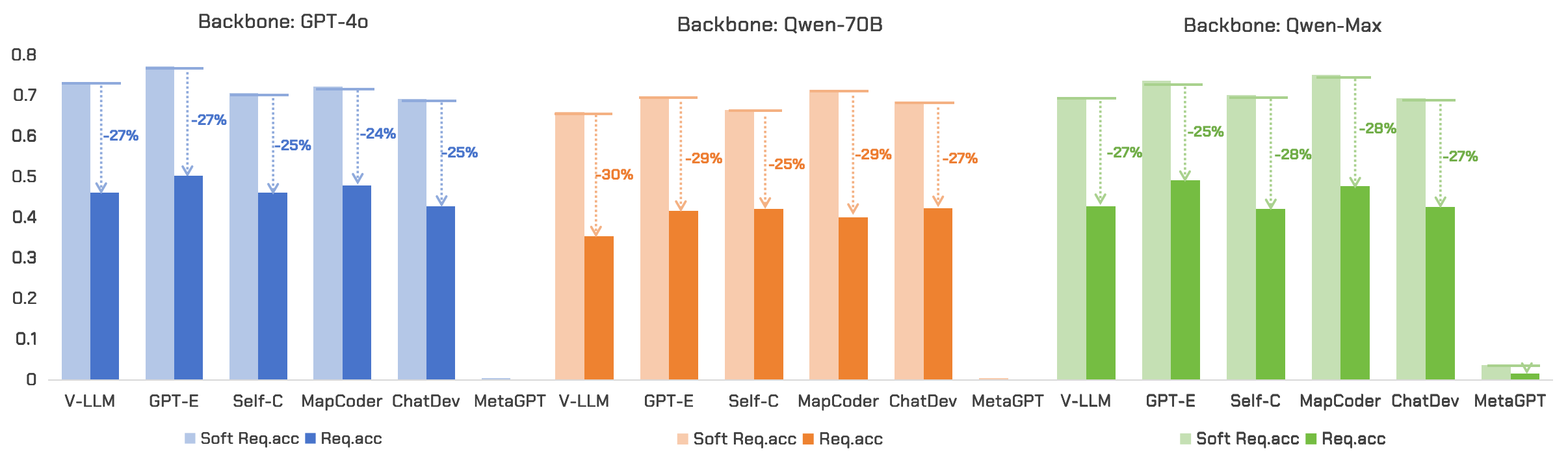}
  \setlength{\abovecaptionskip}{0pt}   
\setlength{\belowcaptionskip}{0pt}
  \caption{Soft-Requirement and Requirement Accuracy for Vanilla LLM (V-LLM), GPT-Engineer (GPT-E), and Self-Collaboration (Self-C). Additional results for other backbones are in Appendix~\ref{appe2e}, Table~\ref{app_tb:soft}.}
  \label{fig:cb}
  \vspace{-3mm}
\end{figure*}

\noindent{\textbf{\underline{Limited efficiency}: Multi-agent methods exhibit excessive interaction rounds and token costs with marginal effectiveness gains.}
As evidenced by the efficiency-oriented metrics in Table~\ref{tab:model_performance}, methods such as Vanilla LLM, GPT-Engineer, and Self-Collaboration maintain high efficiency ($\leq$three times API Calls), benefiting from their streamlined, deterministic workflows. However, regarding the multi-agent frameworks, even static frameworks like MapCoder and MetaGPT require >10 interaction turns per task, amplifying latency and computational expense. Dynamic frameworks (e.g., ChatDev with GPT-4o) reach 15.72 turns on average (minimum of 9), with inefficiency exacerbated by repetitive and uninformative dialogue cycles, as analyzed in detail in Appendix \ref{app:add_efficiency}.

% \begin{figure}[htbp]
% \centering
% \begin{minipage}[t]{0.5\textwidth}
%   \centering
%   \caption{Comparisons across agentic frameworks and the Vanilla LLM (on Req. Acc.). Bars within the gray zone indicate lower performance.}
%   \includegraphics[width=0.98\textwidth]{fig/comp0924_3.png}
%   \label{fig:comp}
% \end{minipage}%
% \hfill
% \begin{minipage}[t]{0.45\textwidth}
%   \centering
%   \captionsetup{type=table}
%   \label{tab:token_turns}
%   \input{tables/turns}
% \end{minipage}
% \end{figure}

% fig/comp0924_3.png

\begin{figure}[htbp]
  \centering
  \includegraphics[width=0.46\textwidth]{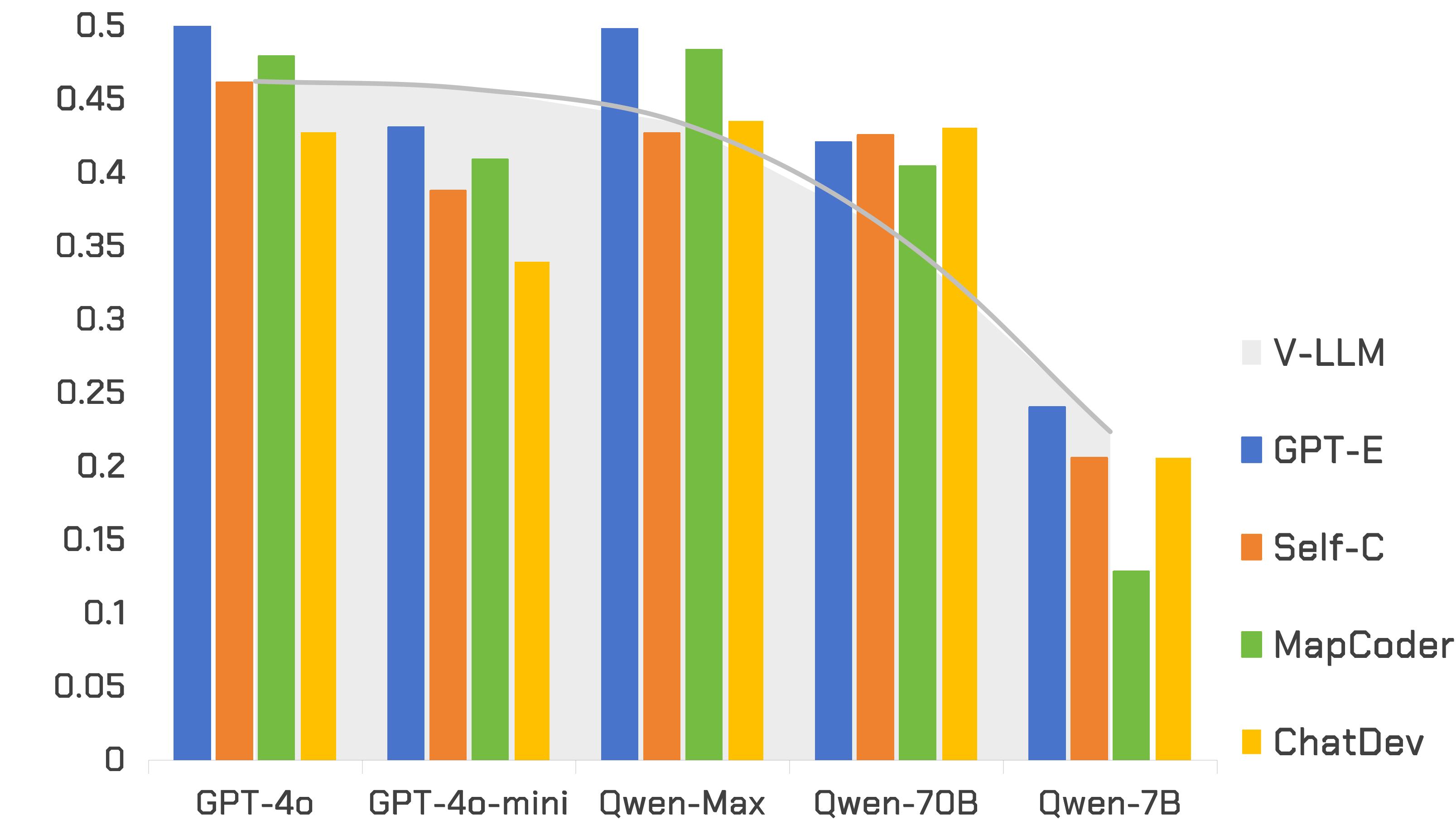}
  \setlength{\abovecaptionskip}{0pt}   
\setlength{\belowcaptionskip}{0pt}
  \caption{{Comparisons across agentic frameworks and the Vanilla LLM (on Req. Acc.).}}
  \label{fig:comp}
  \vspace{-4mm}
\end{figure}

\noindent\textbf{{Agentic frameworks do not consistently improve the effectiveness of base LLMs.}} While agentic frameworks offer the potential for performance gains, their inherent complexity imposes significant demands on the foundational capabilities of the underlying LLM. Figure \ref{fig:comp} shows framework performance varies widely across LLMs, with Vanilla LLMs occasionally surpassing frameworks (and vice versa), revealing architectural overhead. Notably, multi-agent frameworks often underperform single-agent approaches due to the necessary coordination demands among agents, such as task decomposition \citep{xia2024agentless,han2024llm}, role/instruction following \citep{pan2025multiagent,hammond2025multi}, dialogue state/context management \citep{pan2025multiagent,li2023camel,hammond2025multi}, and the increasing interaction costs as shown in Table \ref{tab:token_turns}. Therefore, error accumulation from these complexities can degrade performance, particularly in E2ESD tasks. {Overall, our results indicate that current frameworks rely heavily on strong base LLMs, suggesting that future designs should reduce this reliance to improve robustness across diverse LLM backbones.}
{
\noindent\textbf{\uline{MetaGPT} suffers from communication breakdowns in its multi-agent architecture.}
Even at the function level, we observe evidence of communication failures on HumanEval using the official implementation (Appendix~\ref{humaneval}), which already limits its performance.
At the project level, the same issue becomes more severe, causing MetaGPT to fail on nearly all test cases and requirements, even with strong LLMs (see Section~\ref{indepth}).
}

% \subsection{In-depth Limitation Analysis}
\subsection{{Failure Mode Analysis}}
\label{indepth}
\begin{figure*}[t]
  \centering
  \includegraphics[width=0.95\textwidth]{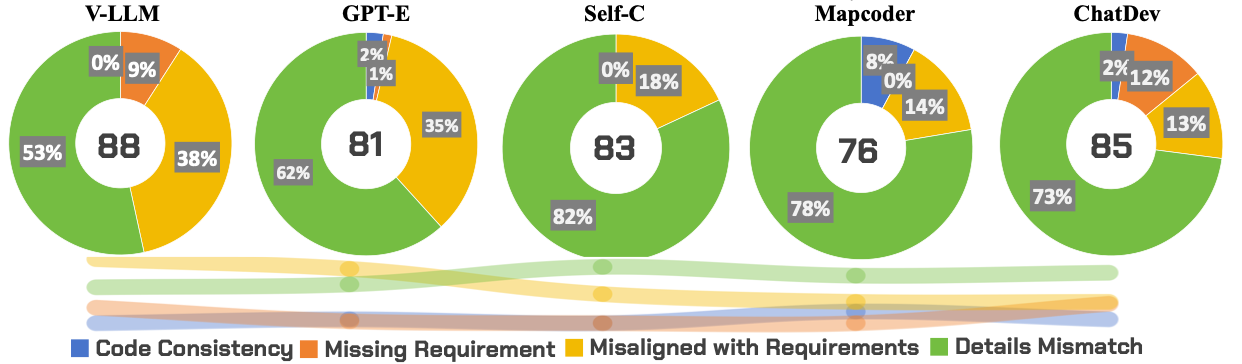}
  \caption{{Requirement-level error distribution across frameworks. Pie charts show error proportions (centers indicate total errors), and trend lines report absolute counts across frameworks.}}
  \label{fig:humanv}
  % \vspace{-5mm}
\end{figure*}

\textbf{Setup.} 
To understand underlying failure causes, we conducted a rigorous human evaluation on 360 projects (randomly sampled from 10 entries across 6 LLMs and 6 frameworks).
% {To better understand the underlying causes of model failures, we conduct a rigorous human evaluation.
% To balance analytical depth with practical constraints, we randomly sample 10 data entries, yielding 360 generated projects across 6 LLM backbones and 6 frameworks.} 
% against the user requirements
Four domain experts evaluate each project along four dimensions: 
\textit{Code Inconsistency} (e.g., missing, conflicting, or empty functions), 
\textit{Requirement Missing} (required features not implemented), 
\textit{Requirement Misaligned} (implementation logic deviates from the requirements), and 
\textit{Detail Mismatch} (mostly correct but with minor errors). 
Figure~\ref{fig:humanv} presents the distribution of failure modes for existing frameworks {on selected LLM backbones (GPT-4o, Qwen2.5-70B, and Qwen-Max)}. Detailed results and case studies are in Appendices \ref{appe2e} and \ref{app_E2ESD_log_case}. Human evaluation protocol and agreement are in Appendix \ref{imp_indepth}.

%   \caption{Statistics of interaction overhead across representative frameworks. \textit{P. Tok.}, \textit{C. Tok.}, and \textit{Turns} refer to average prompt tokens, completion tokens, and dialogue rounds, respectively.}
%   \label{tab:token_turns}
%   \resizebox{\textwidth}{!}{ 
%   \begin{tabular}{llccc}
%     \toprule
%     \textbf{Model} & \textbf{Method} & \textbf{P.Tok} & \textbf{C.Tok} & \textbf{Turns} \\
%     \midrule
%     \multirow{3}{*}{GPT-4o} 
%         & MapCoder   & 16,835.54 & 6,696.46 & 10.00\\
%         & ChatDev    & 53,912.26 & 5,995.65 & 15.72 \\
%         & MetaGPT    & 20,501.28 & 4,387.00 & 10.46\\
%     \midrule
%     \multirow{3}{*}{Qwen-Max} 
%         & MapCoder   & 19,427.28 & 9,410.52 & 10.00 \\
%         & ChatDev    & 46,192.96 & 7,458.87 & 12.50 \\
%         & MetaGPT    & 28,272.87 & 8,724.35 & 11.48 \\
%     \midrule
%     \multirow{3}{*}{Qwen-70B} 
%         & MapCoder   & 22,906.81 & 10,447.55 & 10.00  \\
%         & ChatDev    & 45,167.91  & 8,647.67 & 11.57\\
%         & MetaGPT    & 36,496.70 & 5,568.80 & 13.83 \\
%     \bottomrule
%   \end{tabular}
% }
\begin{table}[htbp]
  \centering
  \resizebox{0.95\linewidth}{!}{%
  \begin{tabular}{llccc}
    \toprule
    \textbf{Model} & \textbf{Method} & \textbf{Tok}$_{\textbf{prompt}}$ & \textbf{Tok}$_{\textbf{compl.}}$ & \textbf{Turns}\\
    \midrule
    \multirow{3}{*}{GPT-4o} 
        & MapCoder   & 16,835.54 & 6,696.46 & 10.00 \\
        & ChatDev    & 53,912.26 & 5,995.65 & 15.72 \\
        & MetaGPT    & 20,501.28 & 4,387.00 & 10.46 \\
    \midrule
    \multirow{3}{*}{Qwen-Max} 
        & MapCoder   & 19,427.28 & 9,410.52 & 10.00 \\
        & ChatDev    & 46,192.96 & 7,458.87 & 12.50 \\
        & MetaGPT    & 28,272.87 & 8,724.35 & 11.48 \\
    \midrule
    \multirow{3}{*}{Qwen-70B} 
        & MapCoder   & 22,906.81 & 10,447.55 & 10.00 \\
        & ChatDev    & 45,167.91 & 8,647.67 & 11.57 \\
        & MetaGPT    & 36,496.70 & 5,568.80 & 13.83 \\
    \bottomrule
  \end{tabular}%
  }
  \setlength{\abovecaptionskip}{0pt}   
\setlength{\belowcaptionskip}{0pt}
  \caption{Statistics of interaction overhead across representative frameworks. Turns = Dialogue Turns}
  \label{tab:token_turns}
  \vspace{-4mm}
\end{table}

% \noindent \textbf{{Analysis of Code Inconsistency}} -- \textbf{Vanilla LLMs generally maintain high code consistency, but integrating them into E2ESD frameworks, especially multi-agent ones, can reduce consistency.} This is likely due to exposure to excessive or irrelevant context and prompts. For example, MapCoder uses analogical prompting \citep{yasunagalarge}, retrieving project exemplars from its memory for few-shot enhancement before tackling a new project. Although intended to provide richer context, these retrieved projects often share only a general domain (e.g., web applications) without functional similarity, introducing noise rather than guidance. In contrast, frameworks that limit the coder agent's input to task-specific requirements and component analyses reduce interference, resulting in more consistent code generation.
\noindent \textbf{Analysis of Code Inconsistency} -- \textbf{{Vanilla LLMs generally maintain high code consistency, whereas this consistency often degrades under multi-agent settings.}}
This degradation is largely caused by exposure to excessive or irrelevant context.
For instance, MapCoder employs analogical prompting \citep{yasunagalarge}, retrieving prior projects from memory for few-shot guidance.
However, these exemplars often share only a coarse domain similarity (e.g., web applications) rather than functional relevance, introducing noise instead of actionable guidance.
In contrast, frameworks that restrict coder inputs to task-specific requirements and component analyses effectively reduce interference, leading to more stable code generation.

% \noindent\textbf{{Analysis of Requirement Missing}} -- \textbf{Flawed multi-agent designs can lead to overlooked requirements.} As Figure~\ref{fig:humanv} shows, \textit{Vanilla LLM} and \textit{ChatDev}, the only frameworks lacking access to requirement analysis during code generation\footnote{GPT-Engineer implicitly analyzes requirements via a ``think step by step'' approach.}, suffer severe missing requirement issues. In ChatDev, although a requirement analysis agent exists, the coding agent communicates only with the Executive Officer, who provides high-level guidance (e.g., HTML structure, JS functionality, CSS styling) but not detailed requirement specifications. This limited visibility increases the risk of missing critical requirements. Ensuring coding agents have direct access to comprehensive requirement analysis is therefore essential for complete software solutions.
\noindent \textbf{Analysis of Requirement Missing} --
\textbf{Flawed multi-agent designs can lead to overlooked requirements.}
As shown in Figure~\ref{fig:humanv}, \textit{Vanilla LLM} and \textit{ChatDev}—the only methods without direct access to requirement analysis during coding\footnote{GPT-Engineer implicitly performs requirement analysis via a ``think step by step'' process.}—exhibit the most severe requirement omissions.
In ChatDev, although a requirement analysis agent exists, the coding agent only receives high-level guidance from the Executive Officer, without access to detailed requirement specifications.
This limited visibility substantially increases the risk of missing essential requirements, highlighting the necessity of exposing coding agents to explicit requirement analyses.

% \noindent \textbf{Analysis of Requirement Misalignment} -- \textbf{Core component analysis helps align code with user requirements.} Figure~\ref{fig:humanv} shows multi-agent frameworks significantly reduce the misalignment compared to others. This improvement likely stems from explicitly integrating core component analysis—specifically, the identification of key structural and functional elements such as HTML layout, JavaScript logic (script.js), and CSS styles (styles.css). This analysis informs the coding agent, allowing a focus on user-driven logic and structure dictated by user requirements. Consequently, the generated code better reflects intended functionality and aligns with user requirements.
\noindent \textbf{Analysis of Requirement Misalignment} --
\textbf{Core component analysis helps align code with user requirements.}
As shown in Figure~\ref{fig:humanv}, multi-agent frameworks significantly reduce requirement misalignment.
This benefit mainly arises from identifying key structural and functional components, such as HTML layout and JavaScript logic, guiding the coding agent toward user-specified logic and structure.
Thus, the generated code more faithfully reflects the intended functionality.

% \noindent\textbf{{Analysis of Detail Mismatch}} -- \textbf{LLMs struggle with fine-grained detail control, which is exacerbated in multi-agent systems due to longer contextual chains.} As shown in Figure~\ref{fig:humanv}, detail mismatches occur across all frameworks, with Vanilla LLM producing the fewest (47) and Self-Collaboration the most (68). We categorize mismatches into three types: (1) \textit{Unhandled logical edge cases}, the most challenging type, often span multiple modules with complex state and parameter dependencies. LLMs struggle with them, and existing frameworks do not handle these dependencies. (2) \textit{Missing error handling or validation logic} and (3) \textit{UI component display inconsistencies}, both mainly arise from the progressive dilution of detail information. As requirements pass through a chain of agents that focus on major functionalities rather than special cases or fine-grained UI constraints, details become diluted by the time they reach the coding agent, leading to overlooked implementation details. Frameworks like MapCoder and ChatDev mitigate this by prepending each agent’s input with the original requirements and prior messages, preserving context and reducing omissions.
\noindent \textbf{Analysis of Detail Mismatch} --
\textbf{LLMs struggle with fine-grained detail control, which is exacerbated in multi-agent systems due to longer contextual chains.}
Figure~\ref{fig:humanv} shows that detail mismatches occur across all frameworks, with Vanilla LLM exhibiting the fewest and Self-Collaboration the most.
We group these errors into three categories:
(1) \textit{Unhandled logical edge cases}, which often span multiple modules and complex state dependencies and remain difficult for all frameworks;
(2) \textit{Missing validation or error handling}; and
(3) \textit{UI display inconsistencies}.
The latter two mainly result from progressive dilution of detailed constraints as requirements pass through multiple agents that prioritize high-level functionality.
Frameworks such as MapCoder and ChatDev partially mitigate this issue by repeatedly prepending original requirements and prior messages, helping preserve critical details.

\begin{figure}[t] 
    \centering
    % \vspace{-3mm}
    \includegraphics[width=0.45\textwidth]{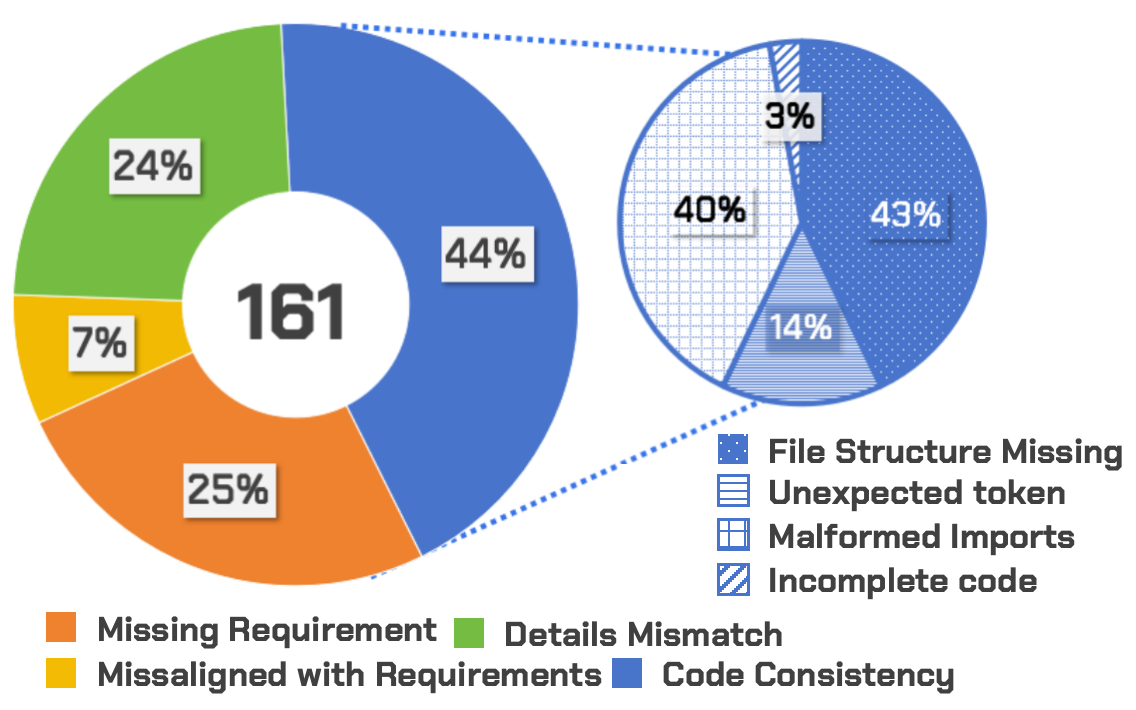} % 图片路径
    \caption{Error distribution of MetaGPT}
    \label{fig:humanv_meta}
    % \vspace{-5mm}
\end{figure}

% \noindent \textbf{{Analysis on \uline{MetaGPT}} -- Communication breakdowns within its multi-agent architecture significantly impair code consistency and requirement fidelity.} As seen in Figure \ref{fig:humanv_meta}, 44\% failures stem from code consistency issues, like missing files and syntax errors. Notably, over 43\% of these are caused by the programmer agent ignoring the architect's file structure, despite explicit constraints. Additionally, 54\% of syntax errors involve malformed imports or unmatched tokens, often resulting from the engineer agent being tasked with both code generation and tool invocation—prompt instructions for the latter outweigh the former by over 10×, introducing noise that disrupts code generation. Furthermore, 25\% of failures are due to missing requirements, with another 31\% stemming from inconsistent or incomplete implementation. This may stem from the product manager frequently rewriting or compressing the original user requirement, despite being instructed to restate it verbatim. We observe similar trends in HumanEval: MetaGPT with GPT-4o only achieves $\sim$50\% pass rate, with $\sim$30\% of completions failing to match required function names. This suggests that an over-engineered agent workflow degrades both adherence to specification and coding reliability.

\noindent \textbf{Analysis of \uline{MetaGPT}} --
\textbf{{Inter-agent communication breakdowns impair code consistency and requirement adherence.}}
As shown in Figure~\ref{fig:humanv_meta}, 44\% of MetaGPT failures arise from code inconsistency, including missing files and syntax errors.
Over 43\% of these occur when the programmer agent ignores the architect’s prescribed file structure.
In addition, 54\% of syntax errors stem from malformed imports or unmatched tokens, likely because the engineer agent is simultaneously responsible for code generation and tool invocation, with tool-related prompts outweighing coding instructions by over 10×.
Moreover, 25\% of failures are caused by missing requirements, and 31\% by incomplete or inconsistent implementations.
This is partly due to the product manager repeatedly rewriting or compressing the original requirements, rather than restating them as instructed.
Similar behavior is observed on HumanEval, where MetaGPT with GPT-4o achieves only $\sim$50\% pass rate, and nearly 30\% of outputs fail to match required function names.
Overall, these results indicate that excessive agent decomposition and communication overhead can undermine both specification adherence and coding reliability.

\section{Conclusion}
{
Our E2EDev represents a pivotal step toward standardizing the evaluation of E2ESD frameworks by providing detailed requirement specifications and reliable evaluation protocols. While LLMs have shown strong capabilities in isolated engineering tasks, our work highlights the significant challenges remaining in achieving fully automated E2ESD, constrained by both inherent model limitations and communication inefficiencies. Consequently, future work should prioritize designing effective, reliable, and cost-efficient E2ESD solutions that can fully realize the potential of LLMs in automating software development.}

% Our E2EDev takes an important step toward standardizing R2E-REP evaluation by introducing detailed requirement specifications and reliable, reproducible evaluation protocols. 
% Although LLMs perform well on code generation and function-level tasks, our evaluation shows that existing frameworks struggle to reliably scale to full R2E-REP pipelines. 
% Future research should therefore focus on developing R2E-REP frameworks that improve robustness and efficiency while reducing coordination and communication overhead.

% \clearpage
\section*{Limitations}
{
\noindent\textbf{Dataset Scale and Construction Complexity.}
}
{Unlike function-level tasks, repository-level coding tasks provide a more comprehensive assessment of LLMs’ capabilities in project construction and requirement fulfillment.}
However, constructing robust repository-level benchmarks remains challenging.
A primary difficulty arises from the diverse and often idiosyncratic implementations that LLMs generate for identical user requirements, which substantially increases the cost and complexity of reliable automated evaluation.
As a result, existing repository-level benchmarks, including E2EDev, are necessarily limited in scale (Table~\ref{tab:benchmark-comparison}).

% \yao{Furthermore, to ensure reliable automated evaluation, our current scope prioritizes front-end web applications, excluding CI/CD workflows and internal back-end logic validation.
% CI/CD processes are omitted as they rely heavily on specific environments and permissions, posing challenges for reproducible evaluation.
% Regarding back-end logic, we align with Requirements Engineering standards (ISO/IEC/IEEE 29148) by adopting a black-box testing approach: we verify system correctness through user-perceptible behaviors (via front-end interactions) rather than inspecting internal implementation details that are invisible to the user.
% }

\noindent\textbf{Scope Validity and Evaluation Protocol.} {Our strategic focus on Web applications is grounded in three methodological imperatives. First, this domain serves as a necessary "lower bound" for assessing E2ESD capabilities; the failures observed in this relatively deployable scenario reflect intrinsic limitations in LLM reasoning or framework design rather than domain-specific artifacts, suggesting that models struggling here are unlikely to succeed in more complex system-level settings. Second, to ensure rigorous reproducibility, we exclude CI/CD workflows and deep backend inspections, which rely heavily on volatile environmental dependencies that hinder consistent measurement. Third, aligned with Requirements Engineering standards (ISO/IEC/IEEE 29148), we adopt a black-box testing approach that verifies system correctness through user-perceptible behaviors via browser-based automation. This protocol allows us to validate whether the integrated system truly satisfies user intent without the noise of environment-specific internal implementation details.
}

{Despite these limitations, our work serves as a reasonable starting point for more rigorous evaluation of E2ESD tasks.
E2EDev contributes a standardized benchmark by providing explicit requirement specifications and reliable evaluation protocols, enabling systematic analysis of requirement fidelity and code consistency.
While LLMs perform well on function-level and isolated programming tasks, our results reveal substantial gaps in scaling these capabilities to fully automated E2ESD settings.
Given the significant resource investment required for benchmark construction, we plan to continuously expand E2EDev and maintain a public leaderboard to support reproducible and longitudinal evaluation.}

% E2EDev evaluates 244 requirements across 46 E2ESD tasks, with each requirement associated with an average of three test cases. Nearly every test case requires approximately 180 lines of code to execute, making manual validation highly labor-intensive (see Appendix~\ref{app:limit} for details). Although limited in size due to annotation constraints, E2EDev provides a robust foundation for evaluating model performance, similar to other high-quality, small-scale benchmarks\citep{chen2021evaluating, he2025deepmath, dong2024bamboo, rein2024gpqa}, as it spans diverse software development tasks and user interaction patterns. As software demands continue to grow, the diversity of development tasks is expected to expand. To keep pace with these changes, E2EDev will be periodically updated to keep pace with these changes.

\section*{Ethical Considerations}
Our work focuses solely on constructing the E2EDev benchmark for end-to-end requirement-to-executable code generation to evaluate the capabilities of LLM-based E2ESD frameworks. The study does not involve human subjects, sensitive personal data, or other ethical risks, and all source data are obtained from open-source communities. Any human annotation or verification of data was performed under controlled conditions. Human annotators were internal CS Master’s and PhD students, with priority given to those who had taken software testing or development courses. They were invited via email, participation was voluntary and unpaid, and written instructions explained the tasks and their intended use. Details are provided in Appendix~\ref{app:benchmark_details}. All experiments and analyses adhere to the ACL Code of Ethics, and no procedures, datasets, or methods in this study raise concerns related to privacy, bias, or safety.

\paragraph{Reproducibility}
To facilitate reproducibility, we publicly release the full E2EDev benchmark source code on GitHub (\url{https://github.com/SCUNLP/E2EDev}), along with the complete benchmark dataset on Hugging Face (\url{https://huggingface.co/datasets/GuanZhiZhao/E2EDev}). Detailed setup and usage instructions are provided in both Appendix~\ref{app:how_to_test} and the repository README. All experiments reported in this paper, including dataset processing, model evaluation, and metric computation, can be fully reproduced using these resources.

\paragraph{Artifacts License and Usage Terms}

All code and data used and created in this work comply with their respective licenses and usage terms. External code and datasets are open-source or publicly available, and their licenses (e.g., CC-BY 4.0 for data, MIT for code) are respected, including proper attribution to the original creators. The E2EDev benchmark we create is released under CC-BY 4.0 for data and MIT for code, and is intended for research purposes. 

\paragraph{LLMs Usage}
LLMs have become helpful tools for efficiently processing data in many recent works. In line with this practice, we used LLMs to accelerate parts of our annotation process, while all annotations were carefully validated by humans (see Appendix~\ref{app:benchmark_details}). Importantly, {all conceptual work and method design were conducted solely by the authors. LLMs were used only to polish the writing of this paper and did not contribute to the benchmark design, evaluation methods, analysis, or any core ideas of our work}.

\section*{Acknowledgments}
This work was supported in part by the National Natural Science Foundation of China (No. U25B201508, No. 62272330, and No.U24A20328); in part by the Singapore Ministry of Education (MOE) Academic Research Fund (AcRF) Tier 1 grant (Proposal ID: 24-SIS-SMU-002).

% Bibliography entries for the entire Anthology, followed by custom entries
% \bibliography{anthology,custom}
% Custom bibliography entries only
% \clearpage
\bibliography{custom}

\appendix

% \appendix

\section{Details on E2EDev Benchmark}
\label{app:benchmark_details}
% Figure \ref{fig:abstract} provides an overview of E2EDev.

% \begin{figure*}[h]
%   \centering
%   \includegraphics[width=0.98\textwidth]{fig/abstract.png}
%   \caption{Overview of E2EDev: a dataset and BDD-based automated evaluation pipeline for E2ESD tasks.}
%   \label{fig:abstract}
%   % \vspace{-3mm}
% \end{figure*}

\subsection{Details on Human Verification for Project Selection}
\label{app:selection}
We select source projects through a two-step process. First, we crawl GitHub repositories using keywords such as ``mini project'', ``software project'', ``web application'', and ``H5'', retaining only those with more than 500 stars. 

To ensure the realism, functionality, and suitability of each project in our benchmark, we conducted a structured manual verification process. In particular, each candidate project was independently reviewed by \textit{five annotators}, all holding at least a bachelor's degree in computer science or a related technical field, with formal training in software development. Before annotation, annotators were informed of the task's complexity, estimated time cost, and confirmed that the task posed no safety risks. All annotators followed a standardized instruction set across three key dimensions: \uline{Executability}, \uline{Functionality}, and \uline{Suitability}, detailed below. 
\begin{itemize}[leftmargin=*]
\item \uline{Executability}.  
Each project must run without major runtime errors under the following standard conditions. Notably, projects were tested locally using lightweight HTTP servers (e.g., \texttt{http-server}, Python’s \texttt{SimpleHTTPServer}). Projects that failed to render the main UI or blocked basic interactions were excluded.
\begin{itemize}
    \item The application must launch and run correctly in the latest stable version of \textit{Google Chrome}.
    \item No deprecated or non-standard dependencies should block execution.
    \item All static assets (HTML, CSS, JS) must load properly.
    \item The browser console should be free of critical errors or warnings affecting core functionality.
\end{itemize}

\item \uline{Functionality}.  
Annotators manually interacted with each component to ensure that it behaved as expected. Examples include the follows. Importantly, this step ensures that all included projects demonstrate reliable and observable behavior, making them suitable for LLM-based evaluation.
\begin{itemize}
    \item Correct handling of inputs and operators in calculators.
    \item Valid drag-and-drop functionality in task managers.
    \item Functional game logic (e.g., win/loss detection in Tic-Tac-Toe).
    \item Visual feedback elements such as animations or theme switching.
\end{itemize}

\item \uline{Suitability}.   
Projects must be feasible for evaluation by LLMs and human annotators within approximately \textit{five minutes}. Projects were excluded if:
\begin{itemize}
    \item They required extensive setup (e.g., database configuration, API keys). This was verified by inspecting the source code.
    \item Their main functionality was overly complex or time-consuming to test in practice.
\end{itemize}

\end{itemize}

% Each project was reviewed in parallel by five annotators working independently. Only projects that passed all three criteria unanimously were included. \chen{verification inter-consistency is required and how you resolve such inconsistency during the verification process}
\paragraph{Verification Protocol \& Inter-Annotator Agreement.}  
To refine and validate our annotation guidelines, we first conducted a pilot annotation round. We randomly selected 10 projects and asked all five expert annotators—each with formal training in software development—to independently label them based on the three key dimensions: \uline{Executability}, \uline{Functionality}, and \uline{Suitability}. Each project was assigned a binary label (1 for accepted, 0 for rejected) only if all three dimensions were satisfied. To ensure objectivity, annotators were not allowed to communicate during the process. After each round, we computed the Fleiss' Kappa score to evaluate inter-annotator agreement. If the score fell below 0.8, annotators convened to review disagreements, refine ambiguous criteria, and update the annotation protocol accordingly. In the first round, the Kappa score was 0.47. After one iteration of guideline refinement, the second round achieved a Kappa of 0.83. The finalized annotation protocol described above was then adopted for the full annotation process. To balance efficiency and agreement validation, we implemented a structured partial-overlap strategy during the main annotation phase. Specifically, we randomly selected 10\% of the full dataset (158 projects) for redundant labeling, with the remaining 90\% divided evenly among the five annotators. This design enabled us to measure agreement while minimizing redundant effort. On the overlapped subset, the final Fleiss' Kappa score reached 0.79, indicating substantial agreement and ensuring the overall reliability of the human verification process.

\paragraph{Final Sections and Statistics.}  
From over \textbf{158} initially shortlisted repositories, only \textbf{46} satisfied all verification criteria and were included in the final \textsc{E2EDev} benchmark.

\subsection{Details on HITL-MAA Implementation}
In this section, we present our semi-automated annotation framework for transforming source web application project to fine-grained requirements paired with executable tests. The process begins with a Test-ID Annotation step, which automatically pre-annotates key components in the source code by assigning meaningful and functionally aligned \textit{test identifiers} (Test-IDs). These identifiers serve as semantic anchors—providing structure-independent but functionally meaningful names for UI components—thus enabling reliable downstream requirement binding and test generation. The pre-annotated code is then passed to our human-in-the-loop multi-agent annotator (HITL-MAA), which refines the annotations and produces executable test cases aligned with each fine-grained requirement.

\label{app:maa}
\paragraph{PreAnnotation}

To address the challenges of annotating key component IDs across multiple source files (HTML and JS), especially considering dynamically generated elements and cross-file dependencies, we designed a \textbf{TestID-Annotation Multi-Agent Group} powered by GPT-4o. The annotation workflow is as follows:

% \begin{lstlisting}[language=Python, basicstyle=\ttfamily\small]
% # TestID Annotation Loop (executed by Multi-Agent Group)

% for file in project_files:
%     if file.type == "html":
%         annotate_interactive_components(file, strategy="add data-testid")
%     elif file.type == "js":
%         annotate_dynamic_elements(file, strategy="inject data-testid")

% # All agents share the same project context to ensure:
% # - Cross-file consistency
% # - Non-conflicting test-ids
% # - Semantic and human-readable test-id names

% # Final annotation is reviewed by human annotators to ensure correctness.
% \end{lstlisting}

\begin{algorithm}[htbp]\small
\caption{TestID Annotation Loop by Multi-Agent Group}
\ForEach{file in \texttt{project\_files}}{
    \If{file.type == ``html''}{
        annotate\_interactive\_components(file, strategy=``add data-testid'')\;
    }
    \ElseIf{file.type == ``js''}{
        annotate\_dynamic\_elements(file, strategy=``inject data-testid'')\;
    }
}
\vspace{1em}
\CommentSty{/* Agents share project context to ensure: */}\;
\Indp
- Cross-file consistency\;
- Non-conflicting test-ids\;
- Semantic and human-readable test-id names\;
\Indm
\vspace{1em}

\CommentSty{/* Final review by human annotators ensures correctness */}\;
\end{algorithm}

After annotation, human reviewers verified the validity of each project to prevent execution errors in GPT-4o-generated code. Using this workflow, no issues were identified across 46 projects. Specific prompt details are provided in Appendix \ref{app_prompt}.
Subsequently, the LLM uses this annotated source as context to generate requirements and test scripts that directly reference these test IDs. For example:

\begin{quote}
\textbf{Requirement:} When the user clicks the button with \texttt{test-id="login-btn"}, the system navigates to the dashboard.\\
\textbf{Python Step Implementation:} \texttt{... click('[test-id="login-btn"]') ...}
\end{quote}

% \vspace{2em}

\subsubsection{HITL-MAA Annotation Process}

\paragraph{Post-processing and Summary Generation.}
For each data entry, we generate a project-level summary based on the validated requirements. This summary includes a description of the overall application logic, key features, and a list of external resources (e.g., APIs, URLs). These resources are appended to the prompt provided to LLMs in downstream benchmarking to avoid execution failures due to inaccessible external dependencies.

\paragraph{Annotation Statistics.}
Approximately 20\% of the extracted requirements and nearly 50\% of the generated test cases required manual refinement by human annotators. This high correction rate of the test cases is not due to inconsistency between the requirement and test case, but rather due to the vagueness in the LLM-generated test descriptions—even when prompted to be detailed. Such vagueness increases the difficulty for the Test Automation Engineer in writing executable Python code.
To mitigate this, we adopt an iterative validation process. For step definition validation using dry-run, the system passes within 3 iterations (max\_iter1 = 3), with an average of 1.47 iterations. For full test script validation, the average number of required iterations is 4.62. In 20\% of the test cases, the maximum iteration limit (max\_iter2 = 6) was reached, prompting human intervention to complete the script refinement. All test scripts included in the dataset are verified to be both executable and correct.

\subsubsection{Protocols for Human-in-the-Loop \& Manual Verification}
\noindent\textbf{Human-in-the-Loop Protocol}. Human involvement in the E2EDev construction process occurs at three key points:
\begin{itemize}[leftmargin=*]
    \item \uline{Requirement Review and Refinement}. Human supervisors validate and refine the user requirements automatically generated by the Requirement Extractor Agent. This ensures accuracy, unambiguousness, and faithfulness to the source project's intended functionality.
    \item \uline{Test Case Refinement and Contribution}: Five human supervisors in software tesing collaboratively assess, refine, and contribute additional test cases and interaction patterns beyond those initially generated. They also ensure logical coherence and semantic alignment with requirements, while filtering out erroneous outputs.
    \item \uline{Test Script Refinement (as a fallback)}: Human supervisors are involved in refining executable test scripts when the Test Automation Engineer Agent reaches its maximum self-correction attempts and cannot autonomously resolve errors or logical inconsistencies.
\end{itemize}

\noindent\textbf{Manual Verification Protocol \& Inner-Agreement}. To ensure the consistency and quality of free-text requirements and test case validation, we implement a \textit{post-hoc gold evaluation} process. Five human annotators with expertise in software testing participate, with one senior expert designated as the \textit{Gold Checker}, responsible for reviewing others’ annotations. This process employs an iterative annotation and feedback mechanism to resolve inconsistencies efficiently. For both requirements and test cases, four regular annotators first perform validation. For requirement annotations, annotators interact with the web application through a browser to directly experience and verify functionality. For test cases, written in natural language as step-by-step operational guides, annotators follow the instructions to ensure consistency and validity against the requirements. Detected discrepancies or errors are revised before submission to the Gold Checker. The Gold Checker evaluates each annotation on three binary criteria: \textit{equivalence}, \textit{completeness}, and \textit{correctness}, where a score of 1 indicates the criterion is met and 0 otherwise. Only when the Fleiss’ Kappa score for all three aspects exceeds 0.7 does the Gold Checker consolidate a final version of the requirement or test case, which may include discarding it if necessary. If the agreement falls below the threshold, the Gold Checker provides detailed feedback, and annotators repeat the annotation process. Newly proposed test cases also undergo gold evaluation. In such cases, all three criteria are initially scored as 0 by the annotators, and the Gold Checker reviews the new case before returning it to all regular annotators for reconsideration and re-annotation. If annotators fail to recognize a new test case, the Gold Checker similarly returns it for re-annotation. This approach balances thoroughness and efficiency, resulting in a reliable, scalable, and reproducible manual annotation protocol suitable for complex free-text requirement and test case validation tasks. It ensures high-quality, consistent annotations for downstream applications.

\noindent\textbf{Reliability Audit \& Scalability Verification.} 
{To further mitigate potential bias from the initial team size and validate the generalizability of our quality standards, we conducted an extended \textit{reliability audit}. 
We expanded the expert pool to \textbf{10 evaluators} (adding five distinct external experts to the original group) and performed a blind review on a stratified random sample of \textbf{100 test cases} from the final dataset. 
This broader panel independently assessed the sample under the same strict criteria. 
The resulting inter-rater agreement yielded a \textbf{Fleiss’ Kappa of 0.81}, demonstrating that our original 5-expert protocol aligns with the consensus of a larger expert body and ensuring the dataset's high quality is statistically robust.}

\begin{algorithm*}[t]\small
\caption{Semi-Automated Annotation Workflow in HITL-MAA}
% \KwInput{Source code annotated with \texttt{test-id}s}
% \KwOutput{Executable and validated end-to-end test scripts with aligned requirements}
\KwIn{Source code annotated with \texttt{test-id}s}
\KwOut{Executable and validated end-to-end test scripts with aligned requirements}
\CommentSty{/* \textbf{Code Analyzer} */}\;
\ForEach{\texttt{HTML} file in project}{
    Analyze frontend framework and interactive components\;
}
\ForEach{\texttt{JS} file in project}{
    Analyze logic functionality and connections with frontend elements\;
}
Summarize the analysis for downstream modules\;

\CommentSty{/* \textbf{Requirement Extractor} */}\;
Extract contextual requirements using both source code and analysis summary\;
\texttt{user\_req\_lst} $\leftarrow$ \texttt{RequirementExtractor.extract(context)}\;

\ForEach{\texttt{req} in \texttt{user\_req\_lst}}{
    Human annotator validates and corrects the extracted requirement\;

    \CommentSty{/* \textbf{Test Case Generator} */}\;
    \texttt{test\_cases} $\leftarrow$ \texttt{TestCaseGenerator.generate(source code, req)}\;
    Human annotator validates generated test cases\;
    \texttt{validated\_cases} $\leftarrow$ validated test cases\;
    \texttt{refined\_req} $\leftarrow$ Refine requirement based on validated test cases to ensure alignment\;

    \ForEach{\texttt{test\_case} in \texttt{validated\_cases}}{
        \CommentSty{/* \textbf{Test Automation Engineer} */}\;
        \texttt{impl} $\leftarrow$ \texttt{write\_test\_script(source code, refined\_req, test\_case)}\;

        \CommentSty{/* \textbf{Step Checker} */}\;
        \For{$i = 1$ \KwTo max\_iter1}{
            Run \texttt{behave --dry-run} to validate step implementation\;
            \If{all steps pass}{
                \textbf{break}\;
            }
            \Else{
                Modify step implementation based on error message\;
            }
        }

        \CommentSty{/* \textbf{Test Runner} */}\;
        \For{$j = 1$ \KwTo max\_iter2}{
            Run full test using \texttt{behave} and parse log\;
            \If{test script passes}{
                \textbf{break}\;
            }
            \Else{
                Modify test script based on failure log\;
            }
        }
        \If{maximum iterations reached and test still fails}{
            Human annotator intervenes to revise and finalize test script\;
        }
    }
}

\CommentSty{/* \textbf{The specific prompt details are provided in Appendix \ref{app_prompt}.} */}\;
% For each data entry, generate a project-level summary based on validated requirements\;
% Include description, key features, and external resources (e.g., APIs, URLs) in prompt to improve LLM's understanding\;

% \CommentSty{/* \textbf{Annotation Statistics} */}\;
% 20\% of requirements and nearly 50\% of test cases required manual refinement\;
% High refinement rate was due not to inconsistency but to vague generation, which increases the burden on test script generation\;
% \texttt{max\_iter1 = 3}, \texttt{max\_iter2 = 6}, average step validation turn = 1.47\;
% Average final test script validation = 4.62 turns; 20\% of scripts reached max iterations and required human intervention\;

\end{algorithm*}

%\clearpage

\subsubsection{Analysis on Human Labor during HITL-MAA}
We conducted a detailed analysis of human effort and behavior during dataset construction using our human-in-the-loop semi-automated annotation framework. Prior to designing this framework, we engaged five experts without a background in software testing to generate requirements and test cases for each source project via conversational interactions with a large language model (LLM) through a chat interface. This manual process proved to be highly time-consuming, labor-intensive, and yielded poor annotation consistency.
During this initial stage, we organized an intensive one-week annotation task where each annotator was assigned 9 independent projects plus 1 additional project overlapping across all annotators to measure inter-annotator agreement. Ultimately, each annotator completed only 5 projects on average. The proportion of executable tests was approximately 80\%, including the consistency check projects. The average number of test cases per project was only 6.5. Aggregating the test cases from all five annotators and excluding semantically redundant cases resulted in just 14 unique test cases. Calculating Fleiss’ Kappa on these 14 cases yielded a low score of 0.23, indicating poor agreement. Overall, the manual annotation process was not only inefficient but also suffered from incompleteness and quality issues.

In contrast, applying our semi-automated annotation framework substantially improved annotation quality, coverage, consistency, and efficiency. Specifically, all generated executable tests were 100\% runnable and logically correct. On average, each project included 3 distinct requirements and 15 test cases, with a maximum of 34 test cases in some projects. Human validation and refinement were necessary for less than 50\% of the annotations: approximately 20\% of requirement refinements and 50\% of test case adjustments. Moreover, manual modifications to test scripts accounted for less than 20\% of the total, demonstrating the high reliability of the automated generation.

Despite incorporating a seemingly time-consuming post-hoc consistency evaluation, the average annotation and verification time per project was only 3.5 hours, with the shortest project completed in 30 minutes. Regarding computational resources, the backbone LLM employed was GPT-4o. On average, annotating one project consumed 140,000 prompt tokens and 22,000 completion tokens. This corresponds to a cost of approximately \$0.48 per project annotation, indicating a highly efficient process.

\begin{table*}[!ht]
\centering
\resizebox{\textwidth}{!}{
\begin{tabular}{lcc}
\toprule
\textbf{Metric} & \textbf{Manual Annotation} & \textbf{Semi-Automated Annotation} \\
\midrule
Average Projects Completed per Annotator & 5 & \textgreater 9 \\
Executable Test Ratio & $\sim$80\% & 100\% \\
Average Test Cases per Project & 6.5 & 15 (max 34) \\
Fleiss’ Kappa (Test Case Agreement) & 0.23 & 0.79 (approx.) \\
Human Refinement (Requirements) & N/A & 20\% \\
Human Refinement (Test Cases) & N/A & 50\% \\
Human Refinement (Test Scripts) & N/A & \textless 20\% \\
Average Annotation + Verification Time & $\gg$ 3.5 hours & 3.5 hours (min 30 min) \\
Average Token Consumption (Prompt / Completion) & N/A & 140k / 22k tokens \\
Estimated Cost per Project & N/A & \$0.48 \\
\bottomrule
\end{tabular}
}
\caption{Comparison of Manual vs. Semi-Automated Annotation}
\label{tab:human-labor-analysis}
\end{table*}

Overall, our semi-automated annotation framework significantly enhances annotation completeness and efficiency, while maintaining high-quality and consistent results, demonstrating an effective human-in-the-loop approach to dataset construction.

\subsection{Case Studies of E2EDev}
\label{ap:real_case_06}
% \begin{figure*}[t]
% \centering
\begin{tcolorbox}[
    breakable,
    colback=white!98!black,
    colframe=black!70!white,
    boxsep=4pt,
    top=2pt,
    title={\textbf{\large E2ESD\_Bench\_06}},
    breakable, 
    coltitle=black,
    colbacktitle=gray!20,
    bottom=2pt
]
\textbf{"""prompt"""}
% \begin{lstlisting}[breaklines=true, basicstyle=\ttfamily\small, aboveskip=1pt, belowskip=1pt, columns=fullflexible]
\begin{lstlisting}[breaklines=true, basicstyle=\ttfamily\small, aboveskip=1pt, belowskip=1pt, columns=fullflexible]
You are tasked with implementing a complete web application using HTML, JavaScript, and CSS. Your implementation must strictly follow the specifications described below.

### SUMMARY
# overview
    The Playable Piano web application allows users to interact with a virtual piano. Users can click on piano keys or press corresponding keyboard keys to play notes. The application includes features for adjusting volume and toggling the visibility of key labels.

# predefined_options
    The web application initializes with a default volume level set to 0.5, and piano key labels are displayed by default upon launch. It features a virtual piano with both black and white keys, each mapped to specific keyboard characters. The black keys correspond to the uppercase letters W, E, T, Y, U, O, and P, while the white keys correspond to A, S, D, F, G, H, J, K, L, and ;. Each key element in the DOM is assigned a unique data-test-id attribute in the format data-test-id=\"piano-key-'x'\", where 'x' is the lowercase version of the corresponding keyboard character (e.g., the key for W uses data-test-id=\"piano-key-'w'\").

# external_resources
    The corresponding audio files for each piano note are stored in the tunes directory, with filenames matching the lowercase key characters followed by .wav, such as tunes/a.wav, tunes/;.wav, tunes/d.wav, tunes/e.wav, and so on, covering all relevant keys including w, e, t, y, u, o, p, a, s, d, f, g, h, j, k, l, and ;.

# external_js_libraries
    No external JavaScript libraries are used in this application.


### Functional Requirements
    Implement the following features as described. For each requirement, make sure the HTML structure, JavaScript behavior, and CSS styles match the specifications exactly.
    
# REQUIREMENTS:
- Requirement 1:
    When the user clicks any visible piano key on the webpage, identified by its data-testid (e.g., 'piano-key-a' for the white key 'a' or 'piano-key-w' for the black key 'w'), the system must play the corresponding sound file from the 'tunes' directory (e.g., 'tunes/a.wav' for key 'a'). The clicked key must receive the 'active' class to visually highlight it, and the highlight must be removed after 150ms by removing the 'active' class.
- Requirement 2:...
- Requirement 3:...
- Requirement 4:...
- Requirement 5:...
\end{lstlisting}

\vspace{2mm}
\hrule height 0.8pt
\vspace{2mm}

\textbf{"""requirement\_summary"""}
\begin{lstlisting}[breaklines=true, basicstyle=\ttfamily\small, aboveskip=1pt, belowskip=1pt, columns=fullflexible]
### overview
    The Playable Piano web application allows users to interact with a virtual piano. Users can click on piano keys or press corresponding keyboard keys to play notes. The application includes features for adjusting volume and toggling the visibility of key labels.
    
### predefined_options
    The web application initializes with a default volume level set to 0.5, and piano key labels are displayed by default upon launch. It features a virtual piano with both black and white keys, each mapped to specific keyboard characters. The black keys correspond to the uppercase letters W, E, T, Y, U, O, and P, while the white keys correspond to A, S, D, F, G, H, J, K, L, and ;. Each key element in the DOM is assigned a unique data-test-id attribute in the format data-test-id=\"piano-key-'x'\", where 'x' is the lowercase version of the corresponding keyboard character (e.g., the key for W uses data-test-id=\"piano-key-'w'\").
    
### external_resources
    The corresponding audio files for each piano note are stored in the tunes directory, with filenames matching the lowercase key characters followed by .wav, such as tunes/a.wav, tunes/;.wav, tunes/d.wav, tunes/e.wav, and so on, covering all relevant keys including w, e, t, y, u, o, p, a, s, d, f, g, h, j, k, l, and ;.

### external_js_libraries
    No external JavaScript libraries are used in this application.
\end{lstlisting}

\vspace{2mm}
\hrule height 0.8pt
\vspace{2mm}

\textbf{"""fine\_grained\_reqs"""}
\begin{lstlisting}[breaklines=true, basicstyle=\ttfamily\small, aboveskip=1pt, belowskip=1pt, columns=fullflexible]
"1": "When the user clicks any visible piano key on the webpage, identified by its data-testid (e.g., 'piano-key-a' for the white key 'a' or 'piano-key-w' for the black key 'w'), the system must play the corresponding sound file from the 'tunes' directory (e.g., 'tunes/a.wav' for key 'a'). The clicked key must receive the 'active' class to visually highlight it, and the highlight must be removed after 150ms by removing the 'active' class.",
"3": ...
"4": ...
"5": ...
"6": ...
\end{lstlisting}

\vspace{2mm}
\hrule height 0.8pt
\vspace{2mm}

\textbf{"""reference\_answer"""}
\begin{lstlisting}[breaklines=true, basicstyle=\ttfamily\small, aboveskip=1pt, belowskip=1pt, columns=fullflexible]
https://github.com/GuanZhizhao/E2EDev/tree/main/E2ESD_Bench_06
\end{lstlisting}

\vspace{2mm}
\hrule height 0.8pt
\vspace{2mm}

\textbf{"""excutable\_tests"""}
\begin{lstlisting}[breaklines=true, basicstyle=\ttfamily\small, aboveskip=1pt, belowskip=1pt, columns=fullflexible]
### 1
# test_case
Feature: Play piano sound and visually highlight the key when clicked
- The system must play the corresponding piano sound and visually highlight the key when the user clicks on a piano key. The highlight should disappear after 150ms.

Scenario: [Normal] User clicks on a white piano key
- Given the webpage is loaded and the piano keys are visible
- When the user clicks on the white piano key with data-testid "piano-key-a"
- Then the system must play the sound
- And the key with data-testid "piano-key-a" must have the "active" class added
- And the "active" class must be removed from the key with data-testid "piano-key-a" after 150ms.
    
# step_code
from behave import given, when, then
from selenium import webdriver
from selenium.webdriver.common.by import By
from selenium.webdriver.support.ui import WebDriverWait
from selenium.webdriver.support import expected_conditions as EC
import time

@given('the webpage is loaded and the piano keys are visible')
def step_given_webpage_loaded(context):
    context.driver = webdriver.Chrome()
    context.driver.get(f"file://{file_path}")
    context.driver.maximize_window()

    WebDriverWait(context.driver, 10).until(
        EC.visibility_of_element_located((By.CSS_SELECTOR, "[data-testid='piano-key-a']"))
    )
    time.sleep(1)  # Allow time for the page to fully load

@when('the user clicks on the white piano key with data-testid "piano-key-a"')
def step_when_user_clicks_piano_key(context):
    piano_key = WebDriverWait(context.driver, 10).until(
        EC.element_to_be_clickable((By.CSS_SELECTOR, "[data-testid='piano-key-a']"))
    )
    piano_key.click()

@then('the system must play the sound')
def step_then_system_plays_sound(context):
    pass

@then('the key with data-testid "piano-key-a" must have the "active" class added')
def step_then_key_has_active_class(context):
    piano_key = context.driver.find_element(By.CSS_SELECTOR, "[data-testid='piano-key-a']")
    class_list = piano_key.get_attribute("class").split()
    assert "active" in class_list, "The 'active' class was not added to the key"

@then('the "active" class must be removed from the key with data-testid "piano-key-a" after 150ms')
def step_then_active_class_removed(context):
    time.sleep(0.15)
    piano_key = context.driver.find_element(By.CSS_SELECTOR, "[data-testid='piano-key-a']")
    class_list = piano_key.get_attribute("class").split()
    assert "active" not in class_list, "The 'active' class was not removed from the key"

def after_scenario(context, scenario):
    if hasattr(context, 'driver'):
        context.driver.quit()
    
# test_case ...
# step_code ...
# test_case ...
# step_code ...

### 3 ...
### 4 ...
### 5 ...
### 6 ...
\end{lstlisting}
\end{tcolorbox}
% \end{figure*}
\subsection{Statistics and Features of E2EDev}
\textbf{Statistics}. Dataset statistics are shown in Table~\ref{tab:bench_statistic}, with example entries provided in the Appendix~\ref{ap:real_case_06}.

\textbf{Features}. E2EDev consists of diverse software development tasks evenly distributed across seven common types of web applications. As shown in Figure \ref{fig:statistic}(a) , the three most frequent categories include: \textit{Mathematics \& Conversion Tools}, such as basic calculators and utilities for tax or BMI calculation; \textit{Account \& Form Management}, covering simple CRUD systems where users submit forms and manage records; and \textit{Mini Games \& Interactive Entertainment}, including lightweight games like Tic-Tac-Toe. E2EDev also captures a wide range of common user interaction patterns. As shown in Figure~\ref{fig:statistic}(b), over 90\% of projects involve basic interactions like clicking and typing, while others include more advanced actions such as sliders, selectors, and drag-and-drop operations. In addition to basic DOM manipulation, the dataset includes non-trivial web application features. For example, nearly one-third of the projects require reasoning over complex mathematical functions, demanding precise function generation to satisfy user needs. About one-quarter of the tasks involve interaction with local storage. Another quarter rely on external APIs for data fetching, audio playback, and more—testing whether models can correctly integrate external resources.
Taken together, the diversity in application types, interaction patterns, and technical complexity demonstrates that E2EDev provides a representative and realistic benchmark for end-to-end web application development.

\label{ap:e2edev_feature}
\begin{figure*}[htbp]
  \centering
\includegraphics[width=1.0\textwidth]{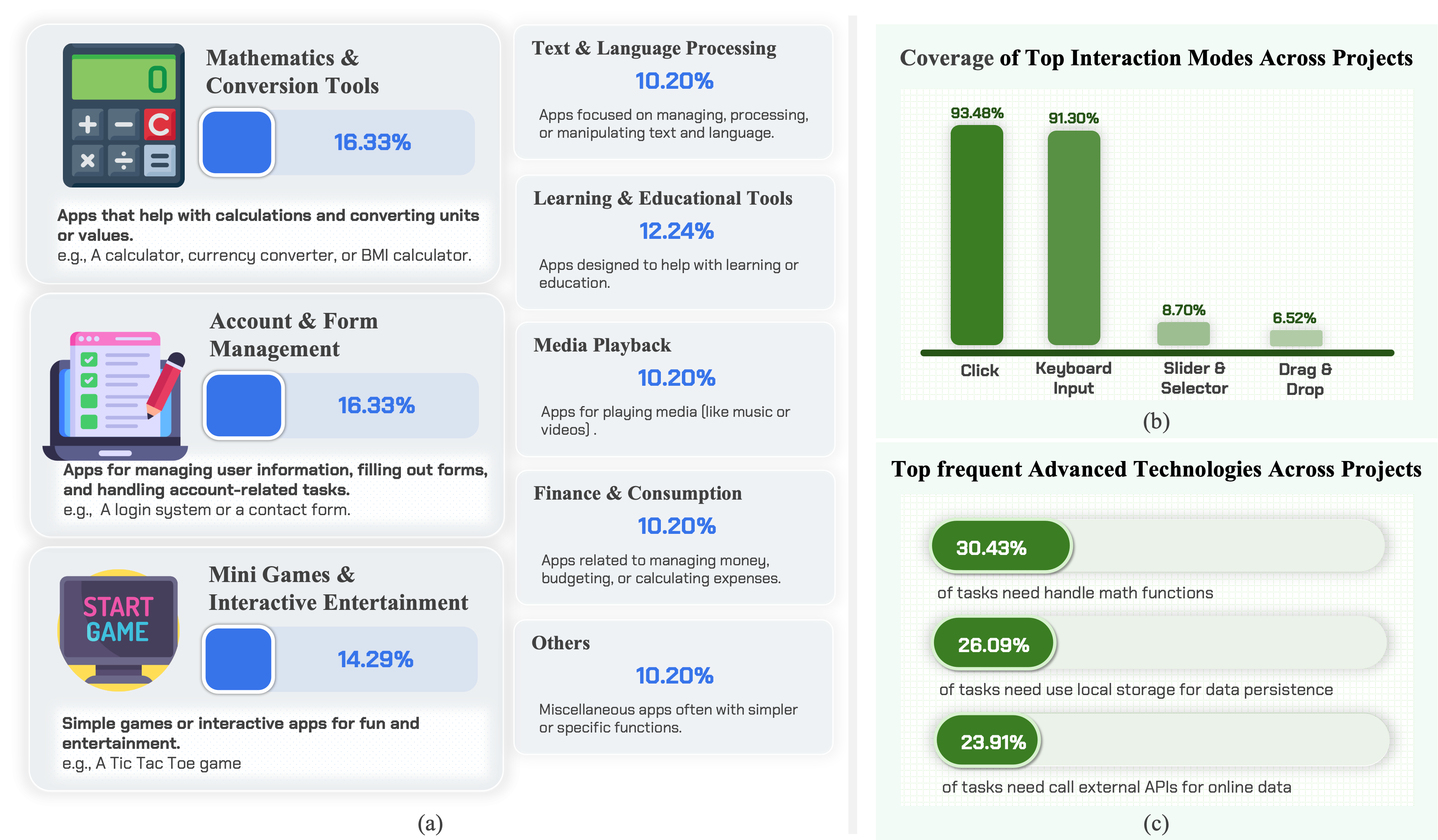}
  \caption{This figure illustrates the features of E2EDev, which encompasses seven common types of web applications. It covers a wide range of user interaction types and incorporates various advanced web technologies.}
  \label{fig:statistic}
\end{figure*}

\section{Implementation Details}
\subsection{Implementation of Baselines}
\label{app:baselines}

%%%%%%% COPY 
We select a diverse set of representative \textbf{requirement-to-executable software development} frameworks across various LLM backbones. These include: \uline{(1) Vanilla LLM} that generates the full codebase via simply prompting LLM using user requirements without external scaffolding. \uline{(2) Single-Agent Framework}, \textit{GPT-Engineer}~\citep{gptengineer}, {which builds the complete codebase in a single reasoning pass with minimal external coordination while maintaining a modular development workflow}. \uline{(3) Multi-Agent Frameworks}, a dominant paradigm, where the development process is decomposed into specialized agent roles based on classic software engineering principles~\citep{sommerville2011software}. %This includes \textit{Self-Collaboration~\citep{dong2024self}}, a streamlined framework for requirement analysis and implementation; \textit{MetaGPT~\citep{hong2024metagpt}}, a structured framework with specialized agents following the Waterfall model~\citep{royce1987managing}; \textit{MAPCoder~\citep{islam2024mapcoder}}, an advanced framework enhanced by incorporating prior project knowledge, and \textit{ChatDev~\citep{qian2024chatdev}}, a dialogue-driven, dynamic framework, where each development stage is co-managed by two agents who collaboratively determine transitions to the next phase.
We benchmark with a wide range of state-of-the-art multi-agent framework, including \textit{Self-Collaboration~\citep{dong2024self}}, \textit{MetaGPT~\citep{hong2024metagpt}}, \textit{MapCoder~\citep{islam2024mapcoder}}, and \textit{ChatDev~\citep{qian2024chatdev}}. {We additionally include two widely used coding agents, \textit{OpenHands}~\citep{wang2024openhands} and \textit{SWE-Agent}~\citep{yang2024swe}, as reference systems.
These agents are designed for general-purpose software engineering scenarios with extensive tool and environment interaction, which differ from the requirement-to-executable software generation setting considered in E2EDev. Their results are reported in the Appendix ~\ref{app:openhands_swe}.}

% \begin{itemize}[leftmargin=*]
% \item \textit{Self-Collaboration~\citep{dong2024self}}: A streamlined framework for requirement analysis and implementation.
% \item \textit{MetaGPT~\citep{hong2024metagpt}}: A structured framework with specialized agents following the Waterfall model~\citep{royce1987managing}.
% \item \textit{MAPCoder~\citep{islam2024mapcoder}}: An advanced framework enhanced by incorporating prior project knowledge.
% \item \textit{ChatDev~\citep{qian2024chatdev}}: A dialogue-driven, dynamic framework, where each development stage is co-managed by two agents who collaboratively determine transitions to the next phase.
% \end{itemize}
Finally, we adopt a range of backbones from two vendors \citep{hurst2024gpt,hui2024qwen2}, covering various scales and architectures: {Claude-Haiku 4.5}, GPT-4o, GPT-4o-mini, Qwen-7B (Qwen2.5-7B-Instruct), Qwen-70B (Qwen2.5-72B-Instruct), and the mixture-of-experts Qwen-Max (Qwen2.5-Max). This enables controlled analysis of framework versus model backbone effects. Implementation details are presented below.

\textbf{Implementations of LLM-based Frameworks.} We reproduced eight LLM-based code generation frameworks using their official GitHub implementations (if available): \textit{Vanilla LLM}, \textit{GPT-Engineer}, \textit{Self-Collaboration Code Generation}, \textit{MapCoder}, \textit{ChatDev}, \textit{MetaGPT}, {\textit{OpenHands}, and \textit{SWE-Agent}}.

\noindent \textbf{Vanilla LLM.} We directly fed the task prompts (i.e., user requirements) along with the following structured instruction:

\begin{lstlisting}[breaklines=true, basicstyle=\ttfamily\small, aboveskip=1pt, belowskip=1pt, columns=fullflexible]
H5_DEVELOPER = """
    I want you to act as an H5 developer on our development team. ..., 
    write HTML, JavaScript, and CSS code ... provide the updated HTML, 
    JavaScript, and CSS code ...
"""
\end{lstlisting}

\noindent \textbf{GPT-Engineer.} GPT-Engineer builds a complete codebase in a single reasoning pass with minimal external coordination, maintaining a modular workflow. We used the official Python packages and executed the frameworks directly with task prompts, without modifying internal workflows.

\noindent \textbf{Self-Collaboration Code Generation.} This framework incorporates software-development methodology by assembling a team of three LLM roles—analyst, coder, and tester—responsible for analysis, coding, and testing. Originally designed for Python-only tasks, we adapted it to our multi-language (HTML/CSS/JavaScript) setting by modifying the system prompts. Specifically, we replaced the default PYTHON\_DEVELOPER role with an {H5\_DEVELOPER} role:

\begin{lstlisting}[breaklines=true, basicstyle=\ttfamily\small, aboveskip=1pt, belowskip=1pt, columns=fullflexible]
H5_DEVELOPER = """
    I want you to act as an H5 developer on our development team. ..., 
    write HTML, JavaScript, and CSS code ... provide the updated HTML, 
    JavaScript, and CSS code ...
"""
\end{lstlisting}

\noindent \textbf{MapCoder.} MapCoder consists of four agents simulating the stages of software development: recalling examples, planning, code generation, and debugging. Similar to Self-Collaboration, it was originally Python-focused; we adapted it to support HTML/CSS/JavaScript by modifying its agent system prompts accordingly.

\noindent \textbf{ChatDev.} ChatDev is a chat-powered software development framework where LLM-driven agents communicate via structured chat chains. Agents contribute to design, coding, and testing phases through multi-turn dialogues, guided by communicative strategies to reduce hallucinations. We implemented it using the official GitHub source without modifications.

\noindent \textbf{MetaGPT.} MetaGPT is a meta-programming framework that incorporates human workflow principles into LLM-based multi-agent collaboration. It encodes Standardized Operating Procedures (SOPs) into prompt sequences, enabling agents with domain expertise to verify intermediate results and reduce errors. Using an assembly-line paradigm, MetaGPT assigns diverse roles to multiple agents to efficiently decompose complex tasks. We used the official implementations and Python packages, executing them directly with task prompts without internal modifications.

\noindent \textbf{OpenHands.}
{OpenHands is an interactive human--agent software development framework that supports
multi-turn collaboration for iterative project construction.
The system relies on continuous human feedback and intervention throughout the
development process, which makes large-scale automated evaluation time-consuming.
To align it as closely as possible with the E2EDev setting, we restrict OpenHands
to a single execution without any human interaction.
We use the official GitHub implementation without modification and report the
results in Appendix~\ref{app:openhands_swe}.}

\noindent \textbf{SWE-Agent.}
{SWE-Agent is designed for repository-level software engineering tasks, where the goal is to resolve GitHub issues by modifying an existing codebase, as evaluated in the SWE-bench setting.
Its input consists of an existing repository and a natural language issue description, and the output is a patched repository rather than a newly generated project.
For our setting, we replace the original prompt template—which instructs the model to read a project from \texttt{\{\{repo\_path\}\}} and modify it according to \texttt{\{\{issue\}\}}—with a new template that directs the model to generate a runnable project directly from the given user requirement.
We implement SWE-Agent using the official GitHub source without further modification, and report its results in Appendix~\ref{app:openhands_swe}.}

\noindent \textbf{Evaluation Coverage.}
{We consider eight LLM-based requirement-to-executable software development frameworks in total.
Six frameworks (\textit{Vanilla LLM}, \textit{GPT-Engineer}, \textit{Self-Collaboration},
\textit{MapCoder}, \textit{ChatDev}, and \textit{MetaGPT}) are evaluated across all six LLM backbones.
Due to their different design goals and substantially higher evaluation cost,
\textit{OpenHands} and \textit{SWE-Agent} are evaluated only on
\textit{Claude-Haiku-4.5} and \textit{GPT-4o},
with their results reported in Appendix~\ref{app:openhands_swe}.}

\subsection{Implementation of Evaluation Metrics}
\label{impl_metric}

We evaluate both \uline{code effectiveness} and \uline{generation efficiency} in E2ESD. On one hand, code effectiveness measures how well the generated software satisfies user requirements. In this regard, we propose three task-level metrics:
\begin{itemize}[leftmargin=*]
\item \textit{Req.Acc:} The average correctness across all requirement tasks per project. For a project with $M$ requirements, if $N$ are fully satisfied (i.e., all associated test cases pass), then:
\begin{equation}
\text{Req.Acc} = \frac{N}{M}
\end{equation}
\item \textit{Test.Acc:} The proportion of all passed test cases across a project:
\begin{equation}
\text{Test.Acc} = \frac{\text{Total Passed Test Cases}}{\text{Total Test Cases}}
\end{equation}
\item \textit{Balanced Score:} To mitigate bias from varying test case granularity per requirement, we define a weighted effectiveness score:
\begin{equation}
\text{Balanced} = \alpha \cdot \text{Req.Acc} + \beta \cdot \text{Test.Acc}
\end{equation}
where $\alpha$ and $\beta$ control the relative importance of requirement-level and test-level correctness, respectively. In this paper, we set $\alpha = 0.6$ and $\beta = 0.4$, reflecting our intuition that requirement-level accuracy (\text{Req.Acc}) plays a more critical role in evaluating the overall performance than test-level accuracy (\text{Test.Acc}). {Due to the limited number of benchmark tasks at the project level, we run each
framework three times on every task.
The effectiveness scores reported in Table~\ref{tab:model_performance} and
Table~\ref{tab:model_performance_additional} are computed as the mean and
standard deviation across runs.}
\end{itemize}
On the other hand, generation efficiency is assessed using three metrics:
\begin{itemize}[leftmargin=*]
\item \textit{Cost (USD):} Estimated based on token usage and API pricing. The total cost is computed as:
\begin{equation}
\begin{split}
\text{Cost} &= (\text{Input Tokens} \times \text{Input Price}) \\
&\quad + (\text{Output Tokens} \times \text{Output Price})
\end{split}
\end{equation}
where the \textit{Input Price} and \textit{Output Price} refer to the per-token charges for input and output tokens, respectively, as defined by the corresponding LLM provider.
The pricing used in our experiments is summarized in Table~\ref{tab:api-pricing}.
\begin{table}[htbp]
\centering
\resizebox{\columnwidth}{!}{%
\begin{tabular}{lcc}
\toprule
\makecell{\textbf{Model}} & 
\makecell{\textbf{Input Price} \\ \textbf{(USD/token)}} & 
\makecell{\textbf{Output Price} \\ \textbf{(USD/token)}} \\
\midrule
\texttt{gpt-4o}         & 0.0025  & 0.0061 \\
\texttt{gpt-4o-mini}    & 0.00015 & 0.00014 \\
\texttt{qwen-7b}        & 0.000069 & 0.00006 \\
\texttt{qwen-70b}       & 0.00055 & 0.0016 \\
\texttt{qwen-max}       & 0.00033 & 0.0013 \\
\bottomrule
\end{tabular}
}
\caption{LLM API Pricing}
\label{tab:api-pricing}
\end{table}
\item \textit{Carbon Footprint (CO\textsubscript{2}):} Computed as the sum of operational and embodied carbon, following \citep{faiz2023llmcarbon}.  We assume NVIDIA A100 as the default chip for estimation purposes.
\begin{equation}
\text{Footprint} = \text{Flops} \cdot \frac{TDP}{\eta} \cdot \text{PUE} \cdot \text{CI} + \text{EC}
\end{equation}
where:
\begin{itemize}
    \item \textbf{FLOPs} $= 2 \times (\text{Input Tokens} + \text{Output Tokens}) \times \text{Active Parameters}$
    \item \textbf{Active Parameters:} The number of model parameters actively involved in the computation for each token. 
    \item \textbf{TDP:} Thermal Design Power of the GPU, measured in watts. For A100 40GB SXM, we assume 400W.
    \item \textbf{$\eta$}: Computational efficiency, measured in FLOPs/Watt. This measures the floating-point operations per watt (FLOPs/W). For A100, we assume 624 TFLOPs (FP16 with sparsity), noting that mixed-precision may be used in practice.
    \item \textbf{PUE:} Power Usage Effectiveness. A metric reflecting datacenter energy overhead. PUE for the Qwen series is 1.2, while for the GPT series it is 1.1. 
    \item \textbf{CI:} Carbon intensity, measured in kgCO\textsubscript{2}/kWh, representing the amount of CO\textsubscript{2} emitted per unit energy consumed\citep{patterson2021carbon}.CI is 625 g/kWh for the Qwen series and 407 g/kWh for the GPT series .
    \item \textbf{EC:} Embodied carbon (CO\textsubscript{2} generated during manufacturing and infrastructure), usually treated as a negligible constant in inference scenarios.
\end{itemize}

% \end{itemize}
\item \textit{Duration (s):} Total wall-clock time for software generation.
\end{itemize}

% \subsection{Implementation of Human Evaluation in Section \ref{indepth}}
% \label{imp_indepth}
% 我们在Section \ref{indepth} 对不同框架在实现software requirement 的层面进行了detail的需求实现失败分析，我们随机从是E2EDev 中随机抽取来10个task 在30个不同方法（5个backbone*6 framworks）。具体我们聘请了5个expert at software testing ,300条数据我们随机抽取了5条数据用语一致性检验，剩余295条数据平均分配给每个annotator进行标注，每个annoatator将随机抽取遵循我们标注的每个需求的表述自然语言文本的guidance 下来执行人工测试。（不用test cases 是因为，需求level 的人工测试量级已经非常大了，64*5=320，每个annotator 已经需要进行320次标注，如果我们细化到test cases 粒度的人工标注的话，将会接新16*64=1024个样本进行标注，而且正如我们HITL-MAA 框架介绍所言，我们其实会根据修改后的test case refine需求，所以需求表述是非常清晰的，甚至包括边界，错误情况应对 see in Appendix \ref{ap:real_case_06}）所以我们的测试过程就是annotator 参考fine-grained requirement的表述来在浏览器上对web application 进行操作并观察他的反馈是否与需求一致。我们的四个错误类型Code Inconsitency, Requiment Missing ,requirement Misaligedn and Detail Mismatch ，annotator 是这样区分的，如果他在街面上根本找不到这个需求的操作入口，例如测试一个管理系统的登陆功能，但却发现该系统连登录界面，登录按钮等可操作组件都没有，那annotator便会在项目源码中找与该功能相关的代码块，如果都没有，那就认定其完全忽略了该需求，如果这时发现他实现了逻辑代码但没实现相关前端组件的话遍人定为是Code Inconsisteny(但是这种概率为零)。其次是Code Inconsitensy 问题，例如html写了响应function name,但js中未实现，反复实现，实现函数为空都属于这类问题，这类问题常伴随前端组件操作无反应，所以annoator操作测试时遇到这种情况，若检查源码却为刚刚所说情况，则标注为该问题。第三是requirement Misaliged 问题，即software实现了该功能但实现逻辑明显与task 需求不一致，实现 虽然功能完成了，但“不是这样用的”，例如ToDo list有个右键删除当前安排的功能需求，但实现的确实也可以删除，但是他用过一个额外的删除按钮来实现该功能。若是这种情况，便属于R euiqment Misalined;最后一种情况Detailed Mismatch 是最难标的，标注过程是，annoator粗略体验功能后发现没用明显问题，但是会通过一些unexpected actions, boundary cases, and defensive handling看功能是否还能正确运行，并且仔细观察feedback在硬性content完全匹配level上观察是否有问题。最终标注结果显示，不同annotator间的一致性Kappas 有0.86，人与机器评估的结果的Kappas 有0.74说明一致性都不错，结果可靠。

\subsection{Implementation of Human Evaluation in Section \ref{indepth}}
\label{imp_indepth}

To support our fine-grained analysis of software requirement satisfaction (Section~\ref{indepth}), we conducted a human evaluation to identify and categorize different types of implementation failures. The goal is to assess not just whether a function exists, but whether it is implemented correctly, meaningfully, and according to the intended specification.

\paragraph{Task Setup.}  
We randomly selected 10 tasks from the E2EDev benchmark and evaluated their implementations under 30 different models (6 backbones $\times$ 6 frameworks), resulting in 360 model-task instances. We recruited 5 professional annotators with expertise in software testing. To ensure labeling consistency, 5 shared examples were used to calibrate judgments, and the remaining 295 instances were evenly distributed across annotators.

Each annotator was asked to assess whether a web application satisfies a specific requirement described in natural language. Importantly, annotators were guided by refined and detailed requirement statements (see Appendix~\ref{ap:real_case_06}), which already include edge cases and expected behaviors. The annotators interact directly with the browser-based application and determine whether the system's observable behavior aligns with the requirement.

\paragraph{Why Requirement-Level Evaluation?}  
We opt for requirement-level evaluation instead of simulating every atomic test case due to scalability and redundancy. Evaluating 64 requirements per annotator is already resource-intensive. Expanding to all test cases would result in over 1,000 evaluations (e.g., 16 test cases per task $\times$ 64 tasks), which is infeasible. Moreover, our HITL-MAA pipeline already aligns test cases with refined requirements. Hence, human evaluation at the requirement level is both efficient and representative.

\paragraph{Error Typology.}  
Based on interactions and source code inspection, annotators classified failures into four categories. Table~\ref{tab:human_error_types} summarizes the definitions and identification criteria.

% \begin{table*}[htbp]
% \centering
% \begin{tabular}{p{3cm}p{12cm}}
% \toprule
% \textbf{Error Type} & \textbf{Definition and Identification Criteria} \\
% \midrule
% \textbf{Requirement Missing} & The required functionality is completely absent. Annotators cannot find relevant UI elements or supporting code. For example, a login feature is required but no login form or logic exists in the codebase. \\ \hline
% \textbf{Code Inconsistency} & UI components refer to expected behaviors (e.g., an `onClick` handler), but the linked code is missing, empty, or non-functional. The interface appears intact, but interaction triggers no effect. \\\hline
% \textbf{Requirement Misaligned} & The core functionality exists, but the way it is implemented deviates from the specification. For example, deleting a to-do item is required via right-click, but is instead implemented through a separate button. \\\hline
% \textbf{Detail Mismatch} & The main function appears to work, but issues arise under edge cases, unexpected actions, or incorrect feedback content. These require nuanced inspection to reveal. \\
% \bottomrule
% \end{tabular}
% \caption{Error types identified during human evaluation}
% \label{tab:human_error_types}
% \end{table*}
% 确保导言区有: \usepackage{tabularx} \usepackage{booktabs}

\begin{table*}[t]
    \centering
    \small % 1. 缩小字体，显得更精致
    \renewcommand{\arraystretch}{1.3} % 2. 增加行间距 (1.3倍)，让文字呼吸，不拥挤
    
    % 3. 使用 tabularx 自动计算宽度，X 列自动换行
    % 第一列 l (左对齐) 自适应宽度，第二列 X 占满剩余空间
    \begin{tabularx}{\textwidth}{l X} 
        \toprule
        \textbf{Error Type} & \textbf{Definition and Identification Criteria} \\
        \midrule
        
        \textbf{Requirement Missing} & 
        The required functionality is completely absent. Annotators cannot find relevant UI elements or supporting code. For example, a login feature is required but no login form or logic exists in the codebase. \\
        \midrule % 4. 统一使用 midrule，或者如果不想要线，可以使用 \addlinespace
        
        \textbf{Code Inconsistency} & 
        UI components refer to expected behaviors (e.g., an `onClick` handler), but the linked code is missing, empty, or non-functional. The interface appears intact, but interaction triggers no effect. \\
        \midrule
        
        \textbf{Requirement Misaligned} & 
        The core functionality exists, but the way it is implemented deviates from the specification. For example, deleting a to-do item is required via right-click, but is instead implemented through a separate button. \\
        \midrule
        
        \textbf{Detail Mismatch} & 
        The main function appears to work, but issues arise under edge cases, unexpected actions, or incorrect feedback content. These require nuanced inspection to reveal. \\
        
        \bottomrule
    \end{tabularx}
    \caption{Error types identified during human evaluation.}
    \label{tab:human_error_types}
\end{table*}

\paragraph{Inter-Annotator Reliability.}  
To assess labeling quality, we compute inter-annotator agreement using Cohen's Kappa. We report a strong agreement score of 0.86 among human annotators and 0.74 between human and automatic evaluation, suggesting both robustness and validity of the human-labeled results.

% \chen{how do you conduct human evaluation}
% \chen{To demonstrate the correlation between our automatic evaluation and human judgment, we conduct a human evaluation (i.e., the \textit{win rate evaluation}) of definitions generated for 100 randomly selected buzzwords in Section \ref{overll}. Due to the resource-intensive nature of human evaluation, our analysis is limited to three representative methods. For each buzzword, two human evaluators compared the definitions generated by different methods across various backbones, considering both SA and SC. Following \citet{10.1145/3488560.3498440}, the evaluators are presented with pairs of anonymized definitions for the same buzzword, without disclosure of the originating model for each definition. Independent evaluations are followed by a discussion to resolve any discrepancies. A "\textit{Win/Lose/Tie}" label is finally assigned if consensus is reached; otherwise, the result is recorded as a "\textit{Tie}". Our experiments reveal inter-annotator agreement rates of 71.50\% and 61.88\% for semantic accuracy (SA) and semantic completeness (SC), respectively. Any remaining discrepancies in evaluation results are classified as Ties.}

\section{Additional Analysis}

\subsection{{OpenHands and SWE-Agent on E2EDev}}
\label{app:openhands_swe}
\begin{table*}[t!]
  \centering
  \scriptsize 
  % [关键调整] 增加列间距 (4pt -> 8pt)，让表格横向更宽，视觉上更舒展
  \setlength{\tabcolsep}{8pt} 
  \renewcommand{\arraystretch}{1.0}

  % =======================================================
  % 颜色定义
  % =======================================================
  \definecolor{rowgray}{HTML}{F5F5F5}
  \definecolor{bgred}{HTML}{FFF0F0}
  \definecolor{bggreen}{HTML}{F0FFF0}
  \definecolor{sdtext}{HTML}{999999}

  % =======================================================
  % 宏定义 (修复双重符号问题)
  % =======================================================
  
  % 1. \sd: 核心样式 (灰色、小号、自带±)
  \newcommand{\sd}[1]{\textcolor{sdtext}{\tiny $\pm$#1}}

  % 2. 样式宏: 
  % [关键修复] 去掉了 #2 前面的 \sd，只负责背景色和加粗
  % 这样 \base{48.69}{\sd{1.30}} 就会输出 "48.69 ±1.30" 而不是 "48.69 ±±1.30"
  
  % Effectiveness - High (Red)
  \newcommand{\high}[2]{\cellcolor{bgred}#1 #2}
  \newcommand{\highb}[2]{\cellcolor{bgred}\textbf{#1} #2}
  
  % Effectiveness - Low (Green)
  \newcommand{\low}[2]{\cellcolor{bggreen}#1 #2}
  \newcommand{\lowb}[2]{\cellcolor{bggreen}\textbf{#1} #2}
  \newcommand{\lowno}[1]{\cellcolor{bggreen}#1}

  % Vanilla (Gray)
  \newcommand{\base}[2]{#1 #2}
  \newcommand{\baseul}[2]{\uline{#1} #2}
  
  % Efficiency (Plain)
  \newcommand{\eff}[2]{#1 #2} 

  % =======================================================

  \resizebox{\linewidth}{!}{%
    \begin{tabular}{ll ccc ccc} 
      \toprule
      \multirow{2}{*}{\textbf{Backbone}} & \multirow{2}{*}{\textbf{Method}} & \multicolumn{3}{c}{\textbf{Effectiveness ($\uparrow$, \%)}} & \multicolumn{3}{c}{\textbf{Efficiency ($\downarrow$)}} \\
      \cmidrule(lr){3-5} \cmidrule(lr){6-8}
      & & Req. Acc. & Test Acc & Bal. Score & Cost (\$) & CO$_{2}$ (g) & Time (s) \\
      \midrule

      % ================= Claude-Haiku 4.5 =================
      \multirow{8}{*}{\textbf{Claude-Haiku 4.5}} 
      & \cellcolor{rowgray}Vanilla LLM & \cellcolor{rowgray}\base{48.69}{\sd{1.30}} & \cellcolor{rowgray}\base{63.08}{\sd{0.50}} & \cellcolor{rowgray}\base{54.45}{\sd{0.98}} & \cellcolor{rowgray}0.015 & \cellcolor{rowgray}0.028 & \cellcolor{rowgray}53 \\
      & GPT-Engineer & \high{\textbf{53.75}}{\sd{0.43}} & \high{\textbf{69.41}}{\sd{1.07}} & \high{\textbf{60.01}}{\sd{0.69}} & 0.016 & 0.038 & 59 \\
      & Self-Collab. & \high{49.01}{\sd{1.74}} & \low{61.50}{\sd{1.26}}  & \low{54.01}{\sd{1.55}}  & 0.014 & 0.033 & 75 \\
      & MapCoder     & \high{\uline{49.61}}{\sd{0.50}} & \high{\uline{65.65}}{\sd{0.44}} & \high{\uline{56.03}}{\sd{0.48}} & 0.099 & 0.081 & 129 \\
      & ChatDev      & \low{44.73}{\sd{1.50}}  & \low{58.09}{\sd{1.03}}  & \low{50.07}{\sd{1.31}}  & 0.174 & 0.188 & 335 \\
      & MetaGPT      & \low{5.39}{\sd{0.76}}   & \low{10.66}{\sd{1.77}}  & \low{7.50}{\sd{1.16}}   & 0.090 & 0.091 & 366 \\
      & Openhands & \low{43.45} & \low{57.77} & \low{49.18} & 0.021 & 0.054 & - \\
      & SweAgent & \low{44.73}{\sd{0.44}} & \low{56.87}{\sd{0.45}} & \low{49.59}{\sd{0.44}} & 0.089 & 0.089 & 148 \\
      \cmidrule(lr){1-8} 

      % ================= GPT-4o =================
      \multirow{8}{*}{\textbf{GPT-4o}} 
      & \cellcolor{rowgray}Vanilla LLM & \cellcolor{rowgray}\base{45.95}{\sd{1.25}} & \cellcolor{rowgray}\base{60.88}{\sd{1.97}} & \cellcolor{rowgray}\base{51.92}{\sd{1.54}} & \cellcolor{rowgray}0.0160 & \cellcolor{rowgray}0.083 & \cellcolor{rowgray}28 \\
      & GPT-Engineer & \high{\textbf{50.83}}{\sd{1.33}} & \high{\textbf{66.59}}{\sd{2.43}} & \high{\textbf{57.13}}{\sd{1.77}} & 0.0198 & 0.132 & 21 \\
      & Self-Collab. & \high{46.83}{\sd{1.12}} & \high{61.15}{\sd{1.47}} & \high{52.56}{\sd{1.26}} & 0.0155 & 0.109 & 37 \\
      & MapCoder     & \high{\uline{47.70}}{\sd{2.57}} & \high{\uline{63.97}}{\sd{2.25}} & \high{\uline{54.21}}{\sd{2.44}} & 0.1091 & 0.750 & 93 \\
      & ChatDev      & \low{42.71}{\sd{2.44}}  & \low{58.93}{\sd{2.66}}  & \low{49.20}{\sd{2.53}}  & 0.1947 & 1.910 & 114 \\
      & MetaGPT      & \low{0.00}{\sd{0.00}}   & \low{0.17}{\sd{0.05}}   & \low{0.07}{\sd{0.02}}   & 0.0951 & 0.794 & 66 \\
      & Openhands & \low{39.25} & \low{55.34} & \low{45.85} & 0.062 & 0.161 & - \\
      & SweAgent & \low{40.83}{\sd{2.23}} & \low{58.05}{\sd{1.91}} & \low{47.72}{\sd{2.10}} & 0.089 & 0.089 & 122 \\
      % \cmidrule(lr){1-8} 

      \bottomrule
    \end{tabular}%
  }
  
  \caption{Benchmark analysis of effectiveness and efficiency.
The \colorbox{rowgray}{grey row} denotes the Vanilla baseline.
\colorbox{bgred}{Red} and \colorbox{bggreen}{green} indicate performance better and worse than Vanilla, respectively.
Standard deviations are shown in small grey text.
For effectiveness, the best result is shown in \textbf{bold} and the second-best is \uline{underlined}.}

  \label{tab:model_performance_additional}
\end{table*}
% 务必在导言区添加: \usepackage{tabularx}

\begin{table*}[t]
    \centering
    \small % 保持字体较小
    % 使用 tabularx 让表格总宽度严格等于页面文字宽度 (\textwidth)
    % X 列会自动换行并调整宽度
    \begin{tabularx}{\textwidth}{l X p{0.18\textwidth} p{0.15\textwidth} p{0.18\textwidth}}
        \toprule
        \textbf{Method} & \textbf{Task Description} & \textbf{Input} & \textbf{Output} & \textbf{Mode} \\
        \midrule
        \textbf{E2EDev (Ours)} & 
        Generates fully runnable software projects from scratch based solely on requirements. & 
        Detailed user requirements & 
        Runnable project & 
        End-to-End Repo Generation \\
        \midrule
        OpenHands & 
        Collaborates via multi-turn dialogue to iteratively build software. & 
        Interactive dialogue & 
        Runnable project & 
        Human-in-the-loop Repo Generation \\
        \midrule
        SWE-Agent & 
        Resolves GitHub issues by modifying existing repositories (SWE-bench). & 
        Existing repo + Issue & 
        Patched repo & 
        Repo Modification \\
        \bottomrule
    \end{tabularx}
    % Caption 放在表格下方
    \caption{\label{tab:framework_comparison} Comparison of task formulations. E2EDev targets ab initio generation, distinct from the collaborative or maintenance-focused nature of OpenHands and SWE-Agent.}
\end{table*}

OpenHands~\citep{wang2024openhands} and SWE-Agent~\citep{yang2024swe} are two representative and widely used LLM-based agents for software engineering tasks. Both systems are designed to support complex development workflows and have shown strong performance in their original evaluation settings. However, their design assumptions and task formulations differ substantially from the setting targeted by E2EDev, {as summarized in Table~\ref{tab:framework_comparison}.}

E2EDev focuses on \textbf{end-to-end requirement-to-executable software development}, where a system is required to generate a complete and runnable software project directly from natural language requirements. This setting emphasizes project-level generation from scratch. In contrast, OpenHands is built around iterative human--agent collaboration, where continuous human feedback guides the development process, while SWE-Agent operates on an existing code repository and targets issue-level modification rather than full project generation. As a result, these systems are not natively aligned with the E2EDev task formulation,
and a direct comparison with E2EDev-oriented frameworks is not conceptually equivalent.

\paragraph{Experimental Setup.}
Due to their substantially higher evaluation cost and different design goals,
we evaluate OpenHands and SWE-Agent only on \textit{GPT-4o} and
\textit{Claude-Haiku-4.5}.
To better align these systems with the E2EDev setting,
OpenHands is restricted to a single execution without human interaction.
For SWE-Agent, we replace its original prompt template—which instructs the model
to load and modify an existing repository—with a new template that directs the model
to generate a runnable software project directly from the given user requirement.
This modification prevents access to any existing repositories and enforces
project generation from scratch.

\paragraph{Results and Discussion.}
The results in Table~\ref{tab:model_performance_additional} show that
OpenHands and SWE-Agent underperform the Vanilla LLM baseline on E2EDev.
This outcome is expected given the mismatch between their original design objectives
and the requirement-to-executable software development task.
In particular, both systems exhibit limitations in directly interpreting and
implementing fine-grained user requirements without relying on interactive guidance
or pre-existing code structures.
Overall, these findings further support our main conclusion that E2EDev poses
distinct challenges beyond those addressed by existing agent-based
software engineering frameworks.

\subsection{Additional Effectiveness Analysis on E2EDev}
\label{appe2e}

\begin{table*}[htbp]
\centering
\resizebox{0.8\linewidth}{!}{%
\begin{tabular}{llccc}
\toprule
Base Model                & Method                             & Req.acc & Soft Req.acc & Delta  \\ 
\midrule
\multirow{6}{*}{claude-haiku 4.5}   & ChatDev                            & 0.4464  & 0.7025       & 0.2718 \\
                           & gpt-engineer                       & 0.5402  & 0.7991       & 0.2589 \\
                           & llmbased                           & 0.4866  & 0.7454       & 0.2589 \\
                           & MapCoder                           & 0.4955  & 0.7723       & 0.2768 \\
                           & MetaGPT                            & 0.0536  & 0.1027       & 0.0491 \\
                           & Self-collaboration-Code-Generation & 0.4911  & 0.7679       & 0.2768 \\ \midrule
\multirow{6}{*}{qwenmax}   & ChatDev                            & 0.4307  & 0.7025       & 0.2718 \\
                           & gpt-engineer                       & 0.4979  & 0.7454       & 0.2475 \\
                           & llmbased                           & 0.4335  & 0.7047       & 0.2712 \\
                           & MapCoder                           & 0.4837  & 0.7605       & 0.2768 \\
                           & MetaGPT                            & 0.0163  & 0.0379       & 0.0216 \\
                           & Self-collaboration-Code-Generation & 0.4268  & 0.7110       & 0.2842 \\ \midrule
\multirow{6}{*}{qwen7b}    & ChatDev                            & 0.1803  & 0.4414       & 0.2611 \\
                           & gpt-engineer                       & 0.2403  & 0.5718       & 0.3315 \\
                           & llmbased                           & 0.2237  & 0.5090       & 0.2853 \\
                           & MapCoder                           & 0.1283  & 0.3967       & 0.2684 \\
                           & MetaGPT                            & 0.0000  & 0.0000       & 0.0000 \\
                           & Self-collaboration-Code-Generation & 0.2058  & 0.4918       & 0.2860 \\ \midrule
\multirow{6}{*}{qwen70b}   & ChatDev                            & 0.4352  & 0.7053       & 0.2701 \\
                           & gpt-engineer                       & 0.4208  & 0.7064       & 0.2856 \\
                           & llmbased                           & 0.3575  & 0.6674       & 0.3099 \\
                           & MapCoder                           & 0.4044  & 0.6949       & 0.2905 \\
                           & MetaGPT                            & 0.0000  & 0.0045       & 0.0045 \\
                           & Self-collaboration-Code-Generation & 0.4257  & 0.6727       & 0.2470 \\ \midrule
\multirow{6}{*}{gpt4omini} & ChatDev                            & 0.3395  & 0.6280       & 0.2885 \\
                           & gpt-engineer                       & 0.4310  & 0.6908       & 0.2598 \\
                           & llmbased                           & 0.4572  & 0.7417       & 0.2845 \\
                           & MapCoder                           & 0.4089  & 0.6884       & 0.2795 \\
                           & MetaGPT                            & 0.0000  & 0.0071       & 0.0071 \\
                           & Self-collaboration-Code-Generation & 0.3876  & 0.6864       & 0.2988 \\ \midrule
\multirow{6}{*}{gpt4o}     & ChatDev                            & 0.4275  & 0.6928       & 0.2653 \\
                           & gpt-engineer                       & 0.5037  & 0.7730       & 0.2693 \\
                           & llmbased                           & 0.4623  & 0.7346       & 0.2723 \\
                           & MapCoder                           & 0.4792  & 0.7228       & 0.2436 \\
                           & MetaGPT                            & 0.0000  & 0.0048       & 0.0048 \\
                           & Self-collaboration-Code-Generation & 0.4614  & 0.7064       & 0.2450 \\ \bottomrule
\end{tabular}
}
\caption{Soft Req. Acc. and Req. Acc. under all LLMs we used. Here, V-LLM refers to the
Vanilla LLM, GPT-E refers to GPT-Engineer, and Self-C refers to Self-Collaboration.}
\label{app_tb:soft}
\end{table*}

% Please add the following required packages to your document preamble:
% \usepackage{multirow}
\begin{table}[htbp]
\centering
\resizebox{\columnwidth}{!}{%
\begin{tabular}{llcccc}
\toprule
Base Model                 & Method   & \multicolumn{1}{l}{CIc} & \multicolumn{1}{l}{MR} & \multicolumn{1}{l}{MwR} & \multicolumn{1}{l}{DM} \\ \midrule
\multirow{6}{*}{gpt4omini} & V-LLM    & 0                                    & 3                                       & 9                                                & 22                                   \\
                           & GPT-E    & 2                                    & 1                                       & 13                                               & 18                                   \\
                           & Self-C   & 1                                    & 0                                       & 12                                               & 20                                   \\
                           & MapCoder & 1                                    & 0                                       & 3                                                & 26                                   \\
                           & ChatDev  & 3                                    & 0                                       & 8                                                & 24                                   \\
                           & MetaGPT  & 36                                   & 15                                      & 0                                                & 3                                    \\ \midrule
\multirow{6}{*}{qwen7B}    & V-LLM    & 12                                   & 5                                       & 20                                               & 11                                   \\
                           & GPT-E    & 3                                    & 2                                       & 21                                               & 18                                   \\
                           & Self-C   & 12                                   & 0                                       & 21                                               & 10                                   \\
                           & MapCoder & 44                                   & 0                                       & 0                                                & 2                                    \\
                           & ChatDev  & 12                                   & 0                                       & 15                                               & 0                                    \\
                           & MetaGPT  & 21                                   & 33                                      & 0                                                & 0                              \\ \bottomrule     
\end{tabular}
}
\caption{Error distribution across frameworks powered by GPT-4o-mini and Qwen-7B. CIc denotes code inconsistency, MR refers to missing requirement, MwR indicates mismatch with requirement, and DM stands for detail mismatch.}
\end{table}

\noindent\textbf{\textit{Does strong standalone performance of a Vallina LLM guarantee its effectiveness as the backbone of an agentic framework?} -- No}  For a more intuitive understanding of this questions, we present a bar chart in Fig.\ref{fig:comp} that clearly highlights the performance gaps between different Agentic Framework powered by powerful LLM Backbones to better indicates which Agentic framework outperperform Valinna LLM, demonstrating their effectiveness, the yellow zone indicates the req.acc of each Vallina LLM (From left(GPT-4o) to the right(Qwen-7B) their performance degrades sequentially). Interestingly, we find that employing a more powerful vanilla LLM as the backbone does not always yield better results on the E2ESD task when integrated into an agentic framework. For example, both GPT-4o-mini exhibit performance drops across several agentic frameworks while it's vallina model performs surprisingly well as a vanilla LLM, with a performance gap to GPT-4o being relatively small. We hypothesize that this is because GPT-4o-mini is a distilled version of GPT-4o and retains many of its generalist strengths, as evidenced by its comparable performance on several benchmarks—except for complex knowledge-intensive tasks such as GPQA, where it shows a clear degradation. In our E2ESD task, although each instance contains 2–10 user requirements, they primarily involve standard software functionalities and do not require specialized knowledge. As a result, GPT-4o-mini performs well as a standalone model. However, it remains a relatively small model, estimated at around 8B parameters~\citep{abacha2024medec}. Like other small models, it suffers from common issues such as performance instability and limited robustness. When used as a backbone within an agentic framework—particularly under heavy task prompts, coupled with additional system-level constraints—its understanding of the task may be compromised. Furthermore, in multi-agent pipelines, the likelihood of error propagation and accumulation increases. This issue is also observed with Qwen-7B. A detailed analysis of this behavior will be provided in the case study section.

\noindent\textbf{\textit{Does equipping an agentic framework with a more powerful LLM backbone necessarily lead to better effectiveness?} -- No. The effectiveness of an agentic framework depends not only on LLM capability, but also on instruction-following alignment.}
As shown in Fig.\ref{fig:comp}, performance varies across agentic frameworks when powered by different LLM backbones. We find that instruction-following ability is key the to a reliable backbone, as the success of an agentic framework relies on each agent following its instructions while staying aligned with others. Among all tested backbones, Qwen-70B (instruction-tuned version) shows the most stable performance, particularly in dynamic systems like ChatDev. It achieves higher requirement accuracy and fewer dialogue turns, as shown in Table\ref{tab:token_turns}, indicating superior alignment and controllable. The detailed analysis of "How Instruction-Following Affects the Agentic Workflow" will follow in the case study section.

\subsection{Additional Efficiency Analysis on E2ESD}
\label{app:add_efficiency}

\begin{figure}[t]
  \centering
  \includegraphics[width=1.0\linewidth]{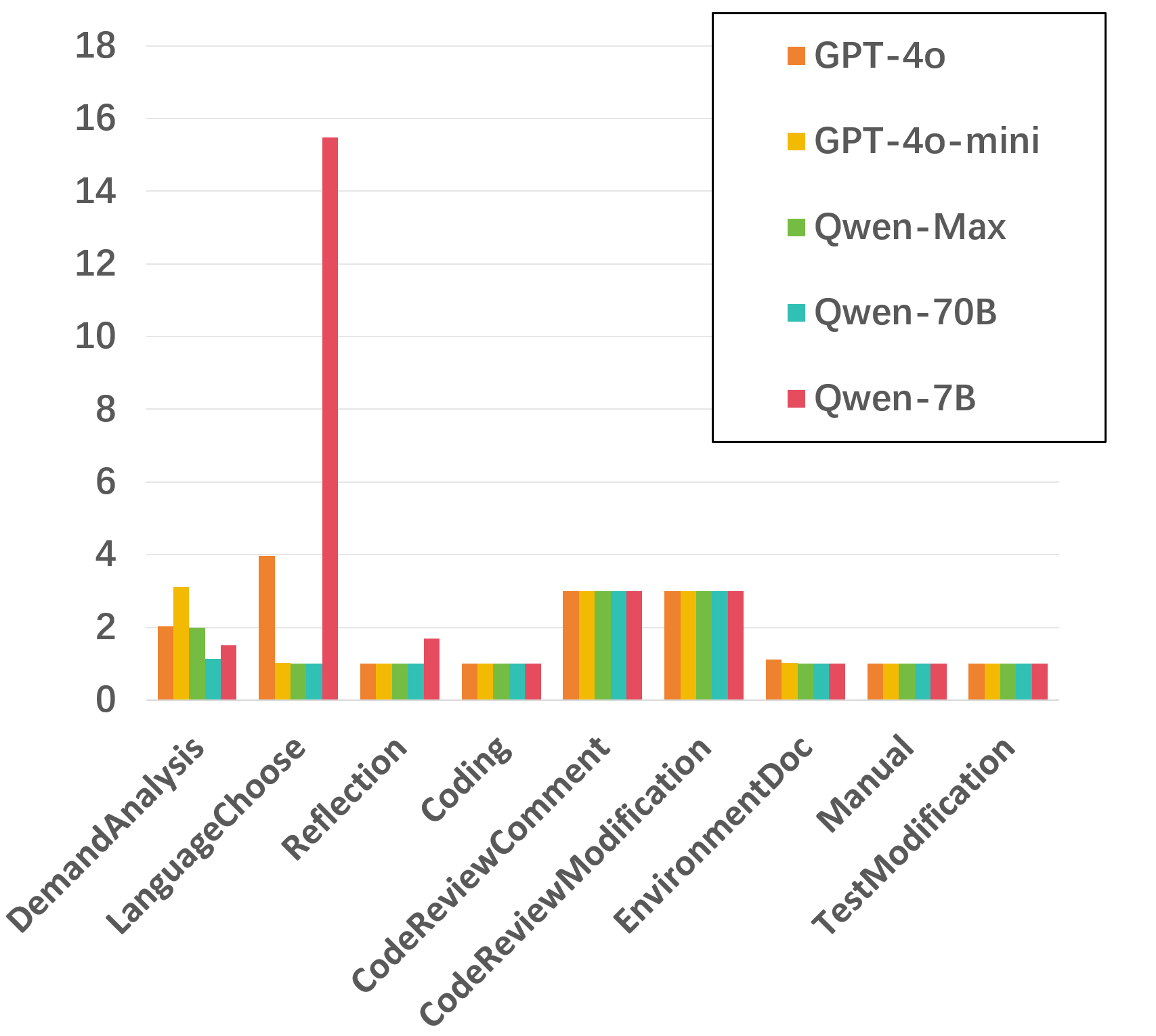}
  \caption{Number of communications in each phase of ChatDev across different backbone models.}
  \label{fig:humanv_ineff}
\end{figure}

\textbf{Inefficiencies in ChatDev's dialogue process are primarily attributed to the models' failure to adhere to flexible stopping signals specified in the prompt, particularly in multi-phase interactions.} As seen in Figure \ref{fig:humanv_ineff}, in Phase 2, the task simply requires selecting a programming language and returning a line in the format "<INFO> *" (where * is the chosen language) indicating the end of the phase. However, models frequently terminate with "<INFO> Finish", violating the instruction and causing unnecessary dialogue iterations. This problem is especially pronounced in smaller models like Qwen-7B, which often reach the 20-turn interaction cap even for such trivial tasks.  Similar inefficiencies are observed in later stages (e.g., Phases 5 and 6, CodeReview), where all models reach the maximum number of turns. Although the prompt explicitly states: “If no bugs are reported, please return only one line like <INFO> Finished,” minor deviations such as surrounding whitespace prevent proper loop termination. Notably, early-stage phases like DemandAnalysis terminate more reliably, likely due to shorter generated outputs and less instruction interference from long context history. These findings highlight a key limitation of ChatDev's current stopping mechanism design: it relies on rigid string-based termination signals that are fragile under imperfect generation and long-context dependencies. Future multi-agent frameworks should consider more adaptive and robust exit strategies, especially under conditions where instruction following is context-sensitive and generation is uncertain.

\subsection{Additional Analysis on HumanEval (Function-level Task)}
\label{humaneval}

The main experimental results show that existing methods struggle to handle project-level E2ESD tasks.
To better understand this limitation, we further evaluate these methods on a simpler, function-level benchmark, HumanEval~\citep{chen2021evaluating}, which isolates code generation from project-level coordination and integration.

As summarized in Table~\ref{tab:model_performance_humaneval}, most methods achieve substantially higher performance on HumanEval than on E2ESD.
This gap indicates that current approaches remain more effective for isolated function-level code generation than for end-to-end project development.

{For MetaGPT, we deliberately evaluate the framework without any HumanEval-specific adaptation.
Under this setting, MetaGPT’s performance is lower than the results reported in the main body of its original paper, but closely matches the authors’ appendix results obtained without benchmark-specific tuning.
This confirms that the observed performance is consistent with MetaGPT’s default framework design.}
{A closer inspection reveals that most performance degradation arises from the multi-agent collaboration process rather than from code generation itself.
During requirement analysis and architectural planning, MetaGPT frequently modifies the target function name or signature specified by the benchmark.
As a result, the generated code no longer matches the expected test interface, causing test failures even when the core logic is otherwise correct.
This behavior mirrors the failure patterns observed on E2EDev, where the framework’s internal analysis stages often alter the original requirements, leading to mismatches between the specified requirements and the generated implementation.}

%\clearpage

% Benchmark analysis on off-the-shelf methods across various LLM backbones regarding their effectiveness and efficiency. Performance of the vanilla LLM is highlighted in \lightgrey{light grey} as a reference. For effectiveness, the highest Req. Acc. value within each group is presented in \textbf{bold}, with the second highest \uline{underlined}.

\begin{table*}[ht]
  \centering
  \resizebox{\linewidth}{!}{%
    \begin{tabular}{llccccc}
      \toprule
      \multirow{2}{*}{\textbf{Backbone LLM}} & \multirow{2}{*}{\textbf{Method}} & \multicolumn{2}{c}{\textbf{Effectiveness ($\uparrow$, \%)}} & \multicolumn{3}{c}{\textbf{Efficiency ($\downarrow$)}} \\
      \cmidrule(lr){3-4} \cmidrule(lr){5-7}
      & & Pass@1 & Pass@1 (ET) & Cost (USD) & Footprint (gCO$_{2}$eq)  & Duration (s) \\
      \midrule
      
      \rowcolor[HTML]{EFEFEF}\multirow{6}{*}{\cellcolor{white}\textbf{GPT-4o}} 
      & Vanilla LLM & 89.02 & 80.49 & 0.0045 & 0.018 & 13 \\
      & GPT-Engineer & \textbf{91.46} & 81.10 & 0.0043 & 0.030 & 28 \\
      & Self-Collaboration & \uline{90.85} & 81.10 & 0.0116 & 0.077 & 48 \\
      & MapCoder & 90.24 & \textbf{81.71} & 0.0542 & 0.338 & 386 \\
      & ChatDev & \uline{90.85} & \textbf{81.71} & 0.0760 & 0.619 & 404 \\
      & MetaGPT & 45.73 & 0.61 & 0.0429 & 0.338 & 109 \\
      \midrule
      
      \rowcolor[HTML]{EFEFEF}\multirow{6}{*}{\cellcolor{white}\textbf{GPT-4o-mini}} 
      & Vanilla LLM & 84.15 & 76.22 & 0.0003 & 0.001 & 14 \\
      & GPT-Engineer & \uline{86.59} & \textbf{79.27} & 0.0003 & 0.001 & 65 \\
      & Self-Collaboration & \uline{86.59} & 75.61 & 0.0006 & 0.003 & 42 \\
      & MapCoder & \textbf{87.20} & \uline{76.83} & 0.0027 & 0.012 & 71 \\
      & ChatDev & 85.37 & 75.61 & 0.0067 & 0.038 & 108 \\
      & MetaGPT & 50.61 & 1.83 & 0.0025 & 0.013 & 325 \\
      \midrule

      \rowcolor[HTML]{EFEFEF}\multirow{6}{*}{\cellcolor{white}\textbf{Qwen-Max}} 
      & Vanilla LLM & 89.63 & \uline{81.71} & 0.0009 & 0.012 & 26 \\
      & GPT-Engineer & \textbf{90.85} & \textbf{82.32} & 0.0006 & 0.014 & 31 \\
      & Self-Collaboration & 84.15 & 76.22 & 0.0017 & 0.038 & 117 \\
      & MapCoder & \uline{90.24} & 81.10 & 0.0080 & 0.177 & 211 \\
      & ChatDev & 89.02 & 80.49 & 0.0104 & 0.302 & 215 \\
      & MetaGPT & 82.32 & 6.10 & 0.0043 & 0.124 & 118 \\
      \midrule
      
      \rowcolor[HTML]{EFEFEF}\multirow{6}{*}{\cellcolor{white}\textbf{Qwen-70B}} 
      & Vanilla LLM & \textbf{87.20} & \textbf{77.44} & 0.0008 & 0.012 & 27 \\
      & GPT-Engineer & \uline{83.54} & 73.17 & 0.0007 & 0.017 & 17 \\
      & Self-Collaboration & 82.31 & \uline{75.00} & 0.0020 & 0.043 & 67 \\
      & MapCoder & \uline{83.54} & 74.39 & 0.0102 & 0.212 & 218 \\
      & ChatDev & 82.23 & 71.34 & 0.0160 & 0.377 & 278 \\
      & MetaGPT & 53.66 & 3.05 & 0.0085 & 0.208 & 229 \\
      \midrule
      
      \rowcolor[HTML]{EFEFEF}\multirow{6}{*}{\cellcolor{white}\textbf{Qwen-7B}} 
      & Vanilla LLM & \textbf{81.71} & \uline{71.95} & 0.0001 & 0.001 & 15 \\
      & GPT-Engineer & \textbf{81.71} & \textbf{75.00} & 0.0001 & 0.001 & 10 \\
      & Self-Collaboration & 71.34 & 64.02 & 0.0002 & 0.004 & 41 \\
      & MapCoder & 76.83 & 68.29 & 0.0011 & 0.023 & 118 \\
      & ChatDev & 73.17 & 65.24 & 0.0027 & 0.065 & 352 \\
      & MetaGPT & 30.49 & 1.22 & 0.0006 & 0.014 & 189 \\
      \bottomrule
    \end{tabular}%
  }
  \caption{Performance of off-the-shelf methods across various LLM backbones on HumanEval and HumanEval-ET. We use Pass@k (with $k=1$) to evaluate effectiveness. Pass@k (ET) refers to results on HumanEval-ET, which includes more test cases per function. For clarity, the highest score within each group is shown in \textbf{bold}, and the second highest is \uline{underlined}.}
  \label{tab:model_performance_humaneval}
\end{table*}

\subsection{Case Studies on LLM-driven Methods for E2ESD}
\label{app_E2ESD_log_case}

% \begin{tcolorbox}[colback=white!98!black,colframe=white!30!black,boxsep=1.1pt,top=6.75pt]
\noindent\makebox[\columnwidth]{\rule{\columnwidth}{0.8pt}}

\textbf{Task:} The application is a web-based calculator that allows users to perform basic arithmetic operations. It includes a display screen for showing inputs and results, a set of buttons for numbers and operators, and a theme toggle feature to switch between dark and light modes.

\medskip

\textbf{Relevant Requirement:}
\begin{itemize}
  \item \textbf{Requirement 2:} When a user interacts with the calculator's operator buttons, the system should append the corresponding operator to the display screen.
  \medskip
  \item \textbf{Edge Case:} If the user enters multiple consecutive operators (e.g., \texttt{+}, \texttt{-}, \texttt{*}, \texttt{/}), only the last one should be retained in the display.
\end{itemize}
\noindent\rule{\columnwidth}{0.8pt}

\medskip

\textbf{Information Given to the Coding Agent:}

\bigskip

% \begin{tcolorbox}[colback=white!98!black,colframe=black!70!white,boxsep=4pt,top=6pt,bottom=6pt]
% % \begin{minipage}[t]{0.3\textwidth}
%     \textbf{Implicit Prompt:}
%     \vspace{2mm}
%     \hrule height 0.8pt
%     \vspace{2mm}
%     \{problem\}
%     \vspace{2mm}
%     % \hrule height 0.8pt
% % \end{minipage}
% % \hfill
% % \begin{minipage}[t]{0.6\textwidth}
%     \vspace{2mm}
%     \hrule height 0.8pt
%     \vspace{2mm}

%     \textbf{Explicit Prompt:} Requests to Ask for Information
%     \vspace{2mm}
%     \hrule height 0.8pt
%     \vspace{2mm}
%     \# Task to Solve\\
%     \{problem\}\\[2mm]
%     \# Instruction for Clarification\\
%     If you need ask for information, please raise clarification question and start your response STRICTLY with: "Clarification Question" followed by your questions.\\[2mm]
%     Otherwise, please provide your answer in \textbackslash\textbackslash boxed\{\}.
% % \end{minipage}
% \end{tcolorbox}

\begin{tcolorbox}[
breakable,
colback=white!98!black,
colframe=black!70!white,
boxsep=4pt,
top=2pt,
title={\textbf{\large Self-Collaboration}},
coltitle=black,
colbacktitle=gray!20,
bottom=2pt
]
\begin{lstlisting}[ breaklines=true, basicstyle=\ttfamily\small, aboveskip=1pt, belowskip=1pt, columns=fullflexible]
{
  "subproblems": [
    ...
    {
      "id": 3,
      "description": "Create JavaScript functionality to append operators to the display, handling empty display and consecutive operator clicks."
    },
    ...
  ],
  "high_level_plan": [
    ...
    {
      "step": 3,
      "description": "Write JavaScript to handle operator button clicks, ensuring proper handling of operator sequences and interaction with numbers on the display."
    },
    ...
  ]
}
\end{lstlisting}
\end{tcolorbox}
\textbf{\textit{Analysis:}} Self-Collaboration explicitly addresses the edge case of consecutive operator inputs in both its subproblem breakdown and high-level plan. This explicit guidance enables the coding agent to implement this logic correctly, resulting in successful execution of the test case when powered by GPT-4o.

\bigskip

\begin{tcolorbox}[
breakable,
colback=white!98!black,
colframe=black!70!white,
boxsep=4pt,
top=2pt,
title={\textbf{\large MapCoder}},
coltitle=black,
colbacktitle=gray!20,
bottom=2pt
]
\begin{lstlisting}[ breaklines=true, basicstyle=\ttfamily\small, aboveskip=1pt, belowskip=1pt, columns=fullflexible]
Recall three (03) relevant and distinct problems (different from the user task).
# Problem 1: To-Do List Application
# Code
# Plan to Solve the problem
# Problem 2: Simple Weather App
# Code
# Plan to Solve the problem
# Problem 3: Simple Quiz Application
# Code
# Plan to Solve the problem


## Planning:

1. **HTML Structure**: ...

2. **CSS Styling**: ...

3. **JavaScript Functionality**:
   - Implement event listeners for number and operator buttons to update the display.
   - Implement the clear button functionality to reset the display.
   - Implement the backspace button functionality to remove the last character from the display.
   - Implement the equal button functionality to evaluate the expression and handle special cases.
   - Implement the theme toggle functionality to switch between dark and light themes.

4. **Testing**: ...
\end{lstlisting}
\end{tcolorbox}
\textbf{\textit{Analysis:}} MapCoder generates solutions based on analogies to unrelated projects, which resemble few-shot prompting. However, none of the referenced examples include edge-case handling for consecutive operators. Moreover, the planning stage does not treat operator input as a separate requirement, leading to insufficient focus during code generation and ultimately causing the failure in this edge case.

\bigskip

\begin{tcolorbox}[
breakable,
colback=white!98!black,
colframe=black!70!white,
boxsep=4pt,
top=2pt,
title={\textbf{\large ChatDev}},
coltitle=black,
colbacktitle=gray!20,
bottom=2pt
]
\begin{lstlisting}[ breaklines=true, basicstyle=\ttfamily\small, aboveskip=1pt, belowskip=1pt, columns=fullflexible]
### Core Components

1. **HTML Structure (`index.html`)**
   ...

2. **JavaScript Functionality (`script.js`)
   - **handleNumberClick**: Appends clicked number to the display.
   - **handleOperatorClick**: Appends clicked operator to the display.
   - **handleClear**: Clears the display.
   - **handleBackspace**: Removes the last character from the display.
   - **handleEquals**: Evaluates the expression on the display.
   - **toggleTheme**: Toggles between dark and light themes.

3. **CSS Styling (`styles.css`)
   ...
\end{lstlisting}
\end{tcolorbox}

\textbf{\textit{Analysis:}} ChatDev does not mention the edge case of consecutive operator input at all. Furthermore, the function \texttt{handleOperatorClick} — which is crucial to this behavior — lacks any instruction regarding handling such cases. As a result, the generated implementation fails to address the edge case, leading to incorrect behavior in testing.

% \smallskip
% \noindent\makebox[\textwidth]{\rule{\linewidth}{0.8pt}}
% \end{tcolorbox}

\section{Introduction on Project/Software Testing}
\label{app:software}
Adhering to software engineering principles, we present the E2EDev benchmark for evaluating the performance of LLM-based frameworks in E2ESD. The benchmark comprises: (1) a detailed list of user requirements for each software project; (2) for every user requirement, multiple test cases are provided, each accompanied by a corresponding executable test script; and (3) an automated testing pipeline built upon the Behave framework. This appendix offers a brief overview of the software engineering principles underlying project/software testing relevant to the benchmark.

\noindent\textbf{Overview}. In modern software engineering, systematic testing is essential for ensuring software quality, reliability, and correctness. Project/software testing refers to a structured process of verifying that a software system meets specified requirements and behaves as expected under various conditions. Testing is generally guided by predefined requirements and specifications and aims to uncover bugs, ensure compliance with user needs, and validate system behavior.

The E2EDev benchmark follows the best practices of behavior-driven development (BDD), where test cases are derived directly from user stories or requirements, as shown in Figure~\ref{fig:pipeline}.. This approach enhances traceability between software specifications and their validations, thereby facilitating more interpretable and maintainable test suites. To this end, we adopt Gherkin as a high-level test specification language and Behave as the test execution framework. These tools allow us to simulate realistic software engineering workflows while enabling automated assessment of LLM-generated artifacts.

\begin{figure*}[htbp]
  \centering
  \includegraphics[width=0.98\textwidth]{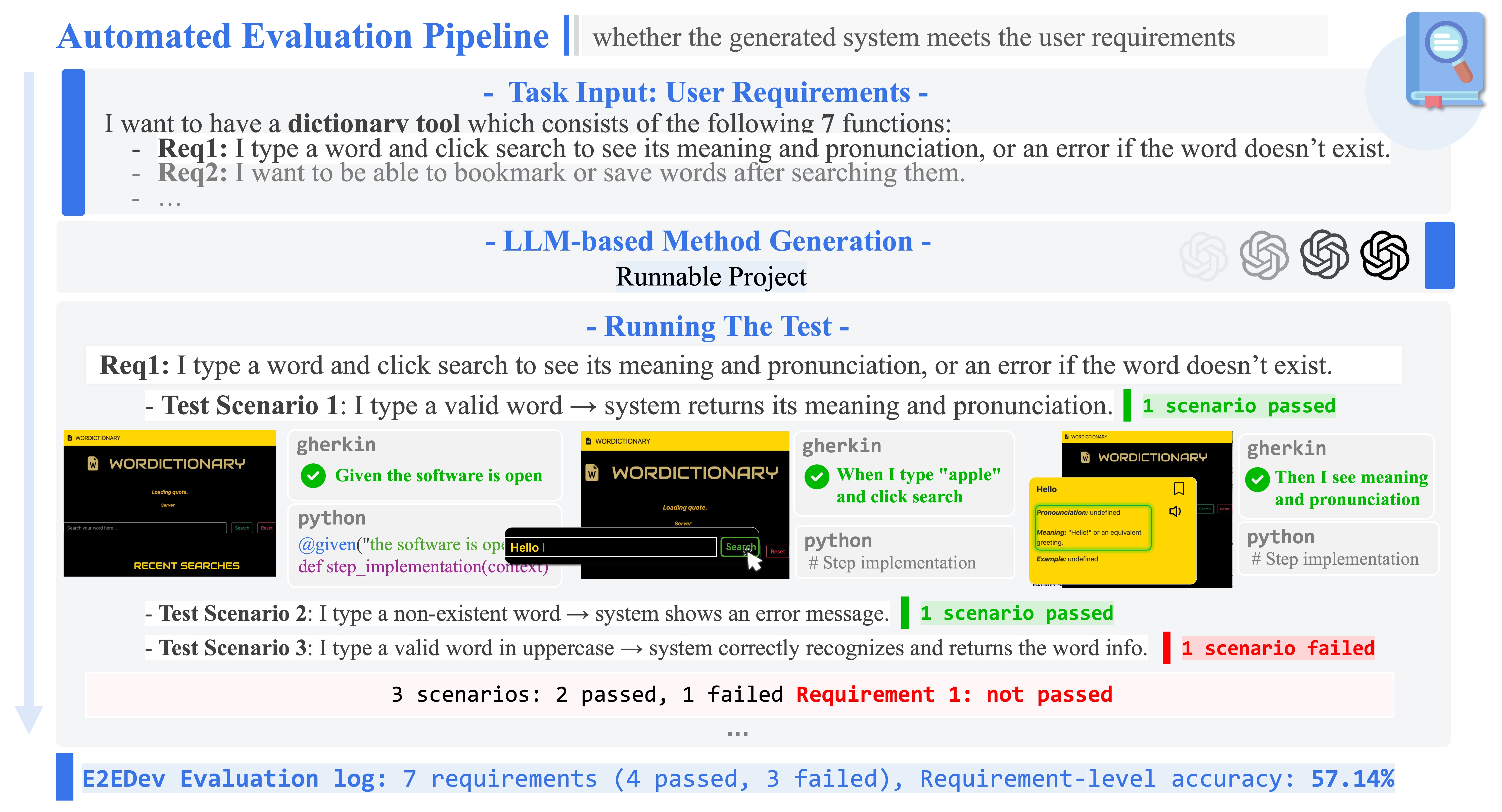}
  \caption{BDD-based automated evaluation pipeline for E2ESD tasks.}
  \label{fig:pipeline}
  \vspace{-3mm}
\end{figure*}

\subsection{Test Case Specification -- Gherkin}

Gherkin is a domain-specific language designed for behavior-driven development (BDD), allowing test cases to be written in a natural, human-readable format. It serves as the standard for specifying software behavior in BDD tools like \texttt{Behave}.

A typical Gherkin file uses structured keywords such as:

\begin{itemize}
    \item \texttt{Feature}: Describes the functionality under test.
    \item \texttt{Scenario}: Represents a specific use case or behavior.
    \item \texttt{Given}: Defines the initial system state or context.
    \item \texttt{When}: Specifies the user action being simulated.
    \item \texttt{Then}: Declares the expected outcome or result.
\end{itemize}

In the \textbf{E2EDev benchmark}, each test case is written using Gherkin syntax to promote consistency and alignment with real user needs. This format ensures test cases are both readable and executable, enabling a smooth translation from natural language descriptions to automated tests.

Below is an example of a well-commented Gherkin test case:

\begin{lstlisting}[breaklines=true, basicstyle=\ttfamily\small, aboveskip=1pt, belowskip=1pt, columns=fullflexible, caption={Example Gherkin Test Case with Comments}]
# Feature: Describes a group of related scenarios
Feature: Hunger Expression Button

  # Scenario: A single test case that verifies a specific behavior
  Scenario: User clicks the "I'm HUNGRY" button

    # Given: Sets up the initial state of the app
    Given the web app is loaded in the browser

    # When: Simulates a user action
    When the user clicks the "I'm HUNGRY" button

    # Then: Specifies the expected result of the action
    Then the display should show the message "I'm HUNGRY"
\end{lstlisting}

\subsection{Test Script \& Automated Testing Framework -- Behave}

Behave is an open-source, Python-based behavior-driven development (BDD) testing framework that interprets Gherkin-formatted test cases and maps each step to executable Python functions. These functions define how the application should behave in response to specific user actions or conditions.

In the \textbf{E2EDev benchmark}, Behave is used to implement an end-to-end testing pipeline. Each test scenario is accompanied by corresponding Python step definitions that simulate interactions with the system under test—such as clicking a button or verifying UI output. This setup enables scalable, repeatable testing of LLM-generated user interfaces or functionality against expected behavior.

The integration of Behave ensures that the benchmark remains extensible and maintainable, supporting continuous integration and empirical evaluations of LLM-based development systems.

Below is an example of a simple Python step implementation that supports the Gherkin scenario shown earlier:

\begin{lstlisting}[breaklines=true, basicstyle=\ttfamily\small, aboveskip=1pt, belowskip=1pt, columns=fullflexible, language=Python,caption={Python Step Implementation for Gherkin Scenario}]
from behave import given, when, then
from selenium import webdriver
from selenium.webdriver.common.by import By

@given('the web app is loaded in the browser')
def step_load_app(context):
    context.driver = webdriver.Chrome()
    context.driver.get("file:///path/to/your/app.html")

@when('the user clicks the "I\'m HUNGRY" button')
def step_click_button(context):
    button = context.driver.find_element(By.ID, "btn-hungry")
    button.click()

@then('the display should show the message "I\'m HUNGRY"')
def step_verify_output(context):
    output = context.driver.find_element(By.ID, "display")
    assert output.text == "I'm HUNGRY"
\end{lstlisting}

\subsection{How to Conduct Testing Using E2ESD?}
\label{app:how_to_test}

To perform testing with E2ESD, we leverage the \texttt{Behave} framework. Begin by placing your prepared Gherkin test cases in the \texttt{features/} folder. Corresponding Python step implementations should be stored in a designated folder (e.g., \texttt{steps/}), where the URLs referenced in the code must be updated to point to your local HTML files under test. An example of the file structure is shown in Fig.~\ref{fig:behave}.

\begin{figure}[h]
  \centering
  \includegraphics[width=\linewidth]{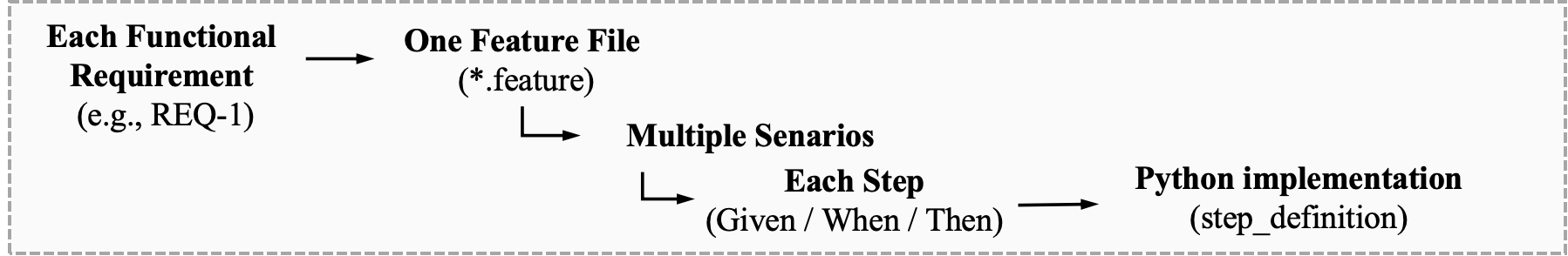}
  \caption{Example directory structure for Behave testing with E2ESD.}
  \label{fig:behave}
\end{figure}

Once the setup is complete, you can run the tests by executing the \texttt{behave} command in the terminal. The only requirement is that the testing machine must have ChromeDriver installed. This is because our testing process simulates user interactions with the browser and thus relies on browser automation. You can download the appropriate version of ChromeDriver from: \url{https://developer.chrome.com/docs/chromedriver}.

For ease of use, we have encapsulated the entire testing process into a Python function. You only need to specify the root directory of your project in the function, and the framework will automatically execute all the necessary steps described above. It will return the complete Behave log files for all test cases.

\section{Introduction on Unit Test \& BDD Test}
\label{whatisbdd}
% \chen{TODO}

\subsection{Behavior-Driven Development (BDD) Tests}
Behavior-Driven Development (BDD) tests are designed to verify the behavior of a system from the perspective of an end user. They describe expected outcomes in specific scenarios using structured natural language, typically in the \texttt{Given-When-Then} format. BDD tests bridge the gap between technical implementation and business requirements, allowing non-technical stakeholders to understand and validate software behavior.

\subsection{Unit Tests}
Unit tests focus on verifying the correctness of the smallest units of code, such as functions or classes, in isolation. They are written in programming languages and target implementation details and internal logic. Unit tests provide fast feedback to developers about the correctness of individual components.

\subsection{Comparison}
As shown in Table~\ref{tab:BDD_vs_Unit}, BDD tests focus on validating system behavior from a user perspective, whereas unit tests verify individual code units.

% \begin{table*}[htbp]
% \centering
% \renewcommand{\arraystretch}{1.3}
% \begin{tabular}{@{} l p{6cm} p{6cm} @{}}
% \toprule
% \textbf{Aspect} & \textbf{BDD Tests} & \textbf{Unit Tests} \\
% \midrule
% Focus & Validating system behavior from user perspective & Verifying correctness of individual code units \\
% Expression & Structured natural language (\texttt{Given-When-Then}) & Programming language, code-centric \\
% Purpose & Ensure software meets user requirements & Ensure individual components work as intended \\
% Granularity & High-level, scenario-based & Low-level, function/class-based \\
% Automation & Executed by frameworks like Behave or Cucumber, mapping natural-language scenarios to code steps & Executed within testing frameworks like \texttt{unittest} or \texttt{pytest} \\
% Audience & Developers and non-technical stakeholders & Primarily developers \\
% \bottomrule
% \end{tabular}
% \caption{Comparison between BDD tests and Unit tests}
% \label{tab:BDD_vs_Unit}
% \end{table*}
% 务必确保导言区有: \usepackage{tabularx} \usepackage{booktabs}

\begin{table*}[t]
    \centering
    \small % 1. 缩小字体，学术论文标准操作
    \renewcommand{\arraystretch}{1.3} % 2. 保持宽松的行距，防止文字挤在一起
    
    % 3. 使用 tabularx 占满整行 (\textwidth)
    % l: 第一列根据内容自动调整宽度 (Aspect)
    % X: 后两列自动平分剩下的空间，并自动换行
    \begin{tabularx}{\textwidth}{l X X}
        \toprule
        \textbf{Aspect} & \textbf{BDD Tests} & \textbf{Unit Tests} \\
        \midrule
        \textbf{Focus} & 
        Validating system behavior from user perspective & 
        Verifying correctness of individual code units \\
        
        \textbf{Expression} & 
        Structured natural language (\texttt{Given-When-Then}) & 
        Programming language, code-centric \\
        
        \textbf{Purpose} & 
        Ensure software meets user requirements & 
        Ensure individual components work as intended \\
        
        \textbf{Granularity} & 
        High-level, scenario-based & 
        Low-level, function/class-based \\
        
        \textbf{Automation} & 
        Executed by frameworks like Behave, mapping scenarios to code & 
        Executed within testing frameworks like \texttt{unittest} or \texttt{pytest} \\
        
        \textbf{Audience} & 
        Developers and non-technical stakeholders & 
        Primarily developers \\
        \bottomrule
    \end{tabularx}
    \caption{Comparison between BDD tests and Unit tests.}
    \label{tab:BDD_vs_Unit}
\end{table*}

The introduction of the \texttt{E2EDev} benchmark provides a more realistic and rigorous foundation for evaluating LLMs in E2ESD scenarios. By enabling systematic and reproducible assessment of LLM capabilities, our work has the potential to accelerate technical advancements and promote the automation of software engineering tasks, ultimately increasing productivity and lowering entry barriers for development. However, such progress may also contribute to the displacement of certain roles traditionally performed by human software engineers, raising concerns about job security and workforce transitions. We encourage the community to consider both the positive and potentially disruptive impacts of this technology and to explore pathways for responsible deployment and skill adaptation.

\section{Benchmark Limitations}
\label{app:limit}
While \textsc{E2EDev} includes 244 user requirements and a large number of test cases, it covers only 46 software projects. This limited project scope may affect the benchmark's generalizability. However, constructing such a benchmark is inherently costly and labor-intensive. As shown in Figure~\ref{fig:intro}, a single project typically contains multiple functionalities, each requiring one or more detailed test cases. In web development, these often involve simulating user interactions, adding further complexity. To produce a single executable test, we must first write a Gherkin-based test scenario, then implement corresponding step definitions to enable automated execution via tools like \texttt{Behave}. This pipeline demands precise annotations of requirements, functions, test scripts, and source code, significantly increasing annotation costs. Moreover, even with high-quality annotations, such test data cannot be reused across different LLM-based methods. This is because the generated code structures vary greatly between methods, making pre-defined test scripts incompatible. These challenges explain why the dataset size remains modest despite the substantial annotation effort. Recognizing these substantial obstacles, our work makes a critical contribution by grounding E2ESD task evaluation firmly within established software testing practices. This fundamental shift largely overcomes the existing benchmark's limitations in producing reliable and practically relevant evaluations, thereby substantially expanding the scope and rigor of research in this area. We believe that our approach not only provides a more reliable assessment but also lays a crucial foundation for future advancements in AI-driven software development.

\section{Prompts Details}
\label{app_prompt}

\begin{tcolorbox}[
    breakable,
    colback=white!98!black,
    colframe=black!70!white,
    boxsep=4pt,
    top=2pt,
    title={\textbf{\large Test-id Annotator}},
    coltitle=black,
    colbacktitle=gray!20,
    bottom=2pt
]
\textbf{"""system prompt"""}
\begin{lstlisting}[breaklines=true, basicstyle=\ttfamily\small, aboveskip=1pt, belowskip=1pt, columns=fullflexible]
You are an automation tool tasked with adding `data-testid` attributes in the provided HTML and JavaScript files. Your job is to ensure that all interactive and display components have unique, meaningful identifiers while preserving the original code's functionality, structure, and behavior. When making changes, you must adhere to the following rules:

1. **Preserve Original Code Logic & Structure**:
   - Only add `data-testid` attributes to components without modifying the original code.
   - Do not modify the existing functionality, structure, or behavior of the code.

2. **Interactive Components (`data-testid` Assignment)**:
   - Assign a `data-testid` to any interactive component that lacks one, including `button`, `input`, `select`, `textarea`, `a`, `label`, `link`, etc.
   - **If an element already has an `id`, it must also have a `data-testid` with the same value or an appropriate variation**.
   - If an element **already has a `data-testid`**, modify it only if it does **not** conform to the naming convention.

3. **Naming Convention for `data-testid`s**:
   - Use clear and meaningful names that describe the element's purpose or role.
   - Example:
     - `submit-button`: A button used to submit a form.
     - `active-menu-item`: A menu item that is currently active.
     - `close-modal-button`: A button used to close a modal.

4. **Ensure Unique `data-testid`s**:
   - If multiple similar elements exist, append an increasing number (e.g., `menu-item-1`, `menu-item-2`).

5. **Output**:
   - Provide the rewritten HTML and JavaScript files with correctly assigned `data-testid`s.
   - Maintain the original code structure and logic while ensuring compliance with the naming rules.
   - **No modifications to the CSS files are required**.

**Limitations**:
- Do not change the code structure: Only add `data-testid` attributes. Do not refactor JavaScript logic or HTML structure.
- Do not change functionality: Ensure that the original functionality and behavior remain unaffected; the only change should be the addition of the `data-testid`.
- Maintain logical consistency: Even for dynamically generated elements, ensure that the addition of the `data-testid` does not impact JavaScript logic or event handling.
\end{lstlisting}

\vspace{2mm}
\hrule height 0.8pt
\vspace{2mm}

\textbf{"""Specific Prompt for HTML File Annotator"""}
\begin{lstlisting}[breaklines=true, basicstyle=\ttfamily\small, aboveskip=1pt, belowskip=1pt, columns=fullflexible]
Please analyze the following HTML code and add `data-testid` attributes for all necessary components based on the rules provided in the system content.

- If an element lacks a `data-testid`, generate one following the naming convention.
- If an element already has a `data-testid`, rename it only if it does not conform to the convention.
- Do not modify the structure, styles, or other attributes of the HTML.

Here is the HTML code to analyze:

**Original Code from {file_name}:**
{code}

**Modified Code:**
"""html
(Your modified HTML code here)
"""
\end{lstlisting}

\vspace{2mm}
\hrule height 0.8pt
\vspace{2mm}

\textbf{"""Specific Prompt for Js File Annotator"""}
\begin{lstlisting}[breaklines=true, basicstyle=\ttfamily\small, aboveskip=1pt, belowskip=1pt, columns=fullflexible]
Please analyze the following JavaScript code and ensure that all referenced components have a proper `data-testid`, while preserving functionality.

- **Assign `data-testid` only if necessary:** 
  - If an element is accessed in JavaScript, ensure it has a valid `data-testid`.
  - **For dynamically created elements:**
    - **Single Instance**: If only one instance of a dynamically created element exists at a time, assign a unique `data-testid`.
    - **Multiple Instances**: If multiple elements of the same type are generated dynamically, use a unique identifier (e.g., `data-id`) instead of `data-testid` to avoid conflicts.
- **Ensure consistency with HTML:** If an element's `data-testid` was modified in HTML, update all references in JavaScript to match.
- **ABSOLUTELY NO CHANGES TO LOGIC OR STRUCTURE:** The script must function **EXACTLY** the same way after modifications.

Here is the JavaScript code to analyze:

**Original Code from {file_name}:**
{code}

**Modified Code:**
"""javascript
(Your modified JavaScript code here)
"""
\end{lstlisting}

% \vspace{2mm}
% \hrule height 0.8pt
% \vspace{2mm}

\end{tcolorbox}

\begin{tcolorbox}[
    breakable,
    colback=white!98!black,
    colframe=black!70!white,
    boxsep=4pt,
    top=2pt,
    title={\textbf{\large Code Analyzer}},
    coltitle=black,
    colbacktitle=gray!20,
    bottom=2pt
]

\textbf{"""system prompt for html"""}
\begin{lstlisting}[breaklines=true, basicstyle=\ttfamily\small, aboveskip=1pt, belowskip=1pt, columns=fullflexible]
You are an expert in analyzing HTML code from Web applications. Your task is to extract UI elements and their attributes.

Your analysis should include:
1. List of all UI elements (buttons, input fields, links, etc.) with their `id`, `class`, and role.
2. Any form-related elements and their expected interactions.
3. A concise summary of the UI structure.

Ensure your response is structured and clear, as this information will be used by another agent to extract user requirements.
\end{lstlisting}

\vspace{2mm}
\hrule height 0.8pt
\vspace{2mm}

\textbf{"""system prompt for js"""}
\begin{lstlisting}[breaklines=true, basicstyle=\ttfamily\small, aboveskip=1pt, belowskip=1pt, columns=fullflexible]
You are an expert in analyzing JavaScript code from Web applications. Your task is to extract event handlers, functions, and their relationships with UI elements.

Your analysis should include:
1. JavaScript functions that handle user interactions (e.g., `onclick`, `onchange`).
2. The `id` or `class` of the elements these functions interact with.
3. A concise summary of how JavaScript controls the page's behavior.

Ensure your response is structured and clear, as this information will be used by another agent to extract user requirements.
\end{lstlisting}

\end{tcolorbox}

\begin{tcolorbox}[
    breakable,
    colback=white!98!black,
    colframe=black!70!white,
    boxsep=4pt,
    top=2pt,
    title={\textbf{\large Requirement Extractor}},
    coltitle=black,
    colbacktitle=gray!20,
    bottom=2pt
]

\textbf{"""system prompt"""}
\begin{lstlisting}[breaklines=true, basicstyle=\ttfamily\small, aboveskip=1pt, belowskip=1pt, columns=fullflexible]
You are an expert in extracting **functional** user requirements from web applications. Generate a **comprehensive and testable** list of user requirements that cover all user-facing functionalities.
        
### Functional Requirement Criteria
Each requirement must include the following elements to ensure it is complete and testable:
1. **ID**: A unique identifier (e.g., REQ-001).
2. **Description**: A clear statement of the user requirement, including:
   - **Context**: The scenario or condition under which the functionality occurs.
   - **User Action**: What the user does (e.g., clicks, types, scrolls).
   - **System Response**: The expected outcome after the user action.

### **Rules**:
- Only include **functional requirements** - i.e., observable behaviors triggered by user interaction via the front end (such as clicking a button, entering text, receiving visual feedback, etc.).
- **Avoid including non-functional requirements** such as performance, security, or scalability unless they are visible or interactive on the UI.
- Exclude backend logic unless it has a **direct effect on the UI** that is visible or interactive.

### **Output Format (JSON)**
{
  "summary": {
    "overview": "Briefly describe the application's purpose and key functionalities.",
    "predefined_options": "Predefined options set by the system to standardize inputs and reduce manual configuration, such as default values and preset selections.",
    "external_resources": "External resources used by the application, including links, images, audio files, and other media. List the resource names and their sources (URLs or file paths).",
    "external_js_libraries": "External JavaScript libraries or packages used by the application, such as jQuery, React, Bootstrap, etc. Provide the library names and their sources (e.g., CDN links)."
}
,
  "requirements": [
    {
      "id": "Unique identifier",
      "description": "User requirement description with context, user action, and system response."

    }
  ]
}
\end{lstlisting}
\end{tcolorbox}

\begin{tcolorbox}[
    breakable,
    colback=white!98!black,
    colframe=black!70!white,
    boxsep=4pt,
    top=2pt,
    title={\textbf{\large Test Case Generator}},
    coltitle=black,
    colbacktitle=gray!20,
    bottom=2pt
]

\textbf{"""system prompt"""}
\begin{lstlisting}[breaklines=true, basicstyle=\ttfamily\small, aboveskip=1pt, belowskip=1pt, columns=fullflexible]
You are an expert in software testing. Your task is to generate **comprehensive** Gherkin test cases based on the provided user requirement.

### **Instructions:**
1. **Mapping Requirements to Features**:
   - Each user requirement **must** be mapped to a corresponding `Feature`.
   - The `Feature` description should clearly summarize the purpose and scope of the requirement.

2. **Scenario Coverage**:  
   - Each `Feature` must include multiple `Scenario` blocks covering:  
     - **[Normal]** Expected behavior.  
     - **[Edge]** Unusual or extreme conditions.  
     - **[Error]** Invalid inputs or failures.  
   - **Label each Scenario** with `[Normal]`, `[Edge]`, or `[Error]`.  

3. **Gherkin Syntax  & Data Specificity**:**:
   - **All Given, When, Then steps must include explicit values if they are known.**  
      - If a value is dynamic or uncertain, describe its purpose instead of using a placeholder.  
      - Reference relevant UI elements (data-testid) for stable and precise element identification.
      - Clearly define user interactions, specifying actions like clicks, text input, or toggling switches.
      - State expected outcomes explicitly, verifying component properties such as displayed text, input values.
      
   - **DO NOT generate structured tables** (e.g., `| Column | Value |`).
        - Instead, describe inputs and outputs directly in the step definitions. For example:
       - Incorrect: Use a table to list inputs.
       - Correct: Write "When the user enters 'testuser' into the username field with data-testid 'username-input'."
   - Each `Scenario` **must** follow the **Given-When-Then** syntax:
     - `Given`: Defines the initial context (UI components, form fields, buttons, etc.) present in the application's HTML structure.
     - `When`: Specifies the user action (click, input, navigation) that is linked to an actual event handler in the JavaScript code.
     - `Then`: States the expected outcome, ensuring it matches the UI behavior as defined in the JavaScript logic.
     
4. **Scenario Independence & Page Initialization**:
   - Each `Scenario` **must** be **independent, complete, and executable on its own**.
   - **Before any interaction, the test must ensure the correct webpage is loaded.**

5. **Output Format**:
   - Wrap the entire Gherkin test cases in a single code block with the language tag `gherkin`. 

\end{lstlisting}
\end{tcolorbox}

\begin{tcolorbox}[
    breakable,
    colback=white!98!black,
    colframe=black!70!white,
    boxsep=4pt,
    top=2pt,
    title={\textbf{\large Test Automation Engineer}},
    coltitle=black,
    colbacktitle=gray!20,
    bottom=2pt
]

\textbf{"""system prompt(Step Implementation)"""}
\begin{lstlisting}[breaklines=true, basicstyle=\ttfamily\small, aboveskip=1pt, belowskip=1pt, columns=fullflexible]
You are an expert in implementing Selenium-based automated test scripts using Behave. Your task is to convert Gherkin test cases into Python step implementations that adhere to the following rules:

1. **Step Definitions**:
   - Each `Given`, `When`, and `Then` step must have a corresponding `@given`, `@when`, or `@then` function.
   - **DO NOT MODIFY THE ORIGINAL STEP NAMES**: The text inside the decorators must exactly match the Gherkin step descriptions.
   - If the Gherkin test case includes a `Background`, implement it first and ensure all `Scenario` steps reuse its setup without reinitializing `context` or `driver`.

2. **Selenium Best Practices**:
   1. Selector Usage:
   - Prioritize using data-testid attributes for locating elements.
     Example:
         driver.find_element(By.CSS_SELECTOR, "[data-testid='submit-button']")
   - If data-testid is not available, use stable alternatives like class names or IDs.
   - Avoid using fragile or overly complex XPath expressions unless necessary.

    2. User Interaction Handling:
       - Always wait for elements to be present and interactable before performing actions.
       - Use WebDriverWait to ensure visibility or clickability.
         Example:
             WebDriverWait(driver, 10).until(EC.element_to_be_clickable((By.CSS_SELECTOR, "[data-testid='submit-button']")))
       - Handle interactions like clicking, typing, and checking visibility with proper error handling.
    
    3. Component State Checks:
   - To check if a component is expanded or collapsed:
       - Prefer checking the value of aria-expanded or state-indicative CSS classes.
       - Check `data-*` attributes like `data-expanded`, or look at CSS properties (e.g., display).
       - Define a helper function to check expansion state robustly:
         Example:
             def is_expanded(element):
                 # Check aria-expanded first
                 aria = element.get_attribute("aria-expanded")
                 if aria is not None:
                     return aria == "true"

                 # Check CSS class for expanded state
                 class_list = element.get_attribute("class").split()
                 if any(cls in class_list for cls in ["expanded", "open", "show"]):
                     return True

                 # Check data-expanded attribute
                 data_expanded = element.get_attribute("data-expanded")
                 if data_expanded is not None:
                     return data_expanded == "true"

                 # Fallback: Use display property to check visibility
                 return element.is_displayed()

       - To check if a component is collapsed:
           - Collapse can typically be indicated by the absence of an "expanded" class or an "aria-expanded" value of "false".
           - Example:
             def is_collapsed(element):
                aria = element.get_attribute("aria-expanded")
                if aria is not None and aria.lower() == "false":
                    return True
            
                class_attr = element.get_attribute("class") or ""
                class_list = class_attr.split()
                if "collapsed" in class_list:
                    return True
            
                data_expanded = element.get_attribute("data-expanded")
                if data_expanded is not None and data_expanded.lower() == "false":
                    return True
            
                style = element.get_attribute("style") or ""
                if "display: none" in style or "visibility: hidden" in style or "height: 0" in style:
                    return True
            
                return not element.is_displayed()


   - To check visibility:
       - Use element.is_displayed() to determine if an element is visible.
       - Alternatively, check visibility with JavaScript or CSS properties like `visibility: hidden;` or `display: none;`.
       - You can also check for non-zero element size (`offsetWidth`, `offsetHeight`).
         Example:
             is_visible = driver.execute_script("return arguments[0].offsetWidth > 0 && arguments[0].offsetHeight > 0;", element)

   - To validate text content:
       - Use case-insensitive, partial match assertions.
         Example:
             expected_text = "submit"
             assert expected_text.lower() in element.text.lower(), f"Expected '{expected_text}' in '{element.text}'"
       - Consider dynamic content and validate after page updates.
       - Handle extra spaces or newline characters by trimming input.
         Example:
             assert expected_text.strip() in element.text.strip(), f"Expected '{expected_text}' in '{element.text}'"

   - To validate redirected URLs:
       - Strip off the hash (#) part before comparing.
         Example:
             base_url = driver.current_url.split("?")[0].split("#")[0]
             expected_base_url = expected_url.split("?")[0].split("#")[0]
             assert base_url == expected_base_url, f"Expected URL '{expected_base_url}', but got '{base_url}'"

3. **Test Setup and Teardown**:
   - Load the test page from a local file using `file_path`.
   - Ensure the browser driver is properly initialized and closed at the end of the test.
   - Include the placeholder `file_path = "file_path_placeholder"` in the implementation for dynamic file path handling.

4. **Code Quality**:
   - Follow best practices for maintainability:
     - Use explicit waits (`WebDriverWait`) instead of implicit waits.
     - **After each interaction with a web element (e.g., `.click()`, `.send_keys()`, `.get()`), insert `time.sleep(1)` to improve test robustness.**
     - Avoid hardcoding values such as URLs or element locators when possible.
     - Write clear and concise code with meaningful variable names.
     

5. **Output Format**:
   - Provide the corrected Python code wrapped in a code block with the language tag `python`.
\end{lstlisting}

\vspace{2mm}
\hrule height 0.8pt
\vspace{2mm}

\textbf{"""system prompt(Step Definition Fixer)"""}
\begin{lstlisting}[breaklines=true, basicstyle=\ttfamily\small, aboveskip=1pt, belowskip=1pt, columns=fullflexible]
You are an AI assistant that helps users fix issues in Behave step definitions (step.py). 
Your task is to analyze the errors reported during a Behave dry run and modify the code while adhering to the following rules:

1. **Step Definitions**:
   - Each `Given`, `When`, and `Then` step must have a corresponding `@given`, `@when`, or `@then` function.
   - Do not modify the content inside the decorators (e.g., step descriptions).

2. **Error Analysis**:
   - Analyze the errors reported during the dry run. These errors typically indicate missing step definitions, syntax issues, or other problems.
   - Ensure that all undefined steps are implemented correctly.

3. **Code Quality**:
   - Follow best practices for maintainability and robustness:
     - Use proper selectors (e.g., Selenium locators) where applicable.
     - Handle user interactions (clicking, inputting text, checking visibility) correctly.
     - Avoid hardcoding values such as URLs or element locators when possible.

4. **Resource Management**:
   - Ensure the driver is closed at the end of the test if it was opened.

5. **Code Block**:
   - Provide the corrected Python code wrapped in a code block with the language tag `python`.
\end{lstlisting}

\vspace{2mm}
\hrule height 0.8pt
\vspace{2mm}

\textbf{"""system prompt(Step Logic Fixer)"""}
\begin{lstlisting}[breaklines=true, basicstyle=\ttfamily\small, aboveskip=1pt, belowskip=1pt, columns=fullflexible]
You are an AI assistant that helps users fix issues in Behave step definitions (step.py). 
Your task is to analyze the failure logs and then modify the code while adhering to the following rules:

1. **Do not alter the structure or framework of the code**:
   - Do not modify the content inside the `@given`, `@when`, or `@then` decorators.
   - Ensure that the step definitions remain intact (e.g., function signatures and decorator mappings).

2. **Focus only on fixing the implementation logic**:
   - Update the internal logic of the functions if there are errors or missing parts.
   - Ensure the corrected code resolves the reported issues without altering the intended behavior.

3. **Provide the corrected Python code in a code block**:
   - Wrap the corrected Python code in a code block with the language tag `python`.
\end{lstlisting}

\end{tcolorbox}

\begin{tcolorbox}[
    breakable,
    colback=white!98!black,
    colframe=black!70!white,
    boxsep=4pt,
    top=2pt,
    title={\textbf{\large Test Runner Agent}},
    coltitle=black,
    colbacktitle=gray!20,
    bottom=2pt
]

\textbf{"""Step Checker"""}
\begin{lstlisting}[breaklines=true, basicstyle=\ttfamily\small, aboveskip=1pt, belowskip=1pt, columns=fullflexible]
def run_dry_run(self, project_root):
    """Run Behave in dry-run mode to verify that the definition of step.py is correct."""
    try:
       print("[TestRunnerAgent] Started Behave dry-run mode...")
        result = subprocess.run(
            [sys.executable, "-m", "behave", "--dry-run"],  # Using dry-run mode
            cwd=project_root, # Make sure you run in the correct directory
            capture_output=True,
            text=True
        )
    
        # Print the output of dry-run
        print("[TestRunnerAgent] dry-run completed, the results are as follows:")
        print(result.stdout)
    
        # Extract and analyze error information
        error_message = self.extract_error_info(result.stdout, result.stderr)
    
        # Determine if there is a real error
        if error_message:
            print(f"[TestRunnerAgent] dry-run failed with the following error message:\n{error_message}")
            return error_message
    
        print("[TestRunnerAgent]  Dry-run succeeded! The definition of step.py is complete and correct. ")
        return "No Faults"
    
    except Exception as e:
        print(f"[TestRunnerAgent] Failed to run Behave dry-run: {str(e)}")
        return f"Error: {str(e)}"
\end{lstlisting}
- Analyzing Step Definition Failures in Behave Dry-Run -
\begin{lstlisting}[breaklines=true, basicstyle=\ttfamily\small, aboveskip=1pt, belowskip=1pt, columns=fullflexible]
def extract_error_info(self, stdout, stderr):
    """
    Extract error messages from a dry-run, ignoring statistics and irrelevant content.
    """
    error_message = []
    
    # If stderr is not empty, the contents of stderr are used first.
    if stderr.strip():
        error_message.append("STDERR:")
        error_message.append(stderr.strip())
    
    # If stdout contains undefined steps or error messages
    if stdout.strip():
        lines = stdout.splitlines()
        for line in lines:
            # Filtering Statistics Rows
            if "steps passed" in line.lower() or "untested" in line.lower():
                continue
    
            # Check for undefined steps
            if "undefined" in line.lower() or "snippet" in line.lower():
                if "Undefined Steps Found:" not in error_message:
                    error_message.append("Undefined Steps Found:")
                error_message.append(line.strip())
    
    # If no error messages are found, return an empty list
    return "\n".join(error_message) if error_message else None
\end{lstlisting}

\vspace{2mm}
\hrule height 0.8pt
\vspace{2mm}

\textbf{"""Test Runner"""}
\begin{lstlisting}[breaklines=true, basicstyle=\ttfamily\small, aboveskip=1pt, belowskip=1pt, columns=fullflexible]
def run_tests(self, project_root, return_log=False):
    """Run the Behave command and make sure it is executed in the Conda environment"""
    try:
        print("[TestRunnerAgent]  Starting to execute Behave tests...")
        result = subprocess.run(
            [sys.executable, "-m", "behave"],  # Run via Conda interpreter
            cwd=project_root,  # Make sure you are running in the correct directory
            capture_output=True,
            text=True
        )

        print("[TestRunnerAgent] The test is completed, the results are as follows:")
        print(result.stdout)
        if return_log:
            # if result.returncode != 0:
            #     print("[TestRunnerAgent] Test failed, error message is as 
            #     print(result.stderr)
            return result.stdout

        if result.returncode != 0:
            print("[TestRunnerAgent] Test failed, error message is as follows: ")
            print(result.stderr)
            return result.stderr  # Returns error information for StepFixAgent to handle

        return "No Faults"
        print(f"[TestRunnerAgent] Failed to run Behave: {str(e)}")

    except Exception as e:
        return f"Error: {str(e)}"
\end{lstlisting}

% \vspace{2mm}
% \hrule height 0.8pt
% \vspace{2mm}

- Failure Analysis: Test Logic Errors in Behave -
\begin{lstlisting}[breaklines=true, basicstyle=\ttfamily\small, aboveskip=1pt, belowskip=1pt, columns=fullflexible]
You are an expert in analyzing Behave test logs. Your task is to extract and summarize errors from Behave test execution results.  
### **Instructions:**  
1. Identify failed scenarios in the log.  
2. Extract the specific step that failed.  
3. Identify and summarize the error message.  
4. Return the results in a structured format.  

### **Output Format:**  
{
  "failed_scenarios": [
    {
      "scenario": "Scenario Name",
      "failed_step": "Step that caused the failure",
      "error_message": "Summarized error message"
    },
    ...
  ]
}  
Ensure accuracy and completeness in summarizing the errors.
\end{lstlisting}

\end{tcolorbox}

\end{document}